\documentclass[11pt,a4paper]{article}
\pdfoutput=1
\usepackage{jheppub}
\usepackage{amsthm,amsbsy,amsfonts,mathrsfs,enumerate,float,wrapfig,amsmath}
\usepackage{subfigure}
\newcommand{\M}{\mathcal{M}}
\newcommand{\N}{\mathcal{N}}
\newcommand{\R}{\mathcal{R}}
\newcommand{\nn}{\nonumber}
\newcommand{\be}{\begin{equation}}
\newcommand{\ee}{\end{equation}}
\newcommand{\bea}{\begin{eqnarray}}
\newcommand{\eea}{\end{eqnarray}}

\title{
5-brane webs for 5d $\mathcal{N}=1$ $G_2$ gauge theories}
\author[a]{Hirotaka Hayashi,}
\author[b,c]{Sung-Soo Kim,}
\author[d]{Kimyeong Lee,}
\author[e]{and Futoshi Yagi}

\affiliation[a]{Department of Physics, School of Science, Tokai University,\\ 4-1-1 Kitakaname, Hiratsuka-shi, Kanagawa 259-1292, Japan}
\affiliation[b]{School of Physics, University of Electronic Science and Technology of China, \\North Jianshe Road, Chengdu 611731, China}
\affiliation[c]{Institute of Fundamental and Frontier Sciences, \\University of Electronic Science and Technology of China, Chengdu 611731, China}
\affiliation[d]{School of Physics, Korea Institute for Advanced Study, \\
85 Hoegi-ro Dongdaemun-gu, Seoul 02455, Korea}
\affiliation[e]{Department of Physics, Technion - Israel Institute of Technology,\\ Haifa 32000, Israel}

\emailAdd{h.hayashi@tokai.ac.jp}
\emailAdd{sungsoo.kim@uestc.edu.cn}
\emailAdd{klee@kias.re.kr}
\emailAdd{fyagi@physics.technion.ac.il}

\abstract{
We propose 5-brane webs for 5d $\mathcal{N}=1$ $G_2$ gauge theories. From a Higgsing of the $SO(7)$ gauge theory with a hypermultiplet in the spinor representation, we construct two types of 5-brane web configurations for the pure $G_2$ gauge theory using an O5-plane or an $\widetilde{\text{O5}}$-plane. Adding flavors to the 5-brane web for the pure $G_2$ gauge theory is also discussed. Based on the obtained 5-brane webs, we compute the partition functions for the 5d $G_2$ gauge theories using the recently suggested topological vertex formulation with an O5-plane, and we find agreement with known results.
}

\begin{document}
\preprint{
\begin{flushright}
\tt 
%preprint number
KIAS-P18002\\
\end{flushright}
}

\maketitle
%=================================================================

%=================================================================

\section{Introduction}
5-brane webs in type IIB string theory have been used to study five-dimensional (5d) superconformal field theories (SCFTs) that are ultraviolet (UV) completions of a certain class of 5d $\mathcal{N}=1$ supersymmetric gauge theories \cite{Seiberg:1996bd,Morrison:1996xf,Douglas:1996xp,Intriligator:1997pq, Aharony:1997ju, Aharony:1997bh}. A 5-brane web configuration provides us with a tool to compute the instanton partition function that captures the BPS spectrum of a 5d theory realized on a 5-brane web as well as with a perspective of qualitative understandings of the SCFTs such as %existence of non-Lagrangian theories, 
global symmetry enhancements and various dualities. %S-duality or base-fiber duality. 

By introducing an orientifold plane like an O7-plane or an O5-plane, %or an ON$^0$-planes, 
5-brane webs can be enriched so that one can describe 5d theories with some other classical gauge group, such as $SO(N), USp(2N)$ \cite{Brunner:1997gk, Bergman:2015dpa, Hayashi:2015vhy, Zafrir:2015ftn}, in addition to the standard classical gauge group $SU(N)$. In recent years there has been some progress on brane configurations with the orientifold planes. For example, in \cite{Hayashi:2015zka,Hayashi:2016abm,Yun:2016yzw}, whether we resolve an O7$^-$-plane into two $[p ,q]$ 7-branes or not in a certain 5-brane web configuration gives an explanation for equivalence proposed in \cite{Gaiotto:2015una} of two theories at the UV fixed point, an $SU(N+1)$ theory with $N_f \le 2N+6$ hypermultiplets in the fundamental representation (flavors) and the Chern-Simons (CS) level $\kappa=\pm\left(N+3-N_f/2\right)$, and an $USp(2N)$ theory with the same number of flavors. In particular, it has been discussed in \cite{Zafrir:2015ftn, Zafrir:2016jpu} that 5-brane configurations with an O5-plane can realize an $SO(N)$ gauge theory even with hypermultiplets in the spinor/conjugate spinor representation. It is also noticeable that one can differentiate the discrete theta angle ($\theta=0,\pi$) of the 5d pure $USp(2N)$ gauge theory \cite{Morrison:1996xf,Douglas:1996xp,Intriligator:1997pq} from 5-brane webs with an O5-plane. The two different theta angles turn out to imply two distinct phase structures for their 5-brane webs, that are characterized by two distinct ``generalized" flop transitions which may be applied to 5-branes intersecting at the same point with an O5-plane \cite{Hayashi:2017btw}. 

There has been also progress along a quantitative side on 5-brane webs with an O5-plane. The conventional topological vertex formalism \cite{Aganagic:2003db, Iqbal:2007ii,Awata:2008ed} enables one to systematically compute the Nekrasov instanton partition function of a 5d theory on a 5-brane web via the correspondence~\cite{Leung:1997tw} between a toric diagram and a certain $(p, q)$ 5-brane web diagram. Although 5-brane webs for $SU(N)$ gauge theories with a large number of flavors or a large CS level often lead to non-toric Calabi-Yau geometries \cite{Benini:2009gi}, the topological vertex formulation is still applicable to reproduce the correct partition function \cite{Hayashi:2013qwa,Hayashi:2014wfa, Hayashi:2015xla, Kim:2015jba, Hayashi:2016abm, Hayashi:2016jak}. %Until recently, however it had not been known how to implement the topological vertex formulation to a 5-brane web with an O5-plane. 
Quite recently, %it was proposed in \cite{Kim:2017jqn} that 
the topological vertex formulation has been further extended to 5-brane webs with an O5-plane \cite{Kim:2017jqn}. Together with a generalized flop transition, %it enables one to employ the %topological vertex 
the new method %to compute the partition functions for $USp(2N)$ gauge theories, where
utilizes a configuration where one-half of the original brane configuration is glued to the other half from the mirror image due to an O5-plane in a specific manner.

The purpose of this paper is to further extend the study of 5-brane webs with an O5-plane and propose 5-brane web diagrams for 5d $\mathcal{N}=1$ gauge theories of an exceptional gauge group $G_2$, using these recent developments on 5-brane webs. We then compute their Nekrasov partition functions based on the topological vertex formalism for 5-brane webs with an O5-plane. 
%
%The purpose of this paper is to further extend the study of 5-brane webs with an O5-plane and propose 5-brane web diagrams for 5d $\mathcal{N}=1$ gauge theories with an exceptional gauge group $G_2$, using these recent developments on 5-brane webs. We then compute their Nekrasov partition functions from the topological vertex formalism for 5-brane webs with an O5-plane. 
%extend 5-brane configurations to study 5d $\mathcal{N}=1$ theory of an exceptional gauge group $G_2$, using these recent developments on 5-brane web. In particular, we propose 5-brane webs for $G_2$ gauge theories. 

%Our strategy is as follows: %As inspired by the brane construction \cite{Zafrir:2015ftn} %for the pure $SO(2N+1)$ gauge theory obtained from the $SO(2N+2)$ gauge theory with one flavor by going to the Higgs branch, 
Our strategy is as follows: A 5-brane web diagram for the $SO(7)$ gauge theory with a hypermultiplet in the spinor representation has been constructed in \cite{Zafrir:2015ftn}. We then consider the Higgs branch of the $SO(7)$ gauge theory with one spinor in terms of the web diagram, which should yield a 5-brane web configuration for the pure $G_2$ gauge theory\footnote{We thank Gabi Zafrir for illuminating discussion about this strategy.}. 

We note that there are two ways to obtain the web diagram for the $SO(7)$ gauge theory with one spinor. One way is to Higgs the $SO(8)$ gauge theory with a hypermultiplet in the vector representation and a hypermultiplet in the spinor or conjugate spinor representation. The other way is to Higgs the $SO(8)$ gauge theory with a hypermultiplet in the spinor representation and a hypermultiplet in the conjugate spinor representation. These two $SO(8)$ gauge theories should be equivalent to each other due to the triality of $SO(8)$, and both Higgsings hence give rises to the $SO(7)$ gauge theory with one spinor, while the resulting brane configurations look different. Further Higgsing of the two types of the diagrams leads to two different 5-brane webs for the pure $G_2$ gauge theory, which therefore gives two different configuration for the same $G_2$ gauge theory. It is possible to add flavors to the pure $G_2$ gauge theory by Higgsing the $SO(7)$ gauge theory with hypermultiplets either in the vector representation or the spinor representation in addition to one spinor. We test our proposal by comparing the area of compact faces that a D3-brane wraps on the 5-brane webs with the tension of a monopole string which can be calculated from the effective prepotential of the theory in question. %, which suggests how D3-brane connects the faces 5-branes in the mirror part due to an $O5$-plane. 
In fact the analysis implies an interesting feature like which faces of a 5-brane web a D3-brane wraps in the presence of an O5-plane.

We then go on to compute the Nekrasov partition function for 5d $\mathcal{N}=1$ $G_2$ gauge theories by applying the recently proposed topological vertex method for 5-brane webs with an O5-plane \cite{Kim:2017jqn}. We check that our partition function for the pure $G_2$ gauge theory reproduces the one-instanton result \cite{Benvenuti:2010pq, Keller:2011ek} and also the two-instanton result \cite{Hanany:2012dm, Keller:2012da, Cremonesi:2014xha}. We also show that the partition function of the $G_2$ gauge theory with one flavor is consistent with flavor decoupling. 

%We then go on to compute the Nekrasov partition function for 5d $\mathcal{N}=1$ $G_2$ gauge theories by applying the recently proposed topological vertex method for 5-brane webs with an O5-plane \cite{Kim:2017jqn}. We will check that our partition function for the pure $G_2$ gauge theory reproduces the one-instanton result \cite{Benvenuti:2010pq, Keller:2011ek} and also the two-instanton result \cite{Hanany:2012dm, Keller:2012da}. We will also see that the partition function of the $G_2$ gauge theory with one flavor is consistent with flavor decoupling. 
%Our partition function passes consistency checks such as the blow up equation \cite{Keller:2012da} and flavor decouplings. % Adding flavor to our brane construction for the $G_2$ theory is straightforward and agrees with the maximal number of flavors to a $G_2$ theory, discussed in then earlier analysis based on instanton operators \cite{Zafrir:2015uaa} and the prepotential \cite{Jefferson:2017ahm}. 

The paper is organized as follows: In section \ref{sec:G2fromO5tilde}, we first discuss a 5-brane web for the $SO(7)$ gauge theory with a hypermultiplet in the spinor representation. %The brane configuration involves an $\widetilde{\text{O5}}$-plane. 
By Higgsing the $SO(7)$ theory with one spinor, we propose a 5-brane web for the pure $G_2$ gauge theory. We check that the proposed diagram is consistent with the effective prepotential of the pure $G_2$ gauge theory. We also present two ways to introduce flavors. 
In section \ref{sec:G2fromO5}, using the triality of $SO(8)$ gauge theory among hypermultiplets in the vector, spinor, and conjugate spinor representations, we propose another 5-brane web for the pure $G_2$ theory through successive Higgsings of the $SO(8)$ gauge theory theory with one spinor and one conjugate spinor. %a hypermultiplet in the spinor representation and a hypermultiplet in the conjugate spinor representation. %We check also that the new diagram is consitent with the analysis of the monopole string tension.  
In section \ref{sec:Nekrasov}, we first review a recent proposal for the topological vertex formulation with an O5-plane and extend it to the cases with an $\widetilde{\text{O5}}$-plane. We then use it to compute the partition functions of 5d $\mathcal{N}=1$ $G_2$ gauge theories with no flavor and with one flavor. %The partition functions for the $G_2$ theory with no flavor and with one flavor are explicitly computed, which are tested on several consistency checks. 
In section \ref{sec:conclusion}, we summarize the results and comment on further directions. \\

{\bf Note added:} We are informed that the authors of \cite{Kim:2018gjo} computed the partition function for 5d $\mathcal{N}=1$ $G_2$ gauge theories using the ADHM-like method, which will appear in arXiv.

%====================================================================
\bigskip

\section{$G_2$ gauge theories from an $\widetilde{\text{O5}}$-plane}\label{sec:G2fromO5tilde}

In string theory, 
a wide class of
5d theories with eight supercharges can be constructed by $(p, q)$ 5-brane webs in type IIB string theory \cite{Aharony:1997ju, Aharony:1997bh, Leung:1997tw} or M-theory on Calabi-Yau threefolds \cite{Morrison:1996xf, Douglas:1996xp, Intriligator:1997pq}. We will make use of the 5-brane web description for constructing 5d $\mathcal{N}=1$ gauge theories in this paper. 
%The main claim of this paper is to propose a general way to compute the Seiberg-Witten curves and also the partition functions of 5d theories realized by 5-brane web diagrams with an O5/$\widetilde{\text{O5}}$-plane. An interesting application of the method is 5d $G_2$ gauge theories. In fact, it turns out that it is possible to realize the $G_2$ theories with flavors by using 5-brane web diagrams with an O5-plane or an $\widetilde{\text{O5}}$-plane. Hence we first obtain 5-brane web diagrams which realize 5d $G_2$ gauge theories with flavors in this section. 
In this section, we 
present 5-brane web diagrams which realize 5d $G_2$ gauge theories by using an $\widetilde{\text{O5}}$-plane. Although D5-branes on top of an O5-plane or an $\widetilde{\text{O5}}$-plane usually generate an $SO/USp$ gauge theory, we will argue that some simple 5-brane web diagram with an $\widetilde{\text{O5}}$-plane can yield a 5d $G_2$ gauge theory in an intriguing way.

\subsection{$SO(7)$ gauge theory with spinor matter}
\label{sec:SO7Spinor}

Before constructing 5-brane webs for $G_2$ gauge theories, 
%we first start from a 5-brane web realization of $SO(7)$ gauge theories using an $\widetilde{\text{O5}}$-plane. 
we first discuss 5-brane web realization of $SO(7)$ gauge theories using an $\widetilde{\text{O5}}$-plane 
%in section \ref{sec:SO7Spinor}.
in this subsection.
By using this construction, 
we will see in section \ref{sec:HiggstoG2} 
that a Higgsing of the 5-brane web diagrams of the $SO(7)$ gauge theories with 
a hypermultiplet in spinor representation
can generate 5-brane webs for the pure $G_2$ gauge theory. 

A 5-brane web diagram for the pure $SO(7)$ gauge theory can be realized using an $\widetilde{\text{O5}}$-plane. 
Naively, a 5-brane web with an $\widetilde{\text{O5}}$-plane 
%is 
may look problematic since the difference of %the D5-brane 
the RR charge between an $\widetilde{\text{O5}}^+$-plane and an $\widetilde{\text{O5}}^-$-plane is fractional, 
which implies the appearance of $(p,q)$ 5-branes with non-integer $p$.
A way out is that an $\widetilde{\text{O5}}$-plane may be thought of as an O5-plane with a half monodromy branch cut associated to a half D7-brane \cite{Zafrir:2015ftn}. 
Namely, an effective description of an $\widetilde{\text{O5}}^-$-plane is an O5$^-$-plane and a half D5-brane plus the half monodromy cut. An $\widetilde{\text{O5}}^+$-plane is also effectively described by an O5$^+$-plane plus the half monodromy cut. 
Since the monodromy created by this cut is half of the original monodromy associated to one full D7-brane,
it changes the potential fractional charge to integer charge.
%and the problem of 
%the jump by the fractional D5-brane charge is resolved. 
%the $(p,q)$ 5-branes with an fractional charge is resolved. 

\begin{figure}
\centering
\subfigure[]{
\includegraphics[width=7cm]{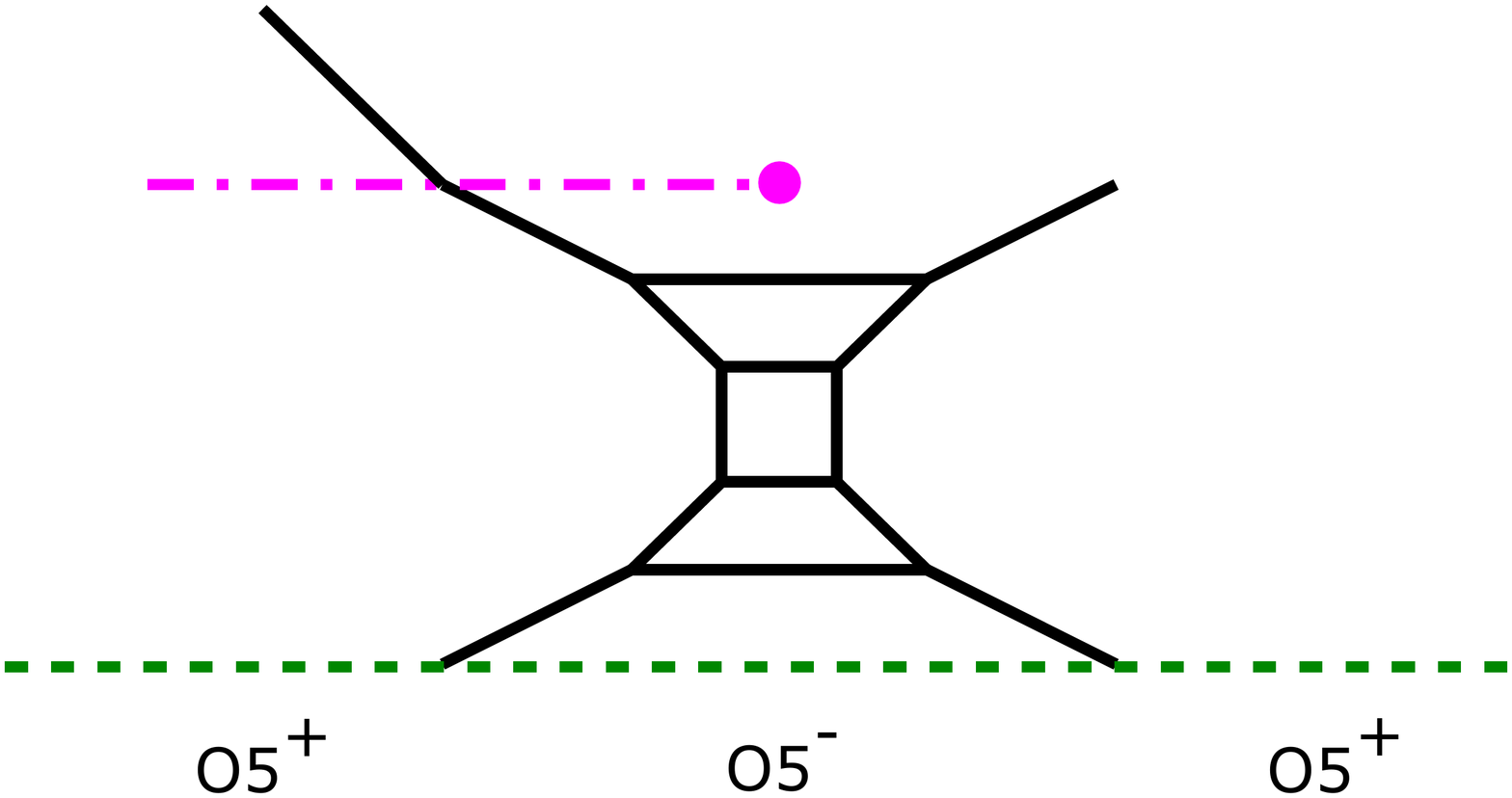}\label{fig:SO8w1flvrHiggsa}}
\subfigure[]{
\includegraphics[width=7cm]{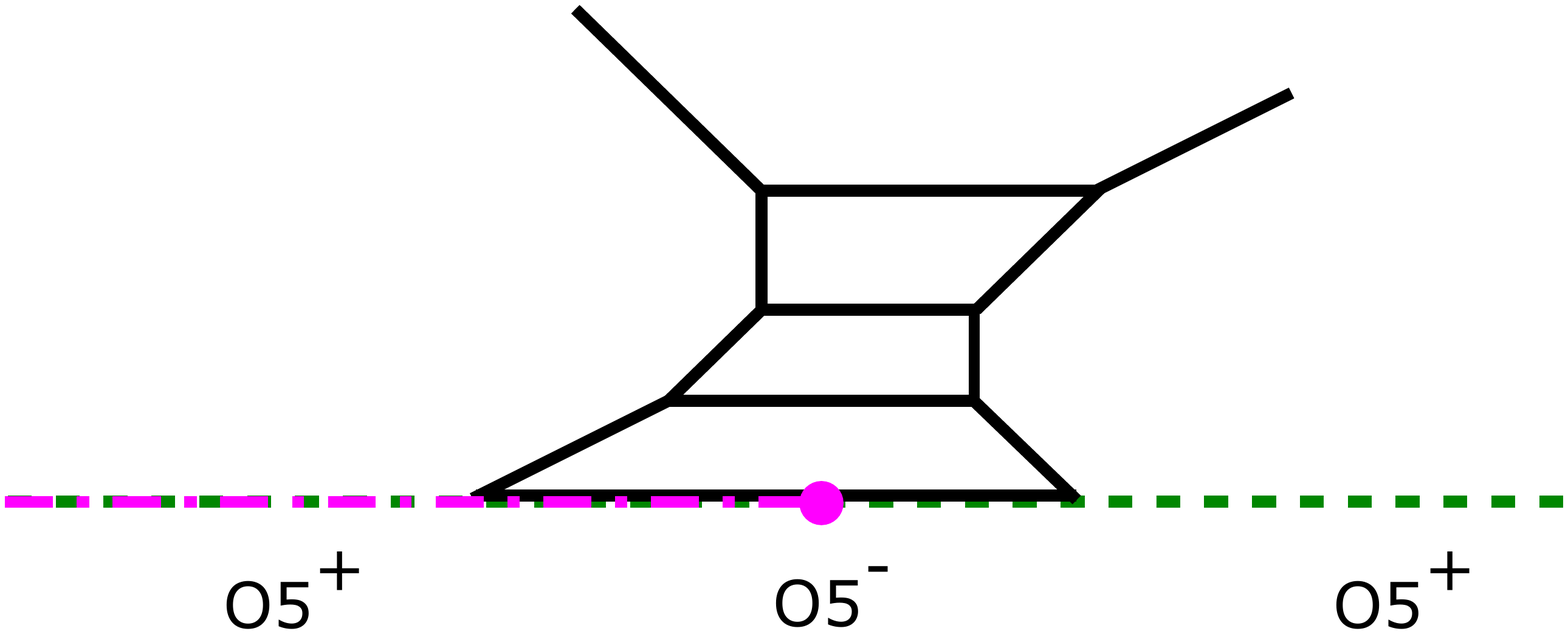} \label{fig:SO8w1flvrHiggsb}}
\subfigure[]{
\includegraphics[width=7cm]{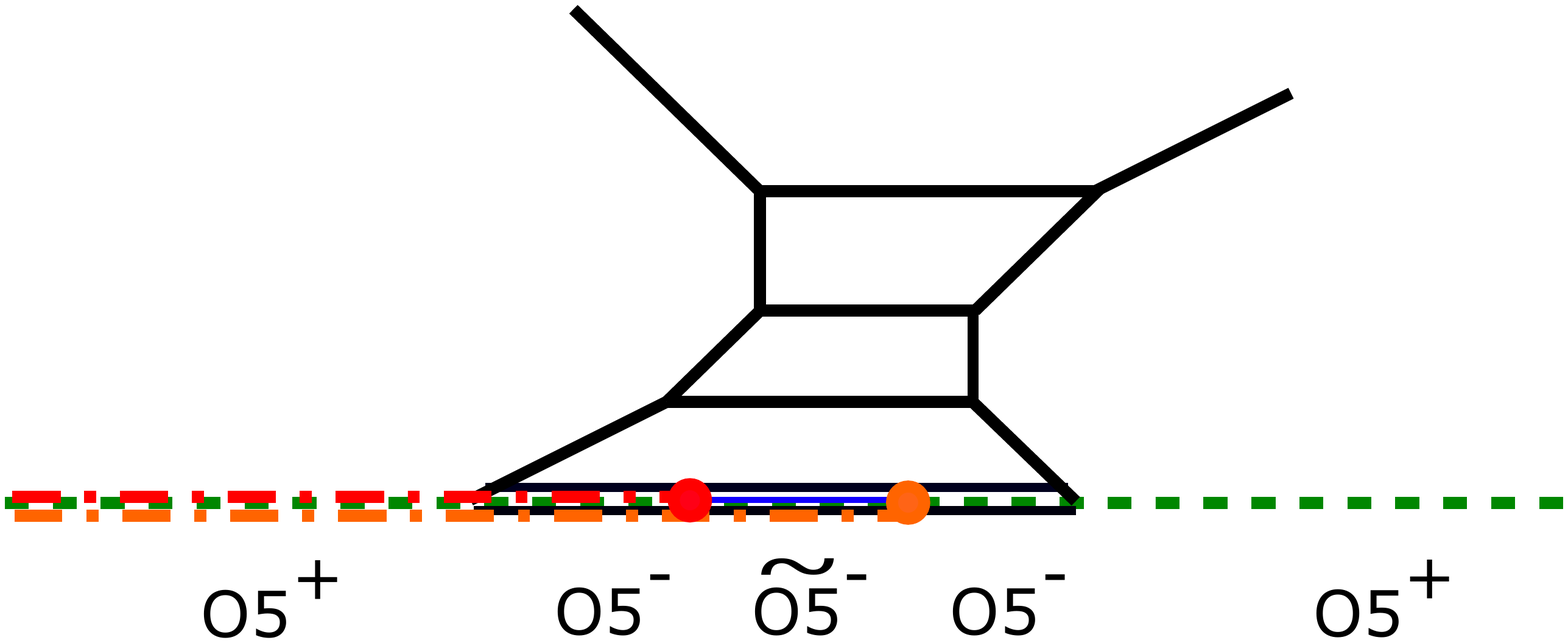} \label{fig:SO8w1flvrHiggsc}}
\subfigure[]{
\includegraphics[width=7cm]{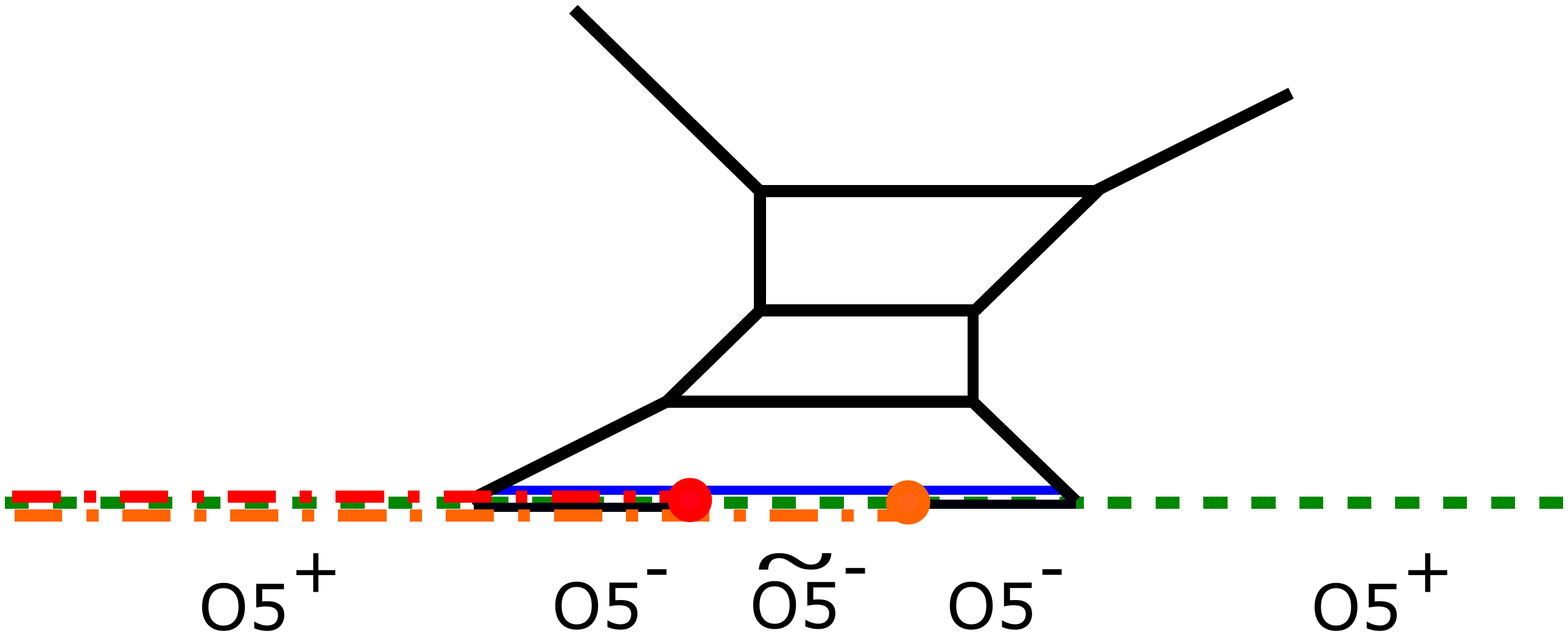} \label{fig:SO8w1flvrHiggsd}}
\caption{A Higgsing procedure of the 5-brane web of the $SO(8)$ gauge theory with one flavor to the 5-brane web of the pure $SO(7)$ gauge theory. (a): Moving the flavor D7-brane to the middle of the diagram. The branch cut is denoted by the dashed line. (b): Lowering the flavor D7-brane as well as the bottom color D5-brane to the O5$^-$-plane. (c): Splitting the D7-branes into two half D7-branes. We have an $\widetilde{\text{O5}}^-$-plane between the half D7-branes and there are effectively three fractional D5-branes between the half D7-branes. We also have monodromy branch cuts for the half D7-branes represented by the red and orange dashed lines. (d): Removing the two fractional D5-branes. We have a half D5-brane denoted by the blue line stretched between the $(2, 1)$ 5-brane and the $(1, -1)$ 5-brane. The diagram gives the pure $SO(7)$ gauge theory.}
\label{fig:SO8w1flvrHiggs}
\end{figure}

The web diagram for the pure $SO(7)$ gauge theory can be derived by
a Higgsing of the $SO(8)$ gauge theory with a hypermultiplet in the vector representation.
Through this process, we will see that the $\widetilde{\text{O5}}$-plane is accompanied by the half monodromy cut.
%It is possible to see the presence of the half monodromy cut by realizing the pure $SO(7)$ gauge theory from a Higgsing of an $SO(8)$ gauge theory with a hypermultiplet in the vector representation. 
A 5-brane web diagram of the $SO(8)$ gauge theory with one flavor is 
constructed with O5$^-$-plane as given in Figure \ref{fig:SO8w1flvrHiggsa}.
%\ref{fig:SO8w1flvr}.
%%%%%%%%%%%%%%%%%%%%%%%%%%%%%%%%%%% 
%%\begin{figure}
%%\centering
%%\includegraphics[width=8cm]{SO8w1flvr.pdf}
%%\caption{A 5-brane web diagram for the 5d $SO(8)$ gauge theory with one flavor. A flavor D7-brane is denoted by a circle. The broken line stands for an O5-plane.}
%%\label{fig:SO8w1flvr}
%%\end{figure}
%%%%%%%%%%%%%%%%%%%%%%%%%%%%%%%%%%% 
%%%%%%%%%%%%%%%%%%%%%%%%%%%%%%%%% 
%%%%%%%%%%%%%%%%%%%%%%%%%%%%%%%%%
%The web diagram of the $SO(8)$ gauge theory is constructed with an O5-plane. 
%We put a D7-brane at the end of the flavor D5-brane in Figure \ref{fig:SO8w1flvr}. 
We put a floating D7-brane to realize one flavor
and we put the branch cut associated with this D7-brane in the left direction.
%A branch cut is associated to the D7-brane and we put the branch cut in the left direction. 
Note that $(p,q)$ charges of the 5-branes change when they go across this cut.
%In order to perform the Higgsing, we first move the D7-brane to the middle of the diagram as in Figure \ref{fig:SO8w1flvrHiggsa}. The flavor D5-brane dsiappears due to the Hanany-Witten transition \cite{Hanany:1996ie}. 
%Then 
In order to perform the Higgsing, we lower this D7-brane 
%and also 
as well as the bottom color D5-brane to the position of the O5$^-$-plane. Since the left part of the web diagram crosses the branch cut, the 5-brane charges change accordingly as in Figure \ref{fig:SO8w1flvrHiggsb}.

On the O5$^-$-plane, the D7-brane can be split into two half D7-branes,
generating an $\widetilde{\text{O5}}^-$-plane in between
 \cite{Evans:1997hk, Giveon:1998sr, Feng:2000eq, Bertoldi:2002nn}. 
 %Note that we have effectively three half D5-branes between the two half D7-branes after the splitting. 
 %Since an O5$^-$-plane crosses the half D7-brane, we should have an $\widetilde{\text{O5}}^-$-plane between the two half D7-branes. 
%Counting the half D5-brane associated to the $\widetilde{\text{O5}}$-plane, we have effectively three half D5-branes. 
Counting the half D5-brane associated to the $\widetilde{\text{O5}}^-$-plane together with the bottom color D5-brane, we have effectively three half D5-branes between these two half D7-branes. 
Note also that half monodromy cut appears between the two half D7-branes.
The 5-brane web after the splitting is depicted in Figure \ref{fig:SO8w1flvrHiggsc}. 

%In this case, 
At this stage, it is possible to move the two half D5-branes between the half D7-branes off the plane of the 5-brane web,
%the motion of the half D5-branes parametrizes 
which degree of freedom corresponds to the one-dimensional Higgs branch.
% of the $SO(8)$ gauge theory with one flavor. 
Thus, removing the two half D5-branes infinitely far away should correspond to the Higgsing of the $SO(8)$ gauge theory with one flavor
down to the pure $SO(7)$ gauge theory. 
After removing the two half D5-branes, 
we see that one half D5-brane 
%should 
stretch between the $(2, 1)$ 5-brane and the $(1, -1)$ 5-brane 
%to preserve the s-rule. 
including the one coming from the $\widetilde{\text{O5}}^-$-plane.
%Then, besides the stretched half D5-brane, 
In addition, we have one half D5-brane connecting the $(2, 1)$ 5-brane to the left half D7-brane and also another half D5-brane connecting the $(1, -1)$ 5-brane to the right half D7-brane as in Figure \ref{fig:SO8w1flvrHiggsd},
which configuration preserves the s-rule.
%The web diagrams of the Higgsing procedure are given in Figure \ref{fig:***}.

%After the Higgsing, we have not only one half D5-brane between the half D7-branes but also a monodromy cut associated to the right half D7-brane between the half D7-branes. 
%Since the branch cut is associated to the half D7-brane brane, its monodromy is half of the original monodromy associated to one full D7-brane. 
%Hence, this Higgsing implies that an $\widetilde{\text{O5}}^-$-plane can be effectively described by an O5$^-$-plane and a half D5-brane plus the half monodromy cut. 
%Due to the half monodromy cut the problem of the jump by the fractional D5-brane charge is resolved. 

\begin{figure}
\centering
\includegraphics[width=8cm]{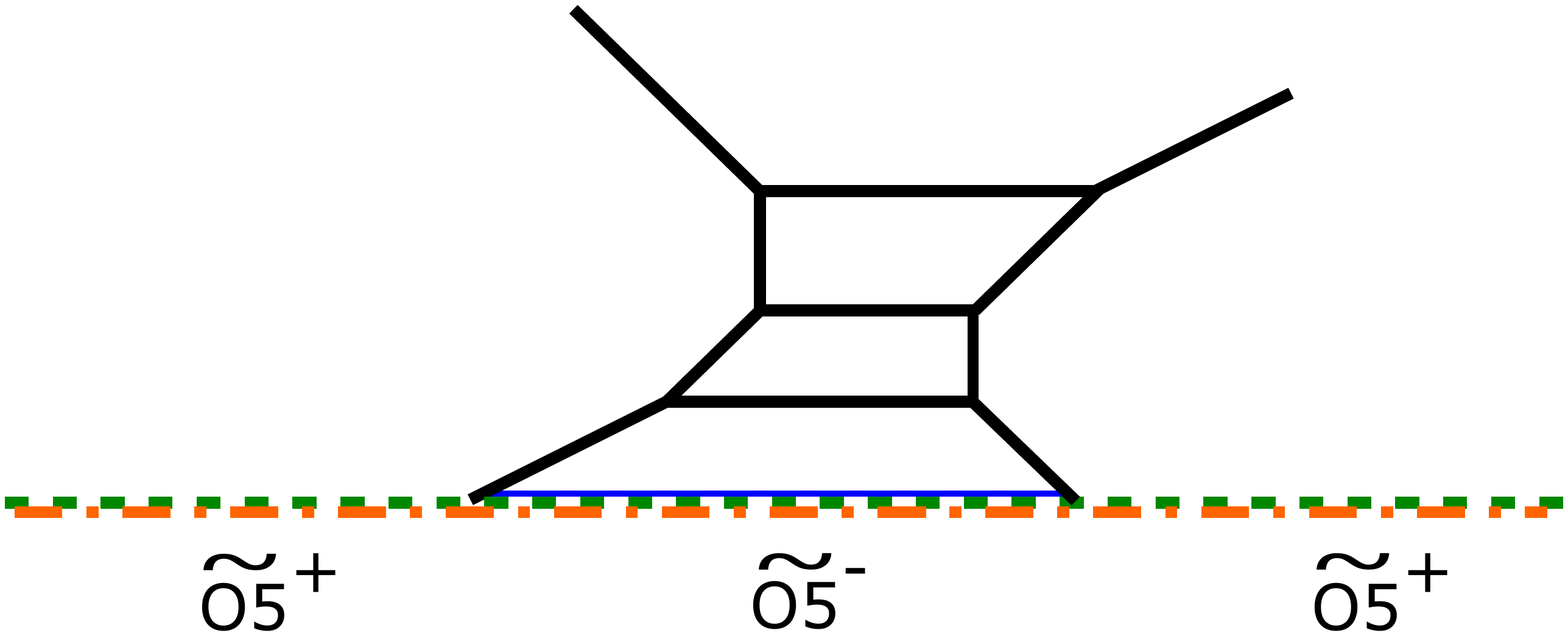}
\caption{Another 5-brane web diagram for the 5d pure $SO(7)$ gauge theory by removing the half D7-branes in the opposite directions. The diagram is constructed with an $\widetilde{\text{O5}}$-plane. The $\widetilde{\text{O5}}$-plane is realized by the O5-plane with the half monodromy cut which is denoted by the orange dashed line. }
\label{fig:pureSO7}
\end{figure}

\begin{figure}
\centering
\includegraphics[width=8cm]{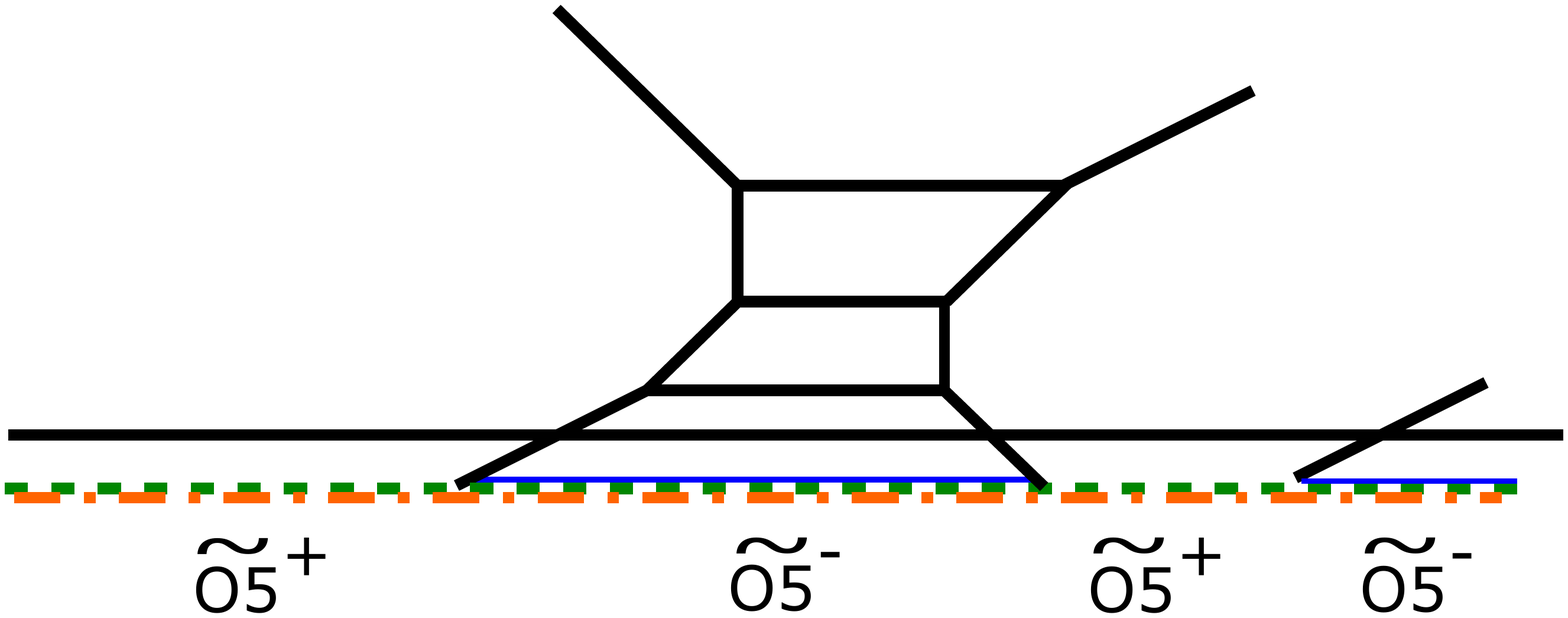}
\caption{A 5-brane web diagram for the $[1] - SO(9) - USp(2) - \left[\frac{3}{2}\right]$ quiver theory.}
\label{fig:SO9USp2quiver}
\end{figure}

\begin{figure}
\centering
\includegraphics[width=8cm]{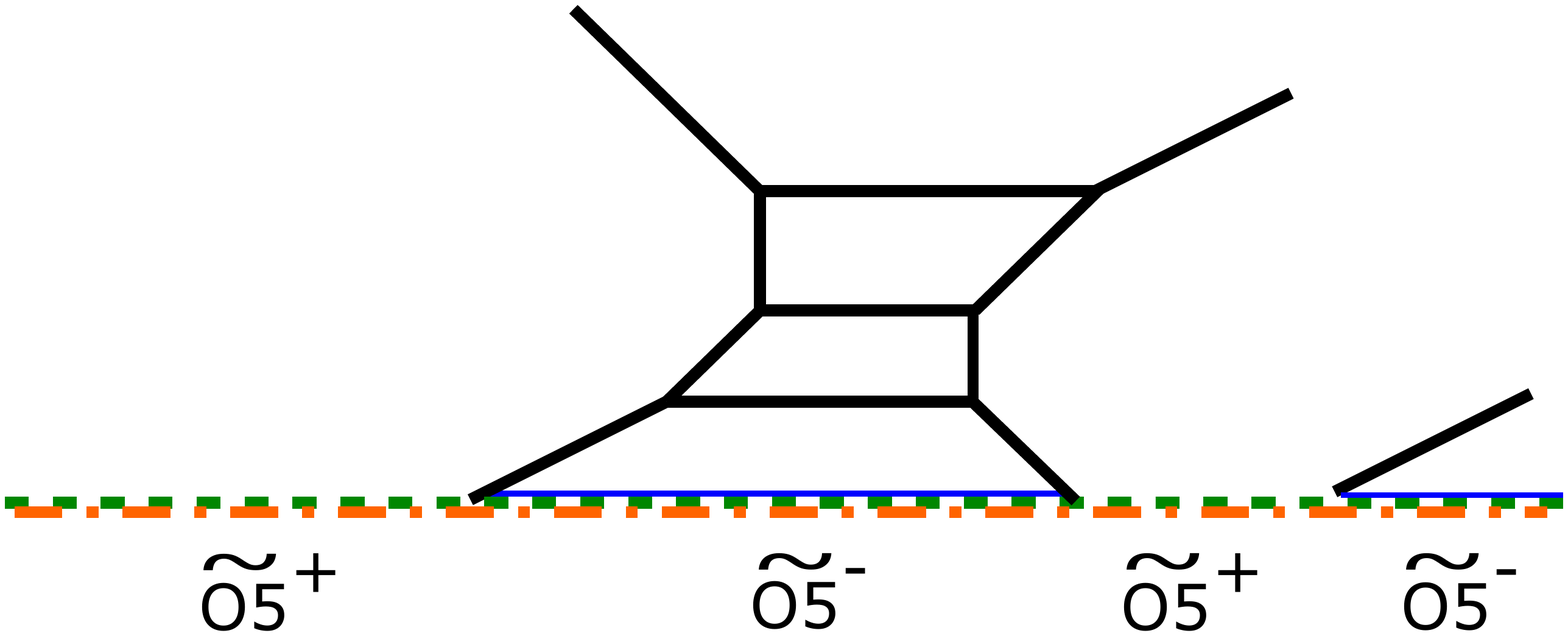}
\caption{A 5-brane web diagram for a $SO(7)$ gauge theory with spinor matter obtained by removing the D5-brane from the diagram for the $[1] - SO(9) - USp(2) - \left[\frac{3}{2}\right]$ quiver theory.}
\label{fig:SO7wspinor}
\end{figure}

Although the 5-brane web diagram in Figure \ref{fig:SO8w1flvrHiggsd} still has an remaining O5-plane, one can change it into an $\widetilde{\text{O5}}^-$-plane by moving the left/right half D7-brane to infinitely left/right respectively. 
No half D5-branes are attached to the half D7-branes 
after they go across the $(2, 1)$ 5-brane or $(1, -1)$ 5-brane
due to Hanany-Witten effect \cite{Hanany:1996ie}.
Note that the half monodromy cut between the two half D7-brane still remains even after moving them to infinity.
That is, we see that the $\widetilde{\text{O5}}$-plane is accompanied by the half monodromy cut.
The 5-brane web diagram after moving the half D7-branes in the opposite directions is given in Figure \ref{fig:pureSO7}. 
This is exactly the web diagram for the pure $SO(7)$ gauge theory, 
which have three color D5-branes with the $\widetilde{\text{O5}}^-$-plane.
%%%%%%%%%%%%%%%%%%%%%%%%%%%%%%%%%
It is straightforward to include a hypermultiplet in the vector representation of $SO(7)$. The vector matter can be introduced by adding a flavor D5-brane to the 5-brane web diagram of the pure $SO(7)$ gauge theory. In fact, one can also introduce spinor matter to the 5-brane web of the $SO(7)$  gauge theory. From the 5-brane web viewpoint, the spinor matter can be realized non-perturbatively \cite{Zafrir:2015ftn}. To see that, let us consider a 5-brane web diagram in Figure \ref{fig:SO9USp2quiver}. 
%%%%%%%%%%%%%%%%%%%%%%%%%%%%%%%%% 
%%%%%%%%%%%%%%%%%%%%%%%%%%%%%%%%% 
The web diagram gives a $[1] - SO(9) - USp(2) - \left[\frac{3}{2}\right]$ quiver theory. Here $[n] - G$ stands for $n$ flavors attached to the $G$ gauge theory. The quiver theory has a Higgs branch associated to moving a D5-brane off the plane of the 5-brane web. The Higgsing yields a 5-brane web diagram given in Figure \ref{fig:SO7wspinor}. 
%%%%%%%%%%%%%%%%%%%%%%%%%%%%%%%%% 
%%%%%%%%%%%%%%%%%%%%%%%%%%%%%%%%% 
The resulting theory might naively look like the pure $SO(7)$ gauge theory. However, the gauge coupling of the ``$USp(0)$'' gauge group is not still turned off. We may expect some additional degrees of freedom whose mass is the inverse of the gauge coupling. In other words, the 5-brane web appears to have ``$USp(0)$'' instantons. Before the Higgsing, we indeed have $USp(2)$ instantons and the instantons carry charges in the spinor representation of 
%$SO(12)$. 
$SO(9)$.
Hence, after the Higgsing, a natural candidate for the ``$USp(0)$'' instantons is a hypermultiplet in the spinor representation of the $SO(7)$ gauge theory. Namely, the 5-brane web in Figure \ref{fig:SO7wspinor} gives rise to an $SO(7)$ gauge theory with a hypermultiplet in the spinor representation.

\subsection{5-brane web for pure $G_2$ gauge theory}
\label{sec:HiggstoG2}

We then move on to construct a 5-brane web diagram for a pure $G_2$ gauge theory. 
By Higgsing the $SO(7)$ gauge theory with one hypermultiplet in the spinor representation yields the pure $G_2$ gauge theory.
Thus, the corresponding process in the 5-brane web diagram for the $SO(7)$ gauge theory with one spinor in Figure \ref{fig:SO7wspinor} should lead to the web diagram for the pure $G_2$ gauge theory.
%An $SO(7)$ gauge theory with one spinor has a Higgs branch and the Higgsing in fact yields a pure $G_2$ gauge theory. 
%Indeed the 5-brane web diagram of the $SO(7)$ gauge theory with one spinor in Figure \ref{fig:SO7wspinor} implies that the theory possesses an $SU(2)$ flavor symmetry associated to parallel external $(2, 1)$ 5-branes. 
The web diagram in Figure \ref{fig:SO7wspinor} implies that the theory possesses an $SU(2)$ flavor symmetry 
associated to parallel external $(2, 1)$ 5-branes.
This global symmetry is expected to act on the hypermultiplet in the spinor representation.
Thus, the distance of the two parallel external $(2, 1)$ 5-branes should be associated to the mass of this spinor 
and the Higgs branch should open up in the massless limit 
at certain subspace in the Coulomb moduli.
%Then, a 
The question is how one can take the massless limit. It seems to be difficult take the massless limit from the diagram in Figure \ref{fig:SO7wspinor} since we need to 
%also shrink 
``flop'' the $\widetilde{\text{O5}}^+$-plane between the $(2, 1)$ 5-brane and the $(1, -1)$ 5-brane. 

\begin{figure}
\centering
\subfigure[]{
\includegraphics[width=7cm]{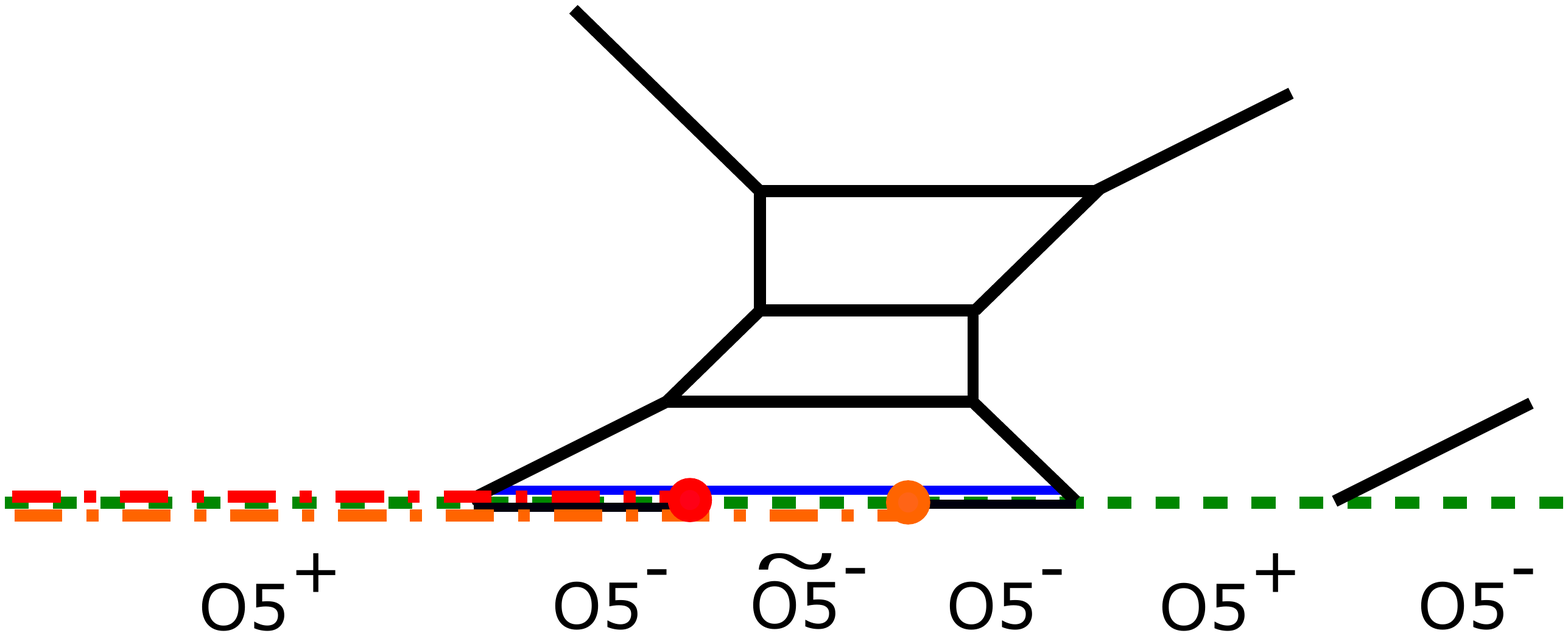} \label{fig:SO7wspinor2a}}
\subfigure[]{
\includegraphics[width=7cm]{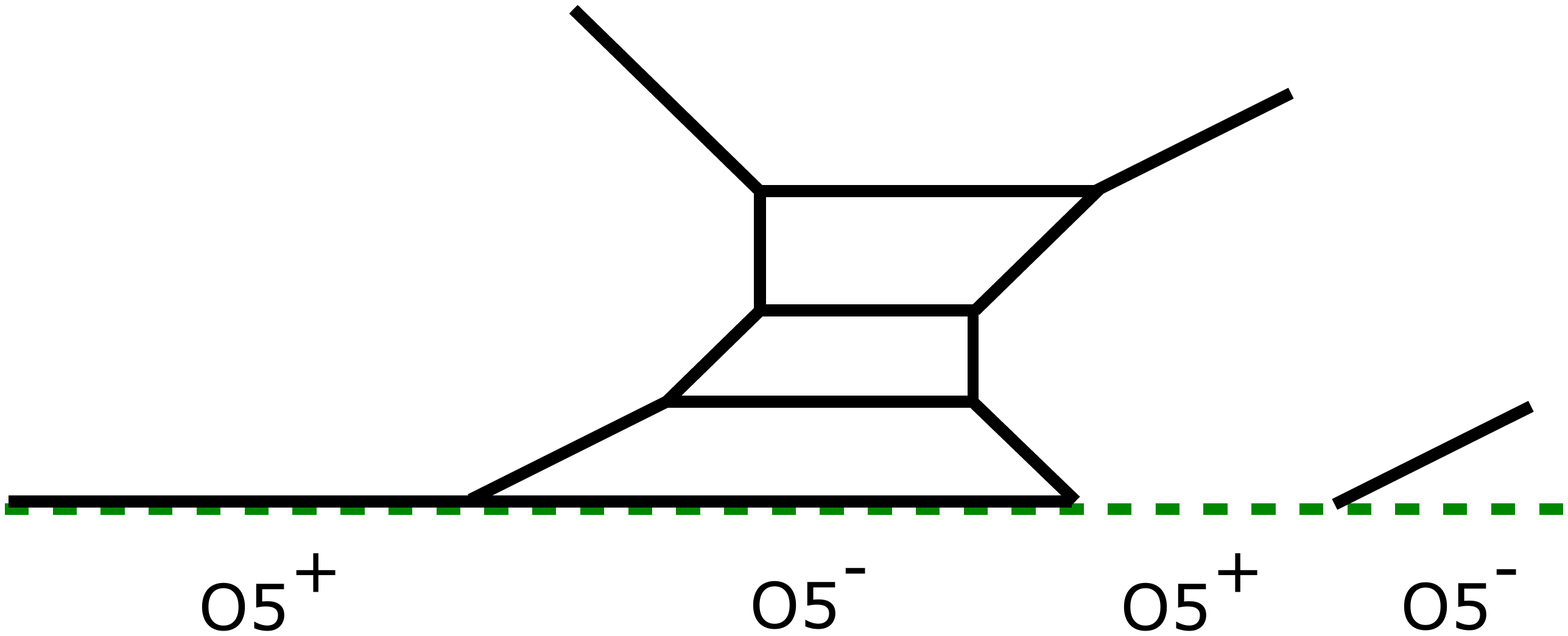} \label{fig:SO7wspinor2b}}
\caption{5-brane web diagrams for the $SO(7)$ gauge theory with one spinor. (a): The 5-brane web diagram which is obtained by Higgsing the diagram of the $SO(8)$ gauge theory with one vector and one spinor. Compared to the diagram in Figure \ref{fig:SO8w1flvrHiggsd}, the $(2, 1)$ 5-brane is attached on the right, yielding the spinor matter. Removing the half D7-branes in the opposite direction gives the diagram in Figure \ref{fig:SO7wspinor}. (b): An equivalent diagram to the one in Figure \ref{fig:SO7wspinor2a} (and also to the one in Figure \ref{fig:SO7wspinor}). We move the half D7-branes in the left direction compared to the diagram in Figure \ref{fig:SO7wspinor2a}. Then the monodromy cut disappears and the diagram is constructed with an O5-plane.}
\label{fig:SO7wspinor2}
\end{figure}

%In order to resolve the issue we consider the ``generalized flop transition'' considered in \cite{Hayashi:2017btw}.
%Then it is useful to consider an equivalent but different 5-brane web diagram for the $SO(7)$ gauge theory with one spinor. 
In order to resolve the issue, it turns out to be useful to consider an equivalent but different 5-brane web diagram for the $SO(7)$ gauge theory with one spinor. 
Analogous to Figure \ref{fig:pureSO7} being obtained from Figure \ref{fig:SO8w1flvrHiggsd},
the diagram in Figure \ref{fig:SO7wspinor} may be obtained by moving the two half D7-branes in the opposite directions from the diagram in Figure \ref{fig:SO7wspinor2a}. 
%%%%%%%%%%%%%%%%%%%%%%%%%%%%%%%%%
%%%%%%%%%%%%%%%%%%%%%%%%%%%%%%%%%
Instead, we can consider another deformation from the diagram in Figure \ref{fig:SO7wspinor2a} by moving both the half D7-branes to infinitely left. The resulting 5-brane web is depicted in Figure \ref{fig:SO7wspinor2b}. After this deformation, the monodromy cuts completely disappears from the diagram and hence we have only an O5-plane. Therefore, a 5-brane web diagram with an $\widetilde{\text{O5}}$-plane has an equivalent 5-brane web diagram without an $\widetilde{\text{O5}}$-plane but with only an O5-plane. The transition is done by moving a half D7-brane from the infinitely right to the infinitely left in the diagram with an $\widetilde{\text{O5}}$-plane. We will make use of this transition in this paper.

\begin{figure}
\centering
\subfigure[]{
\includegraphics[width=4cm]{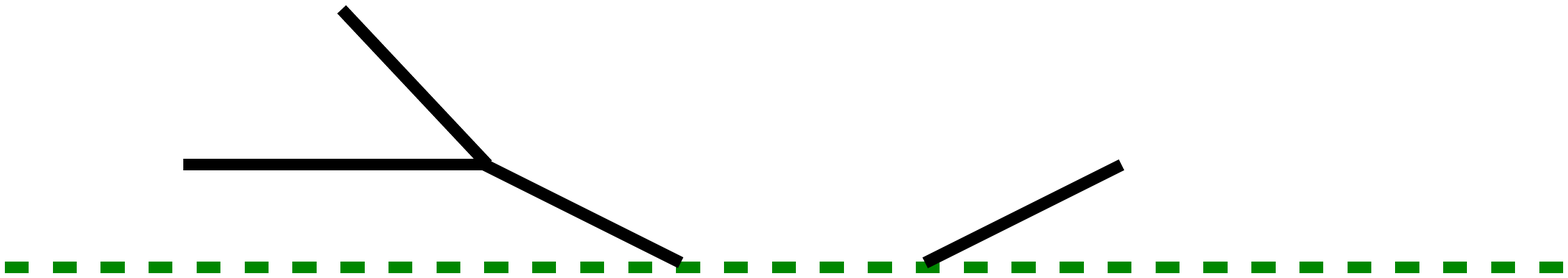} \label{fig:flopO5a}}
\subfigure[]{
\includegraphics[width=4cm]{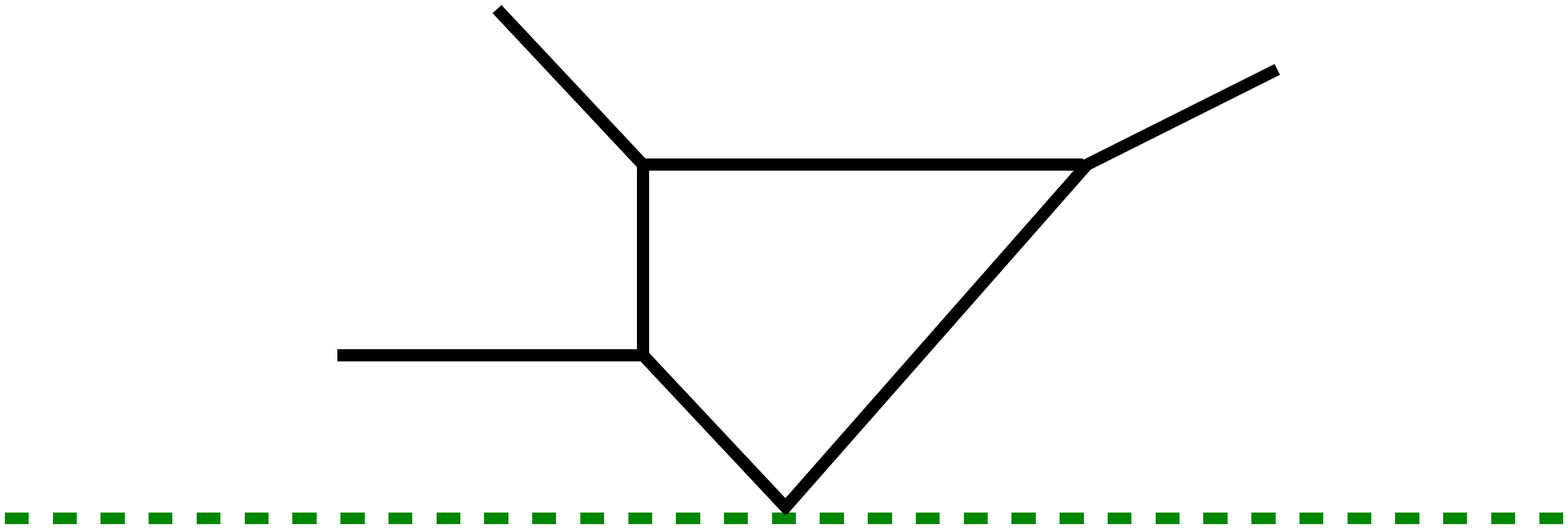} \label{fig:flopO5b}}
\subfigure[]{
\includegraphics[width=4cm]{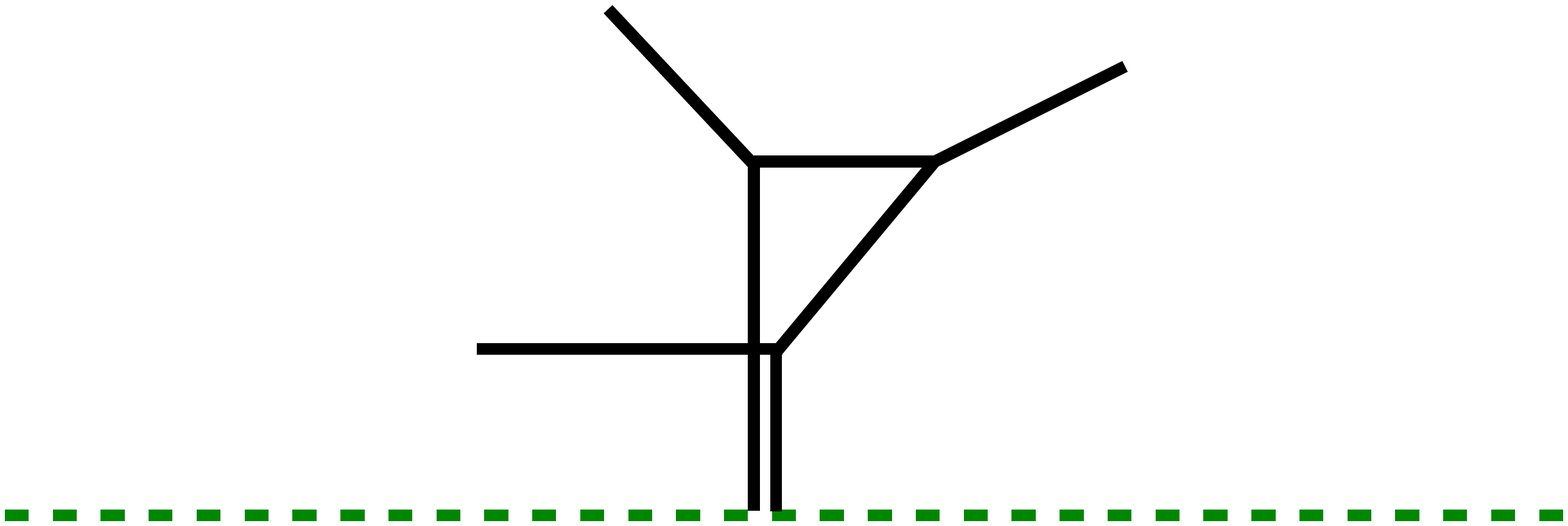} \label{fig:flopO5c}}
\caption{A generalized flop transition for a 5-brane web with an O5-plane. Whether the diagram in (a) is flopped to (b) or (c) depends on the sign of the mass parameter associated to the D5-brane. }
\label{fig:flopO5}
\end{figure}

\begin{figure}
\centering
\subfigure[]{
\includegraphics[width=6cm]{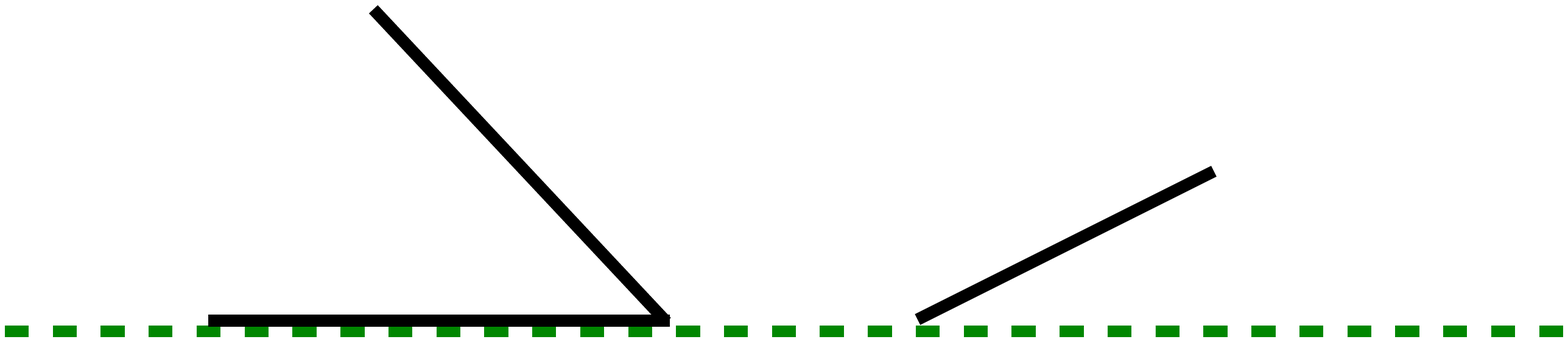} \label{fig:flopO5tildea}}
\subfigure[]{
\includegraphics[width=6cm]{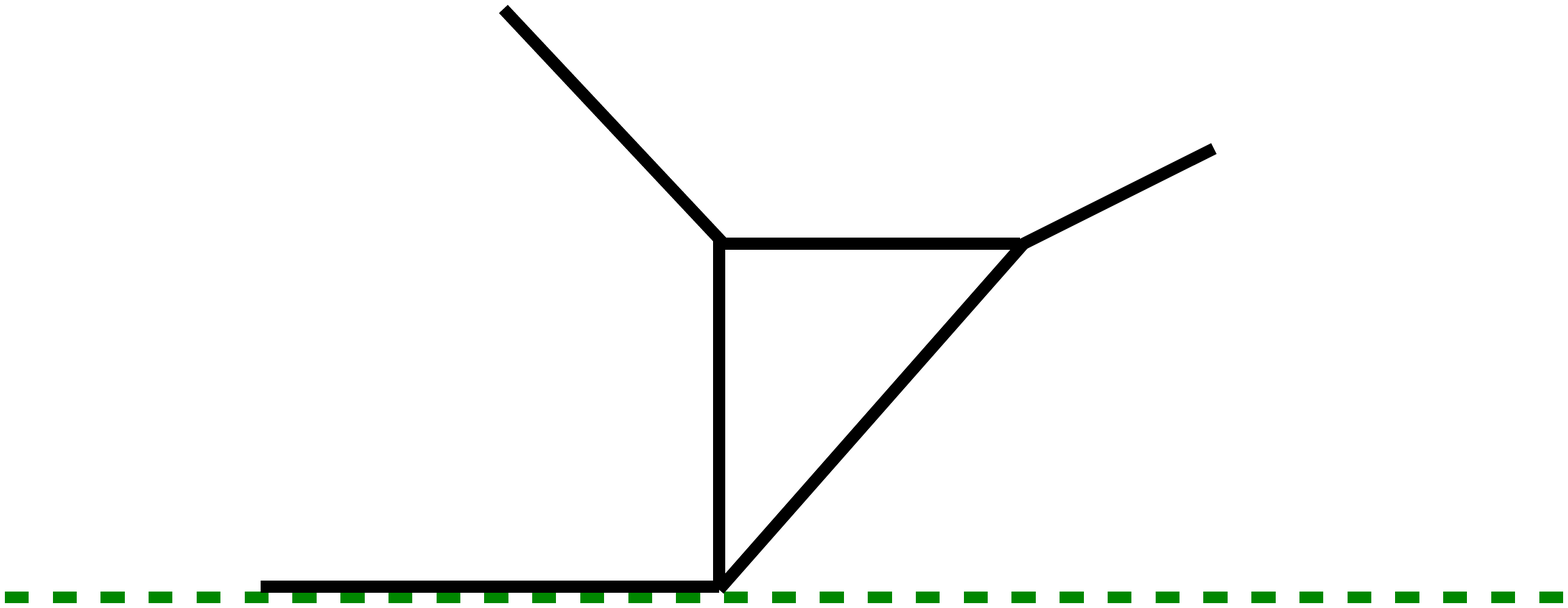} \label{fig:flopO5tildeb}}
\caption{The generalized flop transition in Figure \ref{fig:flopO5} in the case when the height of the D5-brane in Figure \ref{fig:flopO5} is set to zero.}
\label{fig:flopO5tilde}
\end{figure}

\begin{figure}
\centering
\subfigure[]{
\includegraphics[width=6cm]{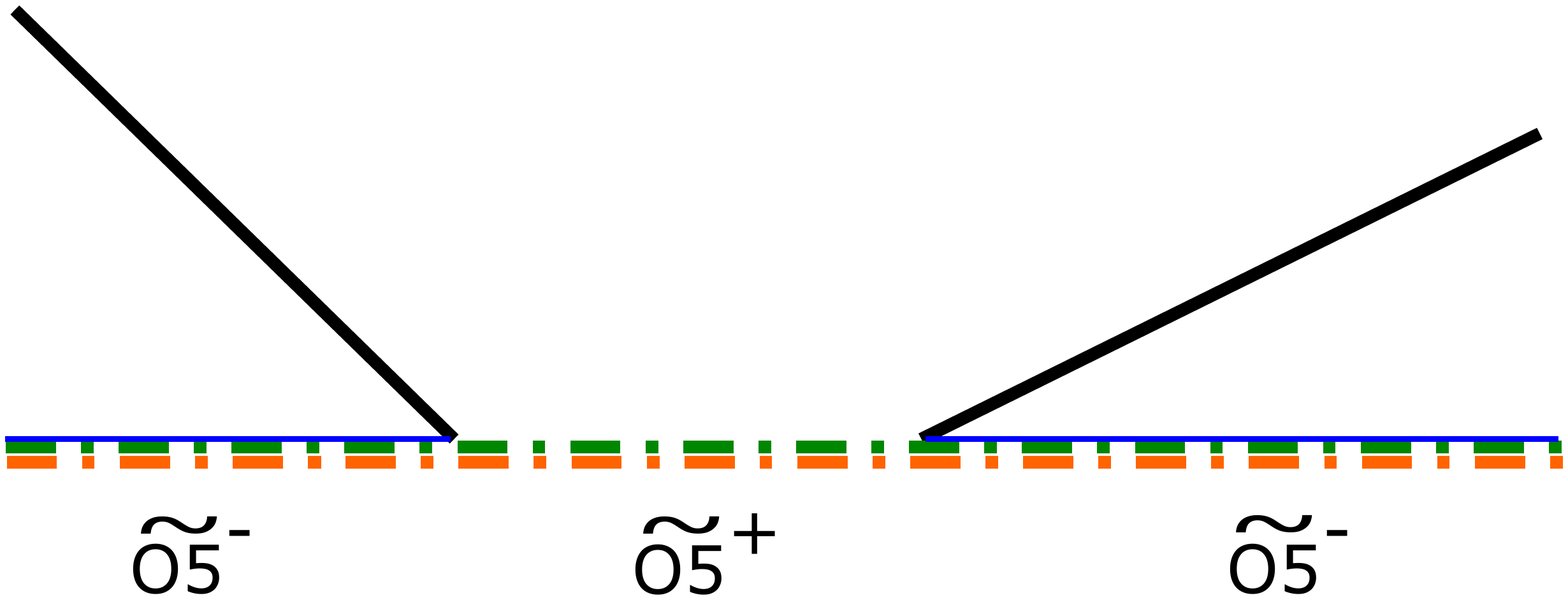} \label{fig:flopO5tilde2a}}
\subfigure[]{
\includegraphics[width=6cm]{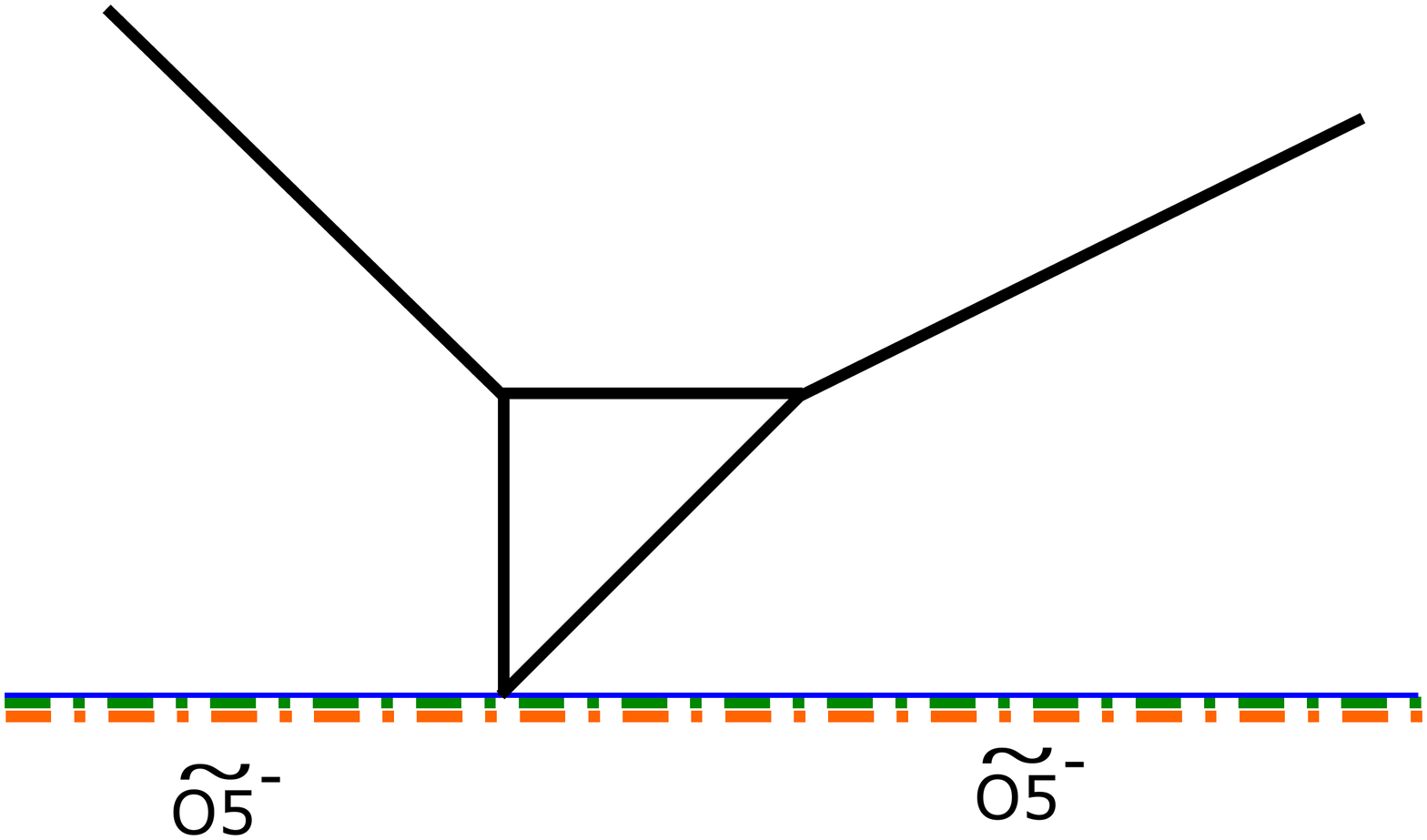} \label{fig:flopO5tilde2b}}
\caption{A generalized flop transition from (a) to (b) which is equivalent to the one in Figure \ref{fig:flopO5tilde}.}
\label{fig:flopO5tilde2}
\end{figure}

By using this diagram, we can take the massless limit of the spinor by
using ``generalized flop transition'' %considered in 
\cite{Hayashi:2017btw}.
Let us focus on the local part of the diagram describing the ``$USp(0)$'' gauge theory. On the left side a $(1, -1)$ 5-brane intersects with an O5-plane together with a full D5-brane. On the other hand, a $(2, 1)$ 5-brane intersects with the O5-plane on the right side. The local structure exactly appears in the 5-brane diagram of the $E_2$ theory, %in \cite{Hayashi:2017btw}. 
%%%%%%%%%%%%%%%%%%%%%%%%%%%%%%%%%%
%%%%%%%%%%%%%%%%%%%%%%%%%%%%%%%%%%
%%%%%%%%%%%%%%%%%%%%%%%%%%%%%%%%%%
and hence we can perform a generalized flop transition for this local part by using the results in \cite{Hayashi:2017btw}.
Through the process of studying the phase diagram of the $E_2$ theory, 
it is proposed that the diagram in Figure \ref{fig:flopO5a}
is flopped either to the one in Figure \ref{fig:flopO5b} or Figure \ref{fig:flopO5c}. %depending on the discrete theta angle of the $USp(0)$ part.
Whether the diagram is flopped to the one in Figure \ref{fig:flopO5b} or Figure \ref{fig:flopO5c} is translated to the sign of the mass parameter of the $USp(0)$ gauge theory, which is the sign of the Coulomb 
%moduli 
branch parameter in our case. 
Although this transition is obtained for the $E_2$ theory, it is natural to assume that this transition is always available regardless of the detail of the remaining diagram to which this subdiagram is attached.
On one hand, by considering a limit where the D5-brane comes down to the position of an O5-plane
as we did in the process of the Higgsing, 
Figure \ref{fig:flopO5a} reduces to Figure \ref{fig:flopO5tildea}, which corresponds to the one appearing in the local part of the diagram in Figure \ref{fig:SO7wspinor2b}. 
On the other hand, by the same limit, Figure \ref{fig:flopO5b} and Figure \ref{fig:flopO5c} both reduce to Figure \ref{fig:flopO5tildeb}. 
Therefore, we propose that the $USp(0)$ part in Figure \ref{fig:SO7wspinor2b} can be flopped to the form in Figure \ref{fig:flopO5tildeb}.
%%%%%%%%%%%%%%%%%%%%%%%%%%%%%%%%%
%%%%%%%%%%%%%%%%%%%%%%%%%%%%%%%%%%
The equivalent flop transition in the presence of an $\widetilde{\text{O5}}$-plane is given in Figure \ref{fig:flopO5tilde2}.
%%%%%%%%%%%%%%%%%%%%%%%%%%%%%%%%%
%%%%%%%%%%%%%%%%%%%%%%%%%%%%%%%%%%

\begin{figure}
\centering
\subfigure[]{
\includegraphics[width=7cm]{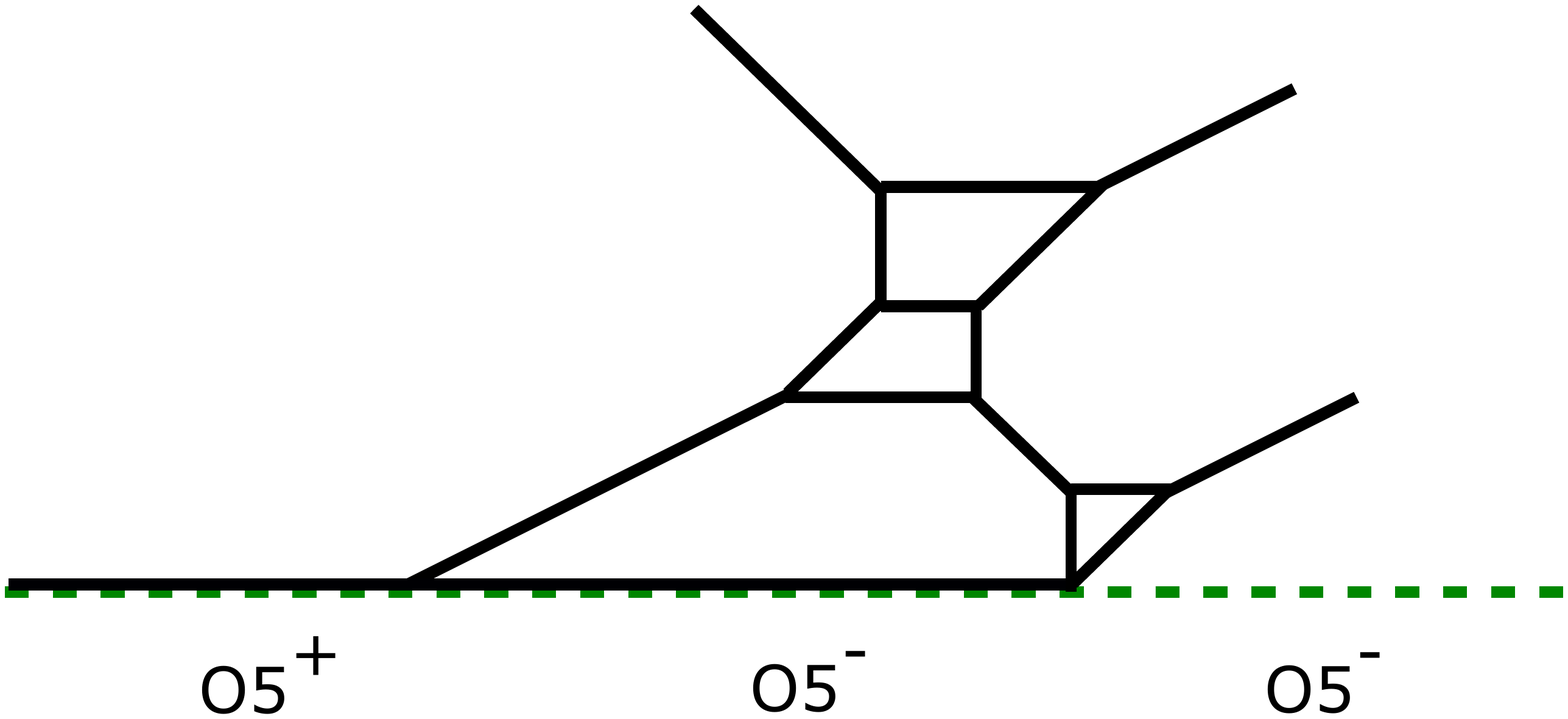} \label{fig:SO7wspinorfloppeda}}
%\subfigure[]{
%\includegraphics[width=7cm]{SO7USp0quivertransformed2.pdf}}
\subfigure[]{
\includegraphics[width=7cm]{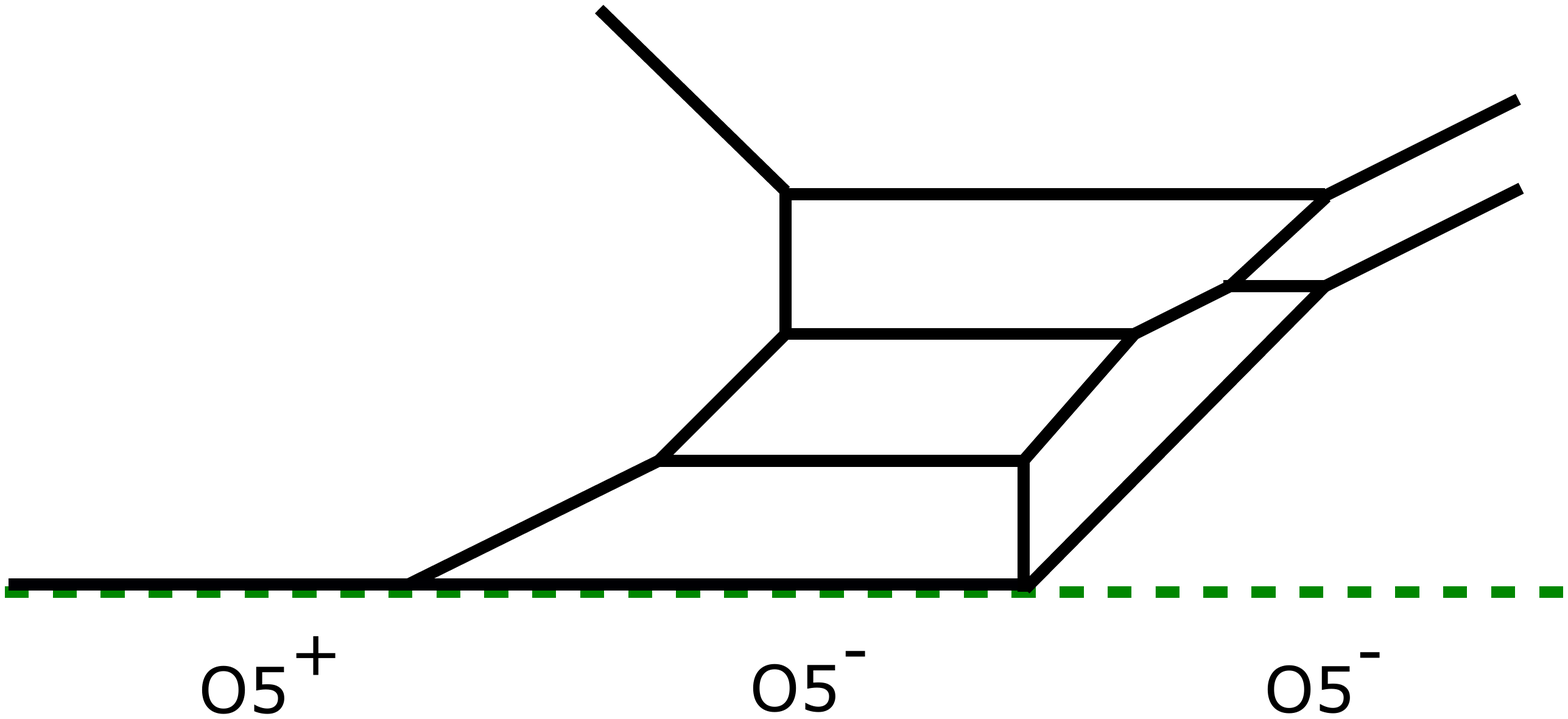} \label{fig:SO7wspinorfloppedb}}
\subfigure[]{
\includegraphics[width=7cm]{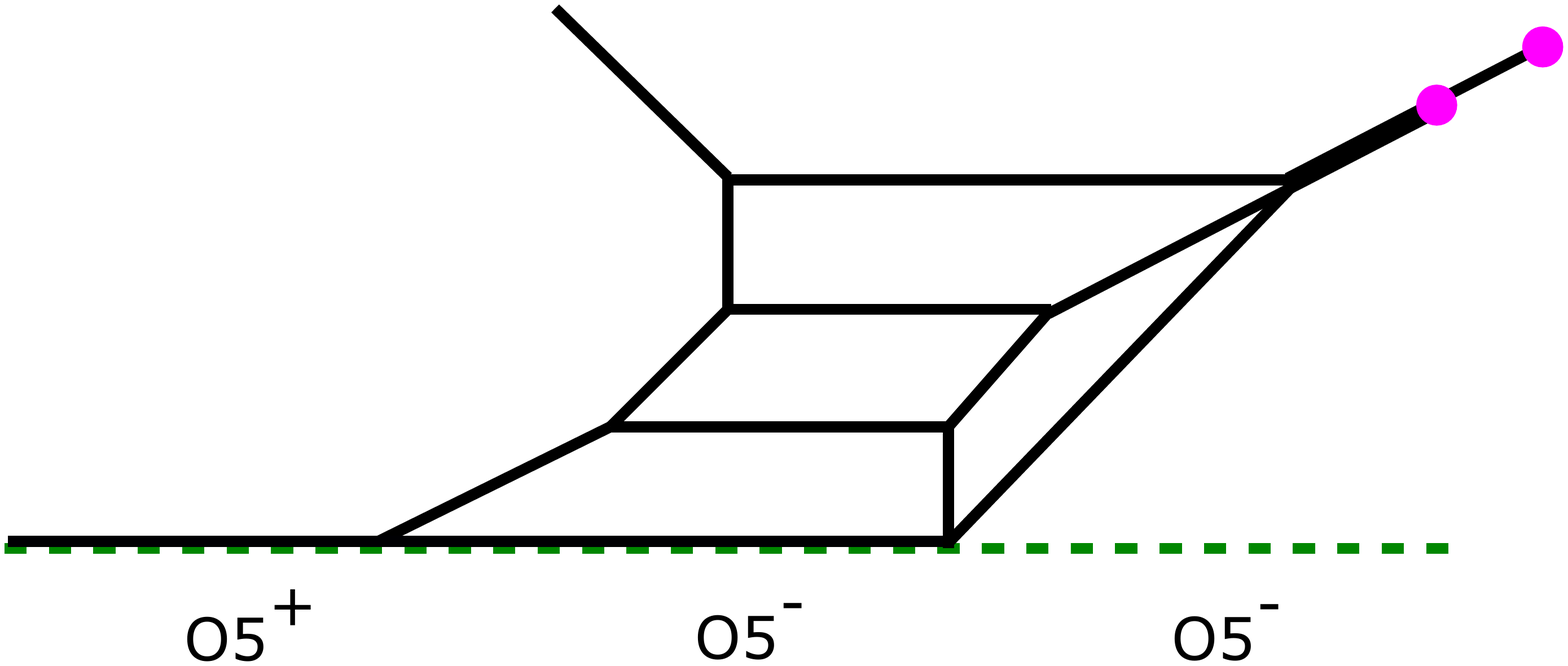} \label{fig:SO7wspinorfloppedc}}
\subfigure[]{
\includegraphics[width=7cm]{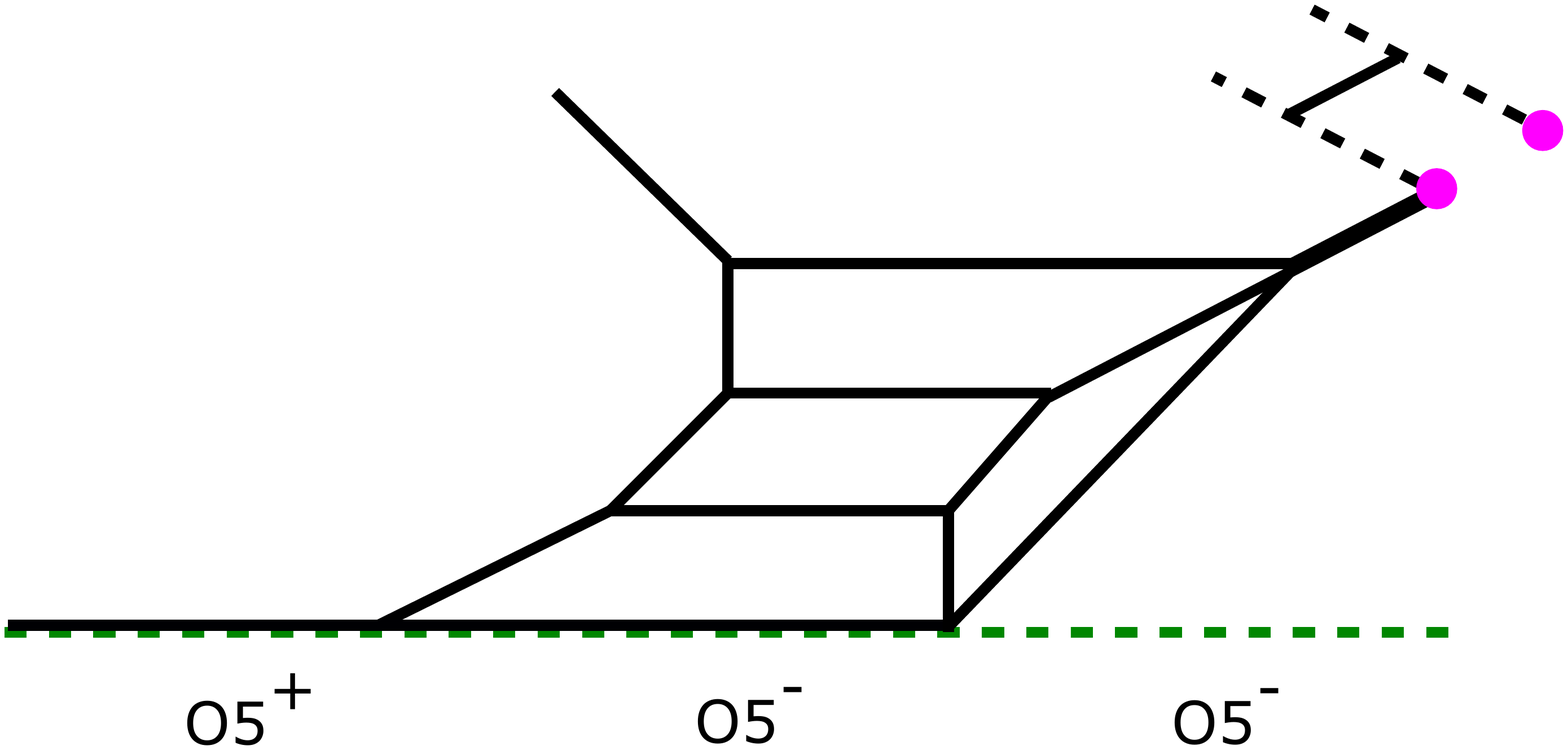} \label{fig:SO7wspinorfloppedd}}
\subfigure[]{
\includegraphics[width=7cm]{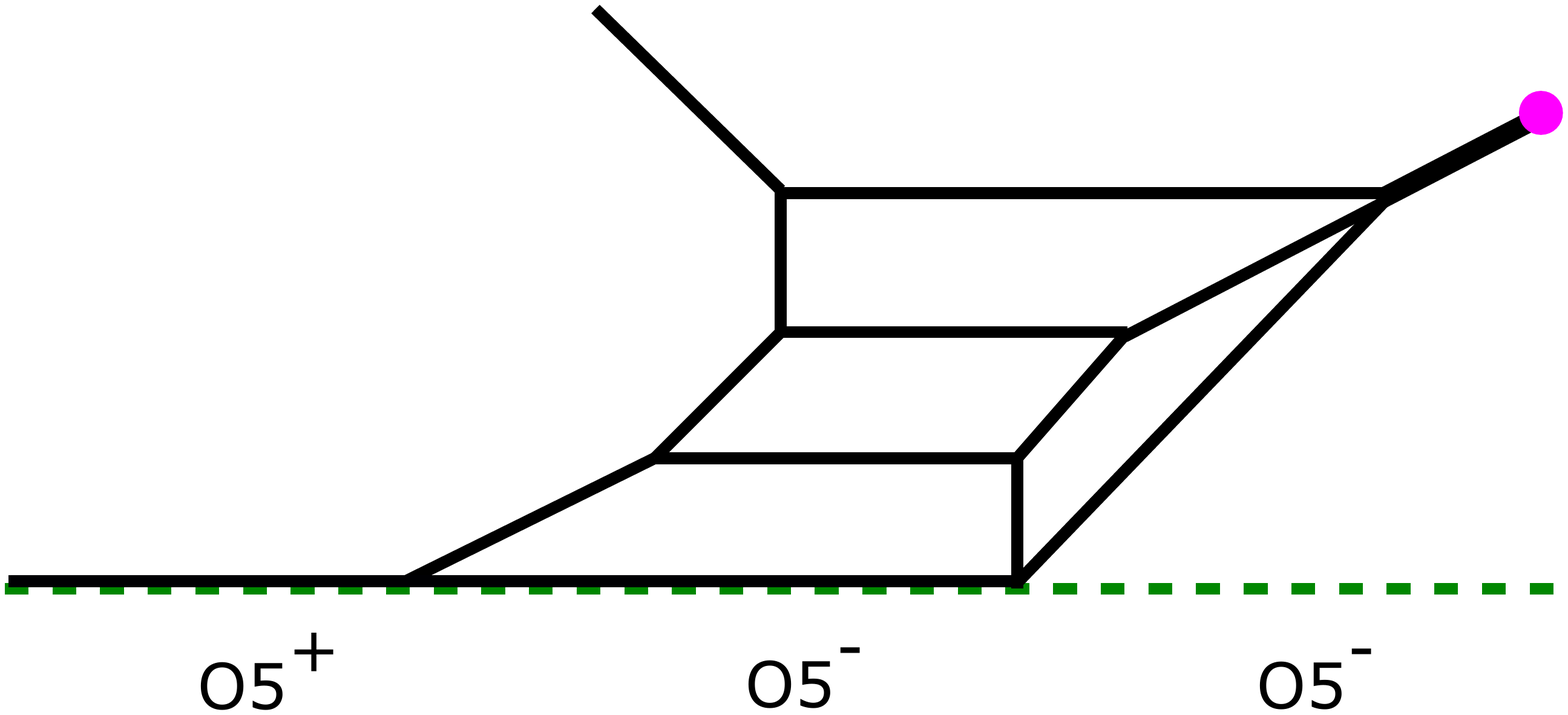} \label{fig:SO7wspinorfloppede}}
\caption{A Higgsing procedure which gives rise to a 5-brane web diagram for the pure $G_2$ gauge theory. (a): We performed the generalized flop transition compared to the diagram in Figure \ref{fig:SO7wspinor2b}. (b): We performed two standard flop transitions to the diagram in Figure \ref{fig:SO7wspinorfloppeda}. (c): Putting the two parallel external $(2, 1)$ 5-branes on top of each other. We also introduced the $(2, 1)$ 7-branes at each end of the external $(2, 1)$ 7-branes. (d): Removing one $(2, 1)$ 5-brane between the $(2, 1)$ 7-branes. (e): Sending the $(2, 1)$ 5-brane to infinity. The diagram yields the pure $G_2$ gauge theory. }
\label{fig:SO7wspinorflopped}
\end{figure}

After considering the transition in Figure \ref{fig:flopO5tilde} for the local $USp(0)$ part in Figure \ref{fig:SO7wspinor2b}, the 5-brane web diagram becomes the one in Figure \ref{fig:SO7wspinorfloppeda}. 
%%%%%%%%%%%%%%%%%%%%%%%%%%%%%%%%%%
%%%%%%%%%%%%%%%%%%%%%%%%%%%%%%%%%%
Hence, we obtain another 5-brane web diagram given in Figure \ref{fig:SO7wspinorfloppeda} which also realizes the $SO(7)$ gauge theory with one spinor. 

From the 5-brane web diagram in Figure \ref{fig:SO7wspinorfloppeda} it is now straightforward to perform the Higgsing:
After two flop transitions, we obtain the diagram in Figure \ref{fig:SO7wspinorfloppedb},
from which we can take the massless limit by putting the parallel external $(2, 1)$ 5-branes on top of each other. 
%In this limit, we also need to tune the Coulomb branch moduli of the $SO(7)$ gauge theory, leaving two Coulomb branch moduli. 
We also tune the Coulomb branch moduli of the $SO(7)$ gauge theory, leaving two Coulomb branch moduli
as in Figure \ref{fig:SO7wspinorfloppedc}.
%In this case, 
By putting a $(2, 1)$ 7-brane at each end of the two parallel external $(2, 1)$ 5-branes,
we can move a segment of a $(2, 1)$ 5-brane between the external $(2, 1)$ 7-branes as in Figure \ref{fig:SO7wspinorfloppedd},
which degrees of freedom corresponds to the Higgs branch of the $SO(7)$ gauge theory with one spinor. 
% and the motion of the D5-brane parametrizes the Higgs branch of the $SO(7)$ gauge theory with one spinor. 
%The chain of the transitions is depicted from Figure \ref{fig:SO7wspinorfloppedb} to Figure \ref{fig:SO7wspinorfloppede}. 
Removing the D5-brane implies that we take a far infrared limit of the $SO(7)$ gauge theory with one spinor at the Higgs branch. Then the resulting 5-brane web diagram should describe the pure $G_2$ gauge theory. Therefore, we conclude that the 5-brane web diagram in Figure \ref{fig:SO7wspinorfloppede} realizes the pure $G_2$ gauge theory. 

\begin{figure}
\centering
\includegraphics[width=8cm]{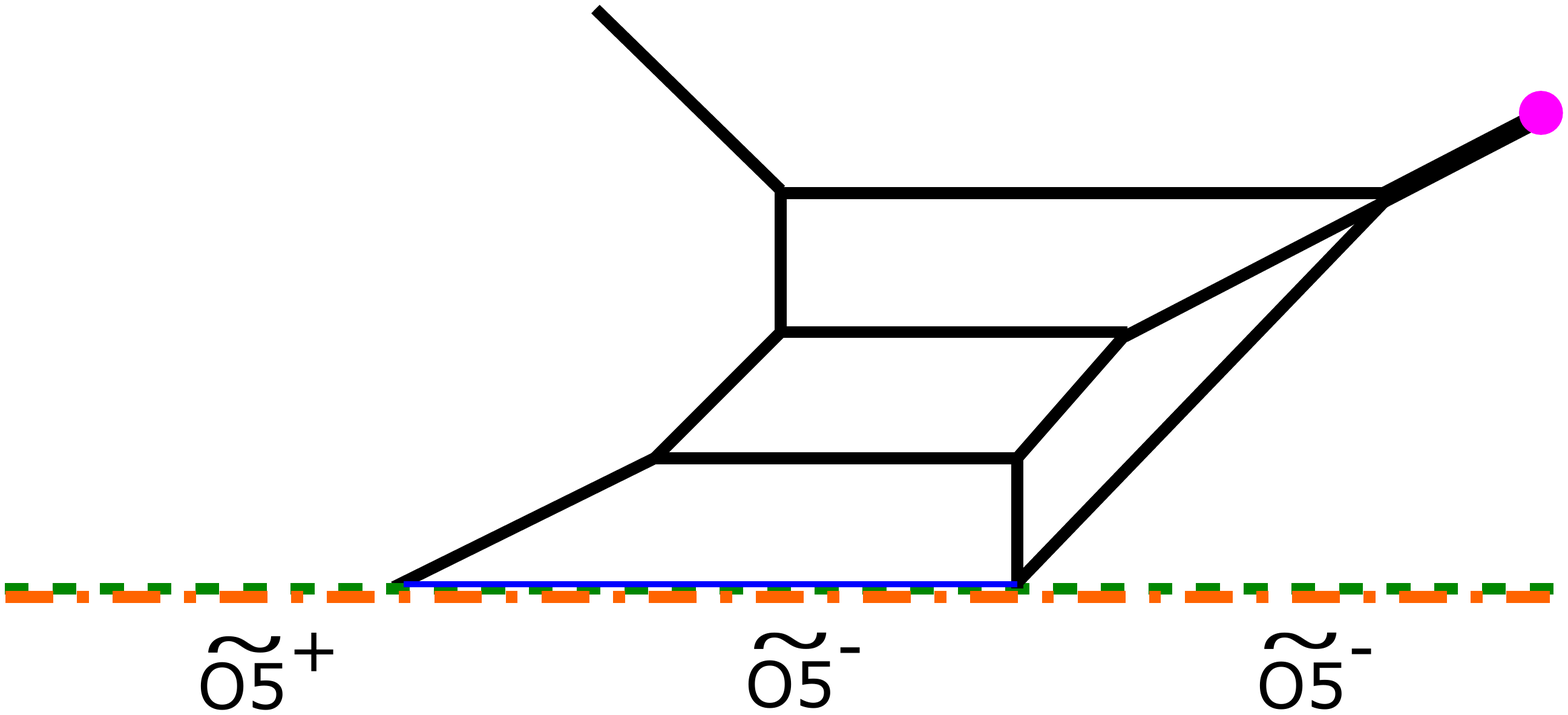}
\caption{Another 5-brane web with an $\widetilde{\text{O5}}$-plane which gives the pure $G_2$ gauge theory. The diagram can be obtained by applying the same Higgsing procedure in Figure \ref{fig:SO7wspinorflopped} to the diagram in Figure \ref{fig:SO7wspinor}.}
\label{fig:pureG2}
\end{figure}

We can write an equivalent diagram with an $\widetilde{\text{O5}}$-plane as in Figure \ref{fig:pureG2}.
%%%%%%%%%%%%%%%%%%%%%%%%%%%%%%%%%%
%%%%%%%%%%%%%%%%%%%%%%%%%%%%%%%%%%
The diagram in Figure \ref{fig:pureG2} is obtained by moving one of the half D7-brane which was sent to the infinitely left to the infinitely right. Then the D5-brane on the O5-plane disappears and we have an $\widetilde{\text{O5}}$-plane realized with the half monodromy cut.

\subsection{Check from effective prepotential}
\label{sec:G2prep1}

%Before considering the application of the 5-brane web in Figure \ref{fig:pureG2}, 
In this subsection,
we %can 
see %an 
evidence that the 5-brane web diagram 
in Figure \ref{fig:pureG2}
yields the 5d pure $G_2$ gauge theory from the analysis effective prepotentials. 

In general, the effective prepotential of a 5d gauge theory with a gauge group $G$ is given by \cite{Seiberg:1996bd, Morrison:1996xf, Intriligator:1997pq}
\begin{equation}
\mathcal{F}(\phi) = \frac{1}{2}m_0h_{ij}\phi_i\phi_j + \frac{\kappa}{6}d_{ijk}\phi_i\phi_j\phi_k + \frac{1}{12}\left(\sum_{r \in \text{roots}}\left|r\cdot \phi\right|^3 - \sum_f\sum_{w \in {\bf R}_f}\left|w\cdot\phi + m_f\right|^3\right), \label{prepotential}
\end{equation}
where $m_0$ is the inverse of the squared gauge coupling, $\phi_i$ are the Coulomb branch moduli, $\kappa$ is the classical Chern-Simons level and $m_f$ is the mass of a hypermultiplet in the representation ${\bf R}_f$ of $G$. $r$ are the roots of $G$ and $w$ are the weights of the representation ${\bf R}_f$. We also used $h_{ij} = \text{Tr}(T_iT_j)$ and $d_{ijk} = \frac{1}{2}\text{Tr}\left(T_i\{T_j, T_k\}\right)$ where $T_i$ are the Cartan generators of $G$.

In the following, we consider the prepotential of the pure $SO(7)$ gauge theory, the $SO(7)$ gauge theory with a spinor and then the pure $G_2$ gauge theory step by step. 

\paragraph{Pure $SO(7)$.}
The first example is the 5d pure $SO(7)$ gauge theory. When we parameterize the Coulomb branch moduli $\phi_i$ in the Dynkin basis, the prepotential of the pure $SO(7)$ gauge theory becomes  
\begin{align}
\mathcal{F}_{SO(7)}(\phi) = &m_0(\phi_1^2-\phi_1\phi_2+\phi_2^2-2\phi_2\phi_3+2\phi_3^2)\cr
&+\frac{4}{3}\phi_1^3 - \frac{1}{2}\phi_1^2\phi_2 - \frac{1}{2}\phi_1\phi_2^2 + \frac{4}{3}\phi_2^3 - 3\phi_2^2\phi_3 + 2\phi_2\phi_3^2 + \frac{4}{3}\phi_3^3, \label{pureSO7prep}
\end{align}
where we chose $[2, -1, 0], [-1, 2, -2], [0, -1, 2]$ for the simple roots for defining the Weyl chamber. The tension of a monopole string may be computed by taking a derivative with respect to a Coulomb branch modulus. The monopole string tension from the prepotential \eqref{pureSO7prep} is
\begin{align}
\frac{\partial \mathcal{F}_{SO(7)}}{\partial \phi_1} =& \frac{1}{2}(2\phi_1 - \phi_2)(2m_0 + 4\phi_1 + \phi_2),\label{pureSO7monopole1}\\
\frac{\partial \mathcal{F}_{SO(7)}}{\partial \phi_2} =& \frac{1}{2}(-\phi_1 + 2\phi_2 - 2\phi_3)(2m_0 + \phi_1 + 4\phi_2 - 2\phi_3),\label{pureSO7monopole2}\\
\frac{\partial \mathcal{F}_{SO(7)}}{\partial \phi_3} =& (-\phi_2 + 2\phi_3)(2m_0 + 3\phi_2 + 2\phi_3). \label{pureSO7monopole3}
\end{align}

It is also possible to compute the tension of a monopole string from 5-brane web diagrams. %which is given by $\frac{\partial \mathcal{F}}{\partial \phi_i}$. 
Monopole strings in a 5d gauge theory are realized by D3-branes which stretch along some face bounded by 5-branes. Therefore, the tension of a monopole string corresponds to the area of a face where a D3-brane can extend. Hence we need to compute the area of faces in the diagram in Figure \ref{fig:pureSO7} for the pure $SO(7)$ gauge theory.

\begin{figure}
\centering
\includegraphics[width=8cm]{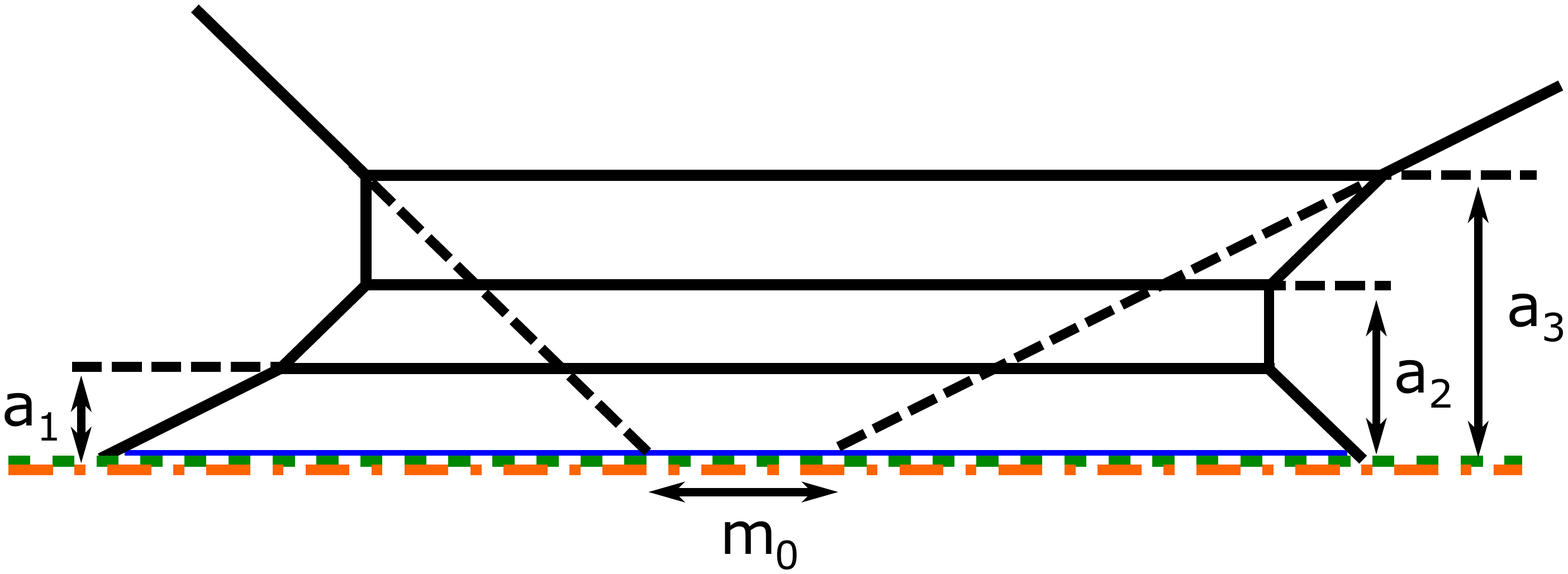}
\caption{A gauge theory parameterization for the pure $SO(7)$ gauge theory. $a_1, a_2, a_3$ are the Coulomb branch moduli and $m_0$ is the inverse of the squared gauge coupling.}
\label{fig:pureSO7parameter}
\end{figure}

For that, we first need to identify the gauge theory parameters with the length of 5-branes in the diagram in Figure \ref{fig:pureSO7}. The height of the color D5-branes is the Coulomb branch modulus and we denote the height of the bottom color D5-brane, the middle color D5-brane and the top color D5-brane by $a_1, a_2, a_3$ respectively. In order to compute the inverse of the squared gauge coupling, we first turn off all the Coulomb branch moduli. Then the external $(2, 1)$ 5-brane and the external $(1, -1)$ directly intersect with the O5-plane and the length of the D5-branes between the $(2, 1)$ 5-brane and the $(1, -1)$ 5-brane on the O5-plane gives $m_0$. Alternatively, one can extrapolate the external $(2, 1)$ 5-brane and the external $(1, -1)$ 5-brane in Figure \ref{fig:pureSO7} to the position of the O5-plane and measure the distance between the external $(2, 1)$ 5-brane and the external $(1, -1)$ 5-brane on the O5-plane. The gauge theory parameterization for the pure $SO(7)$ gauge theory is summarized in Figure \ref{fig:pureSO7parameter}.  
%%%%%%%%%%%%%%%%%%%%%%%%%%%%%%%%%%
%%%%%%%%%%%%%%%%%%%%%%%%%%%%%%%%%%

\begin{figure}
\centering
\includegraphics[width=8cm]{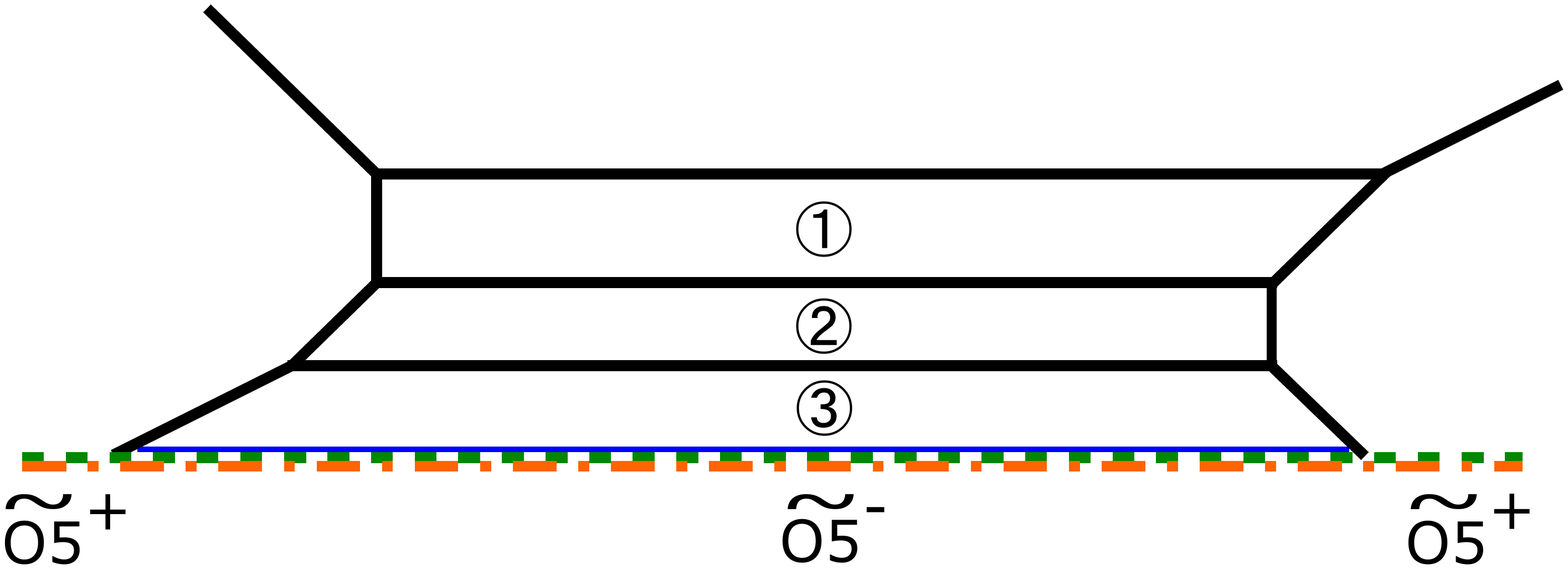}
\caption{Labeling for the three faces in the pure $SO(7)$ diagram.}
\label{fig:pureSO7monopole}
\end{figure}

By using the parameterization depicted in Figure \ref{fig:pureSO7parameter}, we compute the area of faces in the pure $SO(7)$ diagram in Figure \ref{fig:pureSO7}. We label the faces as in Figure \ref{fig:pureSO7monopole}.
%%%%%%%%%%%%%%%%%%%%%%%%%%%%%%%%%%
%%%%%%%%%%%%%%%%%%%%%%%%%%%%%%%%%%
The area of the three faces becomes 
\begin{align}
\textcircled{\scriptsize 1} =&\frac{1}{2}(a_3-a_2)(2m_0+a_2+5a_3),\label{pureSO7monopole4}\\
\textcircled{\scriptsize 2} =&\frac{1}{2}(a_2 - a_1)(2m_0 - a_1 + 3a_2 + 4a_3), \label{pureSO7monopole5}\\
\textcircled{\scriptsize 3} =&\frac{1}{2}a_1(2m_0 + a_1 + 4a_2 + 4a_3). \label{pureSO7monopole6}
\end{align}

We can then compare the are \eqref{pureSO7monopole4}, \eqref{pureSO7monopole5} and \eqref{pureSO7monopole6} with the tension \eqref{pureSO7monopole1}, \eqref{pureSO7monopole2} and \eqref{pureSO7monopole3}. In the computation of the area from the pure $SO(7)$ diagram we parameterized the Coulomb branch moduli $a_1, a_2, a_3$ in the orthonormal basis of $\mathbb{R}^3$. The relation between $a_1, a_2, a_3$ and $\phi_1, \phi_2, \phi_3$ is
\begin{align}
\phi_1 = a_3, \quad \phi_2 = a_2 + a_3, \quad \phi_3 = \frac{1}{2}(a_1 + a_2 + a_3).  \label{SO7relation}
\end{align}
Then the comparison of \eqref{pureSO7monopole4}, \eqref{pureSO7monopole5} and \eqref{pureSO7monopole6} with \eqref{pureSO7monopole1}, \eqref{pureSO7monopole2} and \eqref{pureSO7monopole3} yields 
\begin{align}
\textcircled{\scriptsize 1} =& \frac{\partial \mathcal{F}_{SO(7)}}{\partial \phi_1},\\
\textcircled{\scriptsize 2} =& \frac{\partial \mathcal{F}_{SO(7)}}{\partial \phi_2},\\
2\times\textcircled{\scriptsize 3} =& \frac{\partial \mathcal{F}_{SO(7)}}{\partial \phi_3}.
\end{align}
Therefore, a D3-brane can stretch along the region $\textcircled{\scriptsize 1}$ or the region $\textcircled{\scriptsize 2}$. On the other hand, one needs to double the area of the region $\textcircled{\scriptsize 3}$, which implies that a D3-brane will not end on the $\widetilde{\text{O5}}^-$-plane and will end on a mirror D5-brane.
%This is consistent with the fact that the Langlands dual of an SO(odd) group is an Sp group, which has a long root. \ssk{Is this argument applicable for G2?}

\paragraph{$SO(7)$ with a massless spinor.}
The next example is the $SO(7)$ gauge theory with a hypermultiplet in the spinor representation. For simplicity we consider a case where the spinor matter is massless. In this case, there are several phases where the effective mass of the hypermultiplets vanishes. We here choose a phase where $[0, 0, 1], [0, 1, -1], [1, -1, 1], [-1, 0, 1]$ among the weights of the spinor representation are positive. Then the prepotential of the $SO(7)$ gauge theory with a massless spinor becomes
\begin{align}
\mathcal{F}_{SO(7)_s} =&m_0(\phi_1^2 - \phi_1\phi_2 + \phi_2^2 - 2\phi_2\phi_3 + 2\phi_3^2)\cr
&+\frac{4}{3}\phi_1^3 - \phi_1\phi_2^2 + \frac{4}{3}\phi_2^3 - \phi_1^2\phi_3 + \phi_1\phi_2\phi_3 - 3\phi_2^2\phi_3 + 2\phi_2\phi_3^2 + \phi_3^3. \label{SO7wsprep}
\end{align}
The monopole string tension is then 
\begin{align}
\frac{\partial \mathcal{F}_{SO(7)_s}}{\partial \phi_1} =& (2\phi_1 - \phi_2)(m_0 + 2\phi_1 + \phi_2 - \phi_3), \label{SO7wsmonopole1}\\
\frac{\partial \mathcal{F}_{SO(7)_s}}{\partial \phi_2} =& (-\phi_1 + 2\phi_2 - 2\phi_3)(m_0 + 2\phi_2 - \phi_3),\label{SO7wsmonopole2}\\
\frac{\partial \mathcal{F}_{SO(7)_s}}{\partial \phi_3} =& m_0(-2\phi_2 + 4\phi_3) - \phi_1^2 + \phi_1\phi_2 -3\phi_2^2 + 4\phi_2\phi_3 + 3\phi_3^2. \label{SO7wsmonopole3}
\end{align}

\begin{figure}
\centering
\includegraphics[width=8cm]{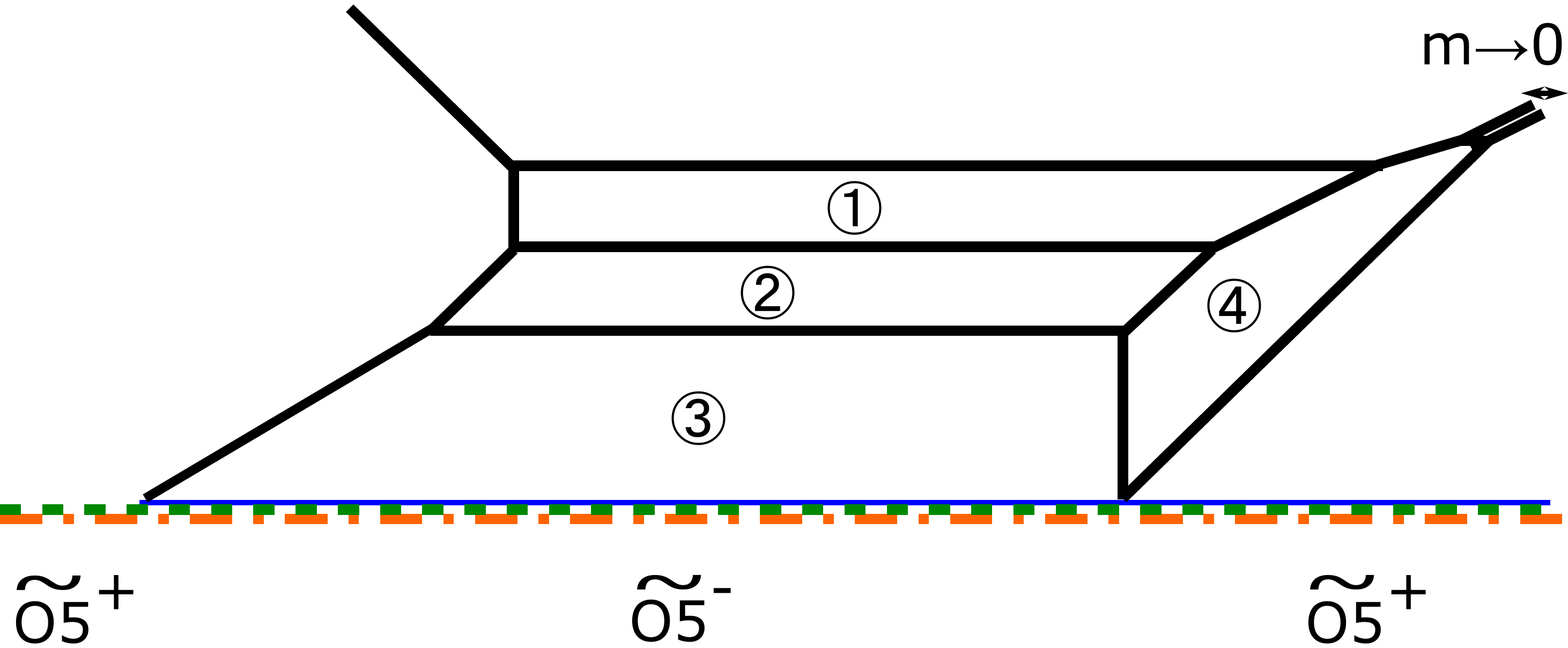}
\caption{A 5-brane web diagram of the $SO(7)$ gauge theory with a massless spinor. The mass is related to the length between the parallel $(2, 1)$ 5-branes and they are on top of each other in the massless limit. We also label the four faces in the diagram.}
\label{fig:SO7wmlspinor}
\end{figure}

\begin{figure}
\centering
\includegraphics[width=8cm]{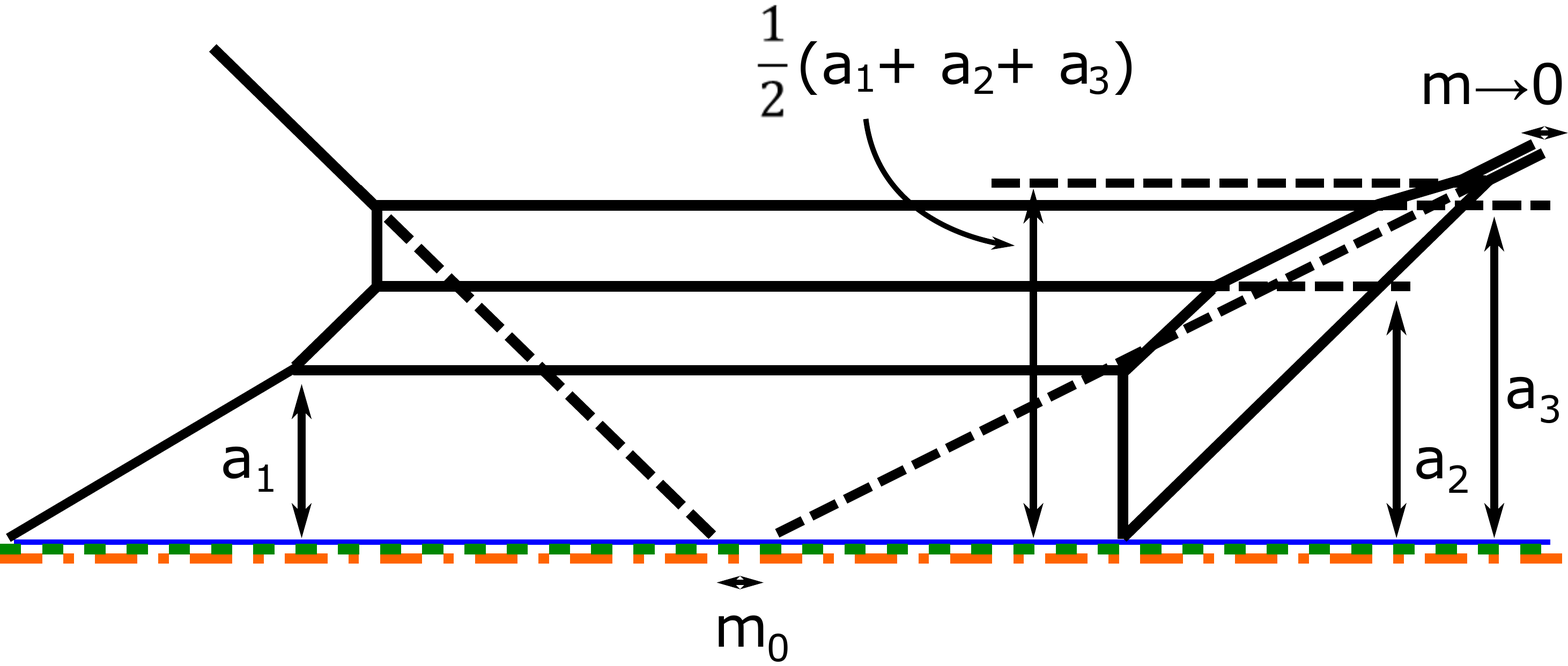}
\caption{A gauge theory parameterization for the $SO(7)$ gauge theory with a massless spinor. $a_1, a_2, a_3$ are the Coulomb branch moduli and $m_0$ is the inverse of the squared gauge coupling.}
\label{fig:SO7wmlspinorparameter}
\end{figure}

Let us then compute the area of faces in a diagram for the $SO(7)$ gauge theory with a massless spinor. We need to use a particular diagram which corresponds to the phase we chose to compute the prepotential \eqref{SO7wsprep}. Such a diagram is depicted in Figure \ref{fig:SO7wmlspinor} and we label the four faces in the diagram. 
%%%%%%%%%%%%%%%%%%%%%%%%%%%%%%%%%%
%%%%%%%%%%%%%%%%%%%%%%%%%%%%%%%%%%
The Coulomb branch moduli $a_1, a_2, a_3$ are again the height of the bottom color D5-brane, the middle color D5-brane and the top color D5-brane respectively. We also extrapolate the external $(2, 1)$ 5-brane and the external $(1, -1)$ 5-brane on top of the $\widetilde{\text{O5}}$-plane and the distance between the extrapolated external 5-branes is the inverse of the squared gauge coupling $m_0$. The gauge theory parameterization is summarized in Figure \ref{fig:SO7wmlspinorparameter}. 
%%%%%%%%%%%%%%%%%%%%%%%%%%%%%%%%%%
%%%%%%%%%%%%%%%%%%%%%%%%%%%%%%%%%%

We can now compute the area of the four faces in Figure \ref{fig:SO7wmlspinor} by the parameterization in Figure \ref{fig:SO7wmlspinorparameter}. The result is 
\begin{align}
\textcircled{\scriptsize 1} =&\frac{1}{2}(a_3-a_2)(2m_0-a_1+a_2+5a_3),\label{SO7wsmonopole4}\\
\textcircled{\scriptsize 2} =&\frac{1}{2}(a_2 - a_1)(2m_0 - a_1 + 3a_2 + 3a_3), \label{SO7wsmonopole5}\\
\textcircled{\scriptsize 3} =&\frac{1}{2}a_1(2m_0 + a_1 + 3a_2 + 3a_3), \label{SO7wsmonopole6}\\
\textcircled{\scriptsize 4} =&\frac{1}{4}(-a_1^2 + 2a_1a_2 - a_2^2 + 2a_1a_3 + 2a_2a_3 - a_3^2). \label{SO7wsmonopole7}
\end{align}
The comparision of \eqref{SO7wsmonopole4}, \eqref{SO7wsmonopole5}, \eqref{SO7wsmonopole6} and \eqref{SO7wsmonopole7} with \eqref{SO7wsmonopole1}, \eqref{SO7wsmonopole2} and \eqref{SO7wsmonopole3} yields relations
\begin{align}
\textcircled{\scriptsize 1} =& \frac{\partial \mathcal{F}_{SO(7)_s}}{\partial \phi_1},\\
\textcircled{\scriptsize 2} =& \frac{\partial \mathcal{F}_{SO(7)_s}}{\partial \phi_2},\\
2\times\textcircled{\scriptsize 3} +  \textcircled{\scriptsize 4} =& \frac{\partial \mathcal{F}_{SO(7)_s}}{\partial \phi_3},
\end{align}
using the relation \eqref{SO7relation}. As in the case of the pure $SO(7)$ gauge theory, we need to double the area of the region $\textcircled{\scriptsize 3}$. In fact, we further need to add the area of the region $\textcircled{\scriptsize 4}$ for the monopole string tension corresponding to $\frac{\partial \mathcal{F}_{SO(7)_s}}{\partial \phi_3}$. This fact becomes important also for the comparison of the monopole tension in the case of the pure $G_2$ gauge theory.

\paragraph{Pure $G_2$.}
Finally we consider the prepotential of the pure $G_2$ gauge theory. When we use the Dynkin basis for parametrizing  the Coulomb branch moduli $\phi_i$, the prepotential of the 5d pure $G_2$ gauge theory becomes 
\begin{equation}
\mathcal{F}_{G_2}(\phi) = m_0(\phi_1^2 - 3\phi_1\phi_2 + 3\phi_2^2) + \frac{4}{3}\phi_1^3 - 4\phi_1^2\phi_2 + 3\phi_1\phi_2^2 + \frac{4}{3}\phi_2^3, \label{G2prep}
\end{equation}
where we chose $[2, -3]$ and $[-1, 2]$ for the simple roots for defining the Weyl chamber.
%This can be obtained either by using \eqref{prepotential} directlyor by replacing the Coulomb branch moduli as $\phi_1 \to \phi_2$, $\phi_2 \to \phi_1$, $\phi_3 \to \phi_2$ in \eqref{SO7wsprep},the latter of which corresponds to the Higgsing from the $SO(7)$ gauge theory to the $G_2$ gauge theory.
Hence the expected monopole tension from the prepotential \eqref{G2prep} is
\begin{eqnarray}
\frac{\partial \mathcal{F}_{G_2}}{\partial \phi_1} &=& (m_0 + 2\phi_1 - \phi_2)(2\phi_1 - 3\phi_2),\label{G2monopole1}\\
\frac{\partial \mathcal{F}_{G_2}}{\partial \phi_2} &=&(- \phi_1 + 2\phi_2)(3m_0 + 4\phi_1 + 2\phi_2). \label{G2monopole2}
\end{eqnarray}
%We note that one can easily see that the Higgsing of the $SO(7)$ prepotential \eqref{SO7wsprep} leads to the $G_2$ prepotential \eqref{G2prep}, as the Higgsing enforces the parameters for the $SO(7)$ gauge theory to be $a_3=a_1+a_2$ in Figure \ref{fig:pureSO7parameter}, or equivalently $\phi_1=\phi_3$, followed by $\phi_1\to \phi_2$ and $\phi_2\to\phi_1$.
%It follows that
%\begin{align}\label{Eq:so7toG2Higgsing}
%	\frac{\partial \mathcal{F}_{G_2}}{\partial \phi_1} &=~ \frac{\partial \mathcal{F}_{SO(7)}}{\partial \phi_2} {}\Big|_{\rm Higgsing},
%	\qquad
%\frac{\partial \mathcal{F}_{G_2}}{\partial \phi_2} = \Big(
%\frac{\partial \mathcal{F}_{SO(7)}}{\partial \phi_1}+\frac{\partial \mathcal{F}_{SO(7)}}{\partial \phi_3}\Big)  {}\Big|_{\rm Higgsing},
%\end{align}
%by the subscript, Higgsing, we mean $\phi_1=\phi_3$ followed by substitution $\phi_1\to \phi_2$ and $\phi_2\to\phi_1$.

\begin{figure}
\centering
\includegraphics[width=8cm]{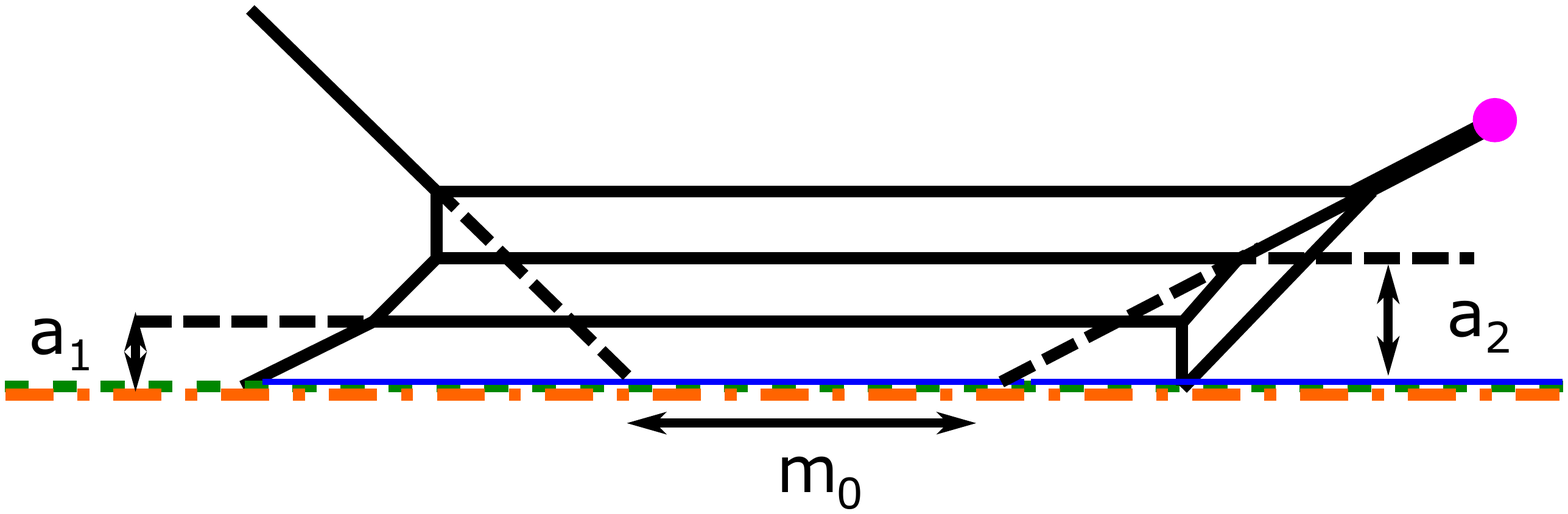}
\caption{A gauge theory parameterization for the pure $G_2$ gauge theory. $a_1, a_2$ are the Coulomb branch moduli and $m_0$ is related to the inverse of the gauge coupling.}
\label{fig:pureG2parameter1}
\end{figure}

\begin{figure}
\centering
\includegraphics[width=8cm]{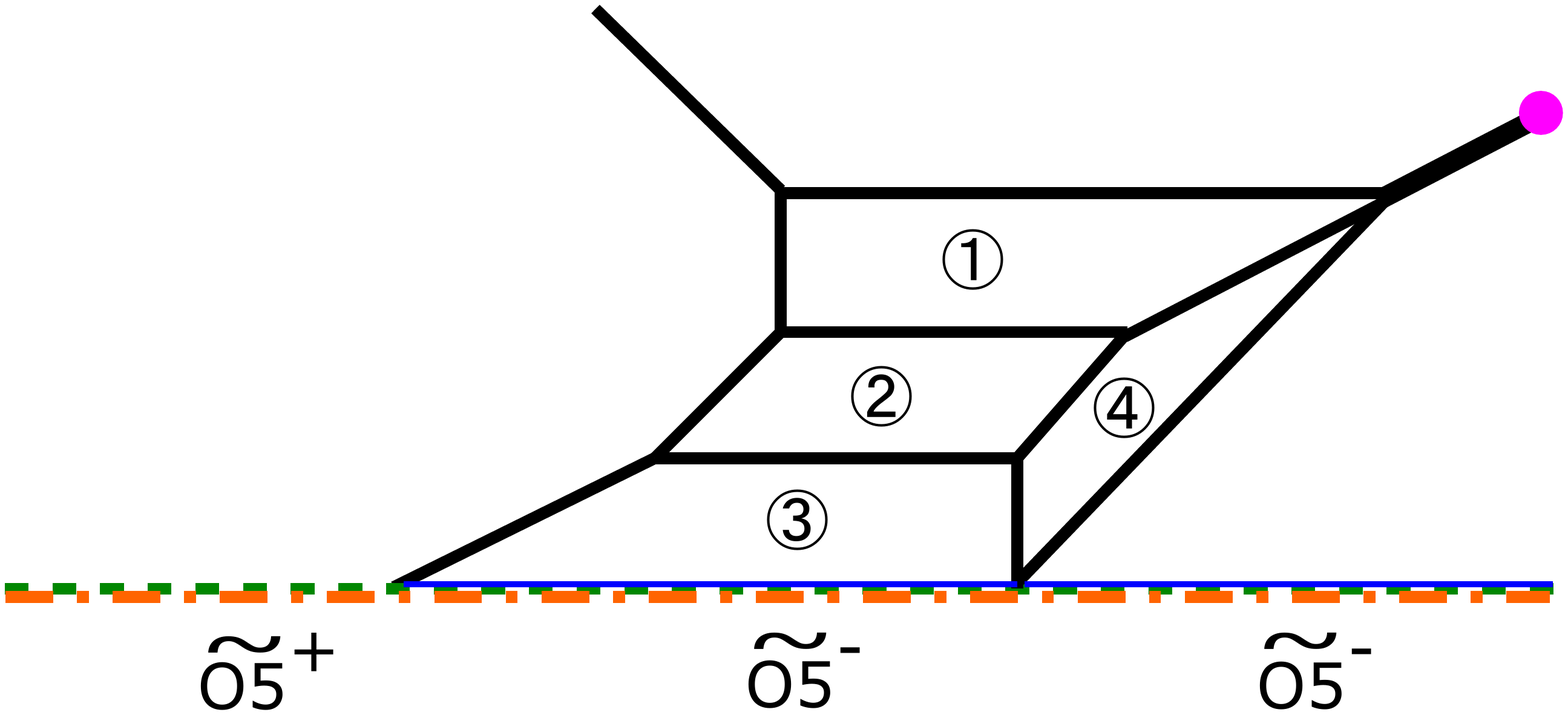}
\caption{Labeling for the four faces in the pure $G_2$ diagram. Note that the regions $\textcircled{\scriptsize 1}$ and $\textcircled{\scriptsize 4}$ are not separated by a 5-brane and are connected to each other.}
\label{fig:pureG2monopole1}
\end{figure}

%Before computing the tension of a monopole string from the 5-brane web of the pure $G_2$ gauge theory from the diagram in Figure \ref{fig:pureG2}, we identify the gauge theory parameters with the length of 5-branes in the diagram. The height of the color D5-branes is related to the Coulomb branch moduli and hence we denote the height of the lowest color D5-brane by $a_1$ and the height of the second lowest color D5-brane by $a_2$ as in Figure \ref{fig:pureG2parameter1}. 
The gauge theory parameterization for the pure $G_2$ gauge theory realized in Figure \ref{fig:pureG2} can be understood in a similar way to the case of the $SO(7)$ gauge theories. We denote the height of the lowest color D5-brane by $a_1$ and the height of the second lowest color D5-brane by $a_2$ as in Figure \ref{fig:pureG2parameter1}. 
%%%%%%%%%%%%%%%%%%%%%%%%%%%%%%%%%%
%%%%%%%%%%%%%%%%%%%%%%%%%%%%%%%%%%
The inverse of the square gauge coupling can be calculated from the distance between the extrapolated external $(2, 1)$ 5-brane and the external $(1, -1)$ 5-brane and it is denoted by $m_0$ in Figure \ref{fig:pureG2parameter1}.
%On the other hand, the inverse of the sqaured gauge coupling is obtained by extrapolating the external $(2, 1)$ 5-brane and the external $(1, -1)$ 5-brane to the potision of an $\widetilde{\text{O5}}$-plane. In other words, we consider a limit where the Coulomb branch moduli become zero. Then the length of D5-branes between the extrapolated $(2, 1)$ 5-brane and the extrapolated $(1, -1)$ 5-brane yields the inverse of the squared gauge coupling which is denoted by $m_0$ in Figure \ref{fig:pureG2parameter1}. 
We are now able to compute the area corresponding to the tension of monopoles strings by using the parameters in Figure \ref{fig:pureG2parameter1}. We label four faces in the pure $G_2$ diagram as in Figure \ref{fig:pureG2monopole1}. 
%%%%%%%%%%%%%%%%%%%%%%%%%%%%%%%%%%
%%%%%%%%%%%%%%%%%%%%%%%%%%%%%%%%%%
The area of the four regions is
\begin{eqnarray}
\textcircled{\scriptsize 1} &=& a_1(m_0 + 2a_1 + 3a_2),\\
\textcircled{\scriptsize 2} &=& (a_2 - a_1)(m_0 + a_1 + 3a_2),\\
\textcircled{\scriptsize 3} &=& a_1(m_0 + 2a_1 + 3a_2),\\
\textcircled{\scriptsize 4} &=& a_1a_2.
\end{eqnarray}

We can deduce which area we should compare with the monopole string tension \eqref{G2monopole1} and \eqref{G2monopole2} from the analysis of the $SO(7)$ gauge theory with a massless spinor. 
%In that case, 
For the $SO(7)$ gauge theory with a spinor, the area corresponding to the monopole string tension is $\textcircled{\scriptsize 1}$, $\textcircled{\scriptsize 2}$, $2\times\textcircled{\scriptsize 3} + \textcircled{\scriptsize 4}$. After the Higgsing,
% the diagram of the $SO(7)$ gauge theory with a spinor, 
the region $\textcircled{\scriptsize 1}$ is combined with $\textcircled{\scriptsize 4}$, hence the area corresponding to the monopole string tension for the pure $G_2$ gauge theory should be $\textcircled{\scriptsize 2}$ and $\textcircled{\scriptsize 1} + 2\times \textcircled{\scriptsize 3} + \textcircled{\scriptsize 4}$. Such an area is given by
\begin{eqnarray}
\textcircled{\scriptsize 2} &=& (a_2 - a_1)(m_0 + a_1 + 3a_2), \label{G2monopole3}\\
\textcircled{\scriptsize 1} +2\times\textcircled{\scriptsize 3} +  \textcircled{\scriptsize 4}  &=&a_1(3m_0 + 6a_1 + 10a_2). \label{G2monopole4}
\end{eqnarray}

We compare the tension \eqref{G2monopole1} and \eqref{G2monopole2} computed from the prepotential \eqref{G2prep} with the tension \eqref{G2monopole3} and \eqref{G2monopole4} from the area of the 5-brane web in Figure \ref{fig:pureG2monopole1}. Note that the Coulomb branch moduli $a_1, a_2$ are related to $\phi_1, \phi_2$ by
\begin{equation}
2\phi_1 - 3\phi_2 = a_2 - a_1, \quad -\phi_1 + 2\phi_2 = a_1. \label{G2relation}
\end{equation}
Using the relation \eqref{G2relation}, we can see that 
\begin{align}
\textcircled{\scriptsize 2} =&\frac{\partial \mathcal{F}_{G_2}}{\partial \phi_1},\label{Eq:G2monotension1}\\
\textcircled{\scriptsize 1} +2\times\textcircled{\scriptsize 3} +  \textcircled{\scriptsize 4}  = &\frac{\partial \mathcal{F}_{G_2}}{\partial \phi_2}.\label{Eq:G2monotension2}
\end{align}
%The analysis of the prepotential presents further support for the claim that the diagram in Figure \ref{fig:pureG2} realizes the 5d pure $G_2$ gauge theory.

We note that the Higgsing of the $SO(7)$ gauge theory with a spinor enforces the parameters for the $SO(7)$ gauge theory to be $a_3=a_1+a_2$ in Figure \ref{fig:SO7wmlspinorparameter}. This means that the $SO(7)$ Coulomb branch moduli satisfy $\phi_1=\phi_3$. With the proper map between the $SO(7)$ Coulomb branch moduli $\phi_i^{SO(7)}$ and the $G_2$ Coulomb moduli $\phi^{G_2}_j$ given by $\phi_1^{SO(7)}\to \phi^{G_2}_2$ and $\phi_2^{SO(7)}\to\phi_1^{G_2}$, one can easily see that the prepotential for 5d $SO(7)$ gauge theory with a spinor \eqref{SO7wsprep} becomes  the $G_2$ prepotential \eqref{G2prep}. It follows that the tensions of the monopole strings are also consistent with the Higgsing
\begin{align}\label{Eq:so7toG2Higgsing}
\frac{\partial \mathcal{F}_{SO(7)_s}}{\partial \phi_2} {}\Big|_{\rm Higgsing} &=~ 	\frac{\partial \mathcal{F}_{G_2}}{\partial \phi_1},
	\qquad
\Big(
\frac{\partial \mathcal{F}_{SO(7)_s}}{\partial \phi_1}+\frac{\partial \mathcal{F}_{SO(7)_s}}{\partial \phi_3}\Big)  {}\Big|_{\rm Higgsing} =  \frac{\partial \mathcal{F}_{G_2}}{\partial \phi_2},
\end{align}
or in other words, through the Higgsing, \eqref{SO7wsmonopole3} $\to$ \eqref{Eq:G2monotension1} and  agrees with \eqref{SO7wsmonopole1} $+$\eqref{SO7wsmonopole3}, $\to$ \eqref{Eq:G2monotension2}.
%where, by the subscript, Higgsing, we mean $\phi_1=\phi_3$ followed by substitution $\phi_1\to \phi_2$ and $\phi_2\to\phi_1$.
The analysis of the prepotential therefore presents further support for the claim that the diagram in Figure \ref{fig:pureG2} realizes the 5d pure $G_2$ gauge theory.

\subsection{Adding flavors to $G_2$}
\label{sec:addingF}

\begin{figure}
\centering
\subfigure[]{
\includegraphics[width=7cm]{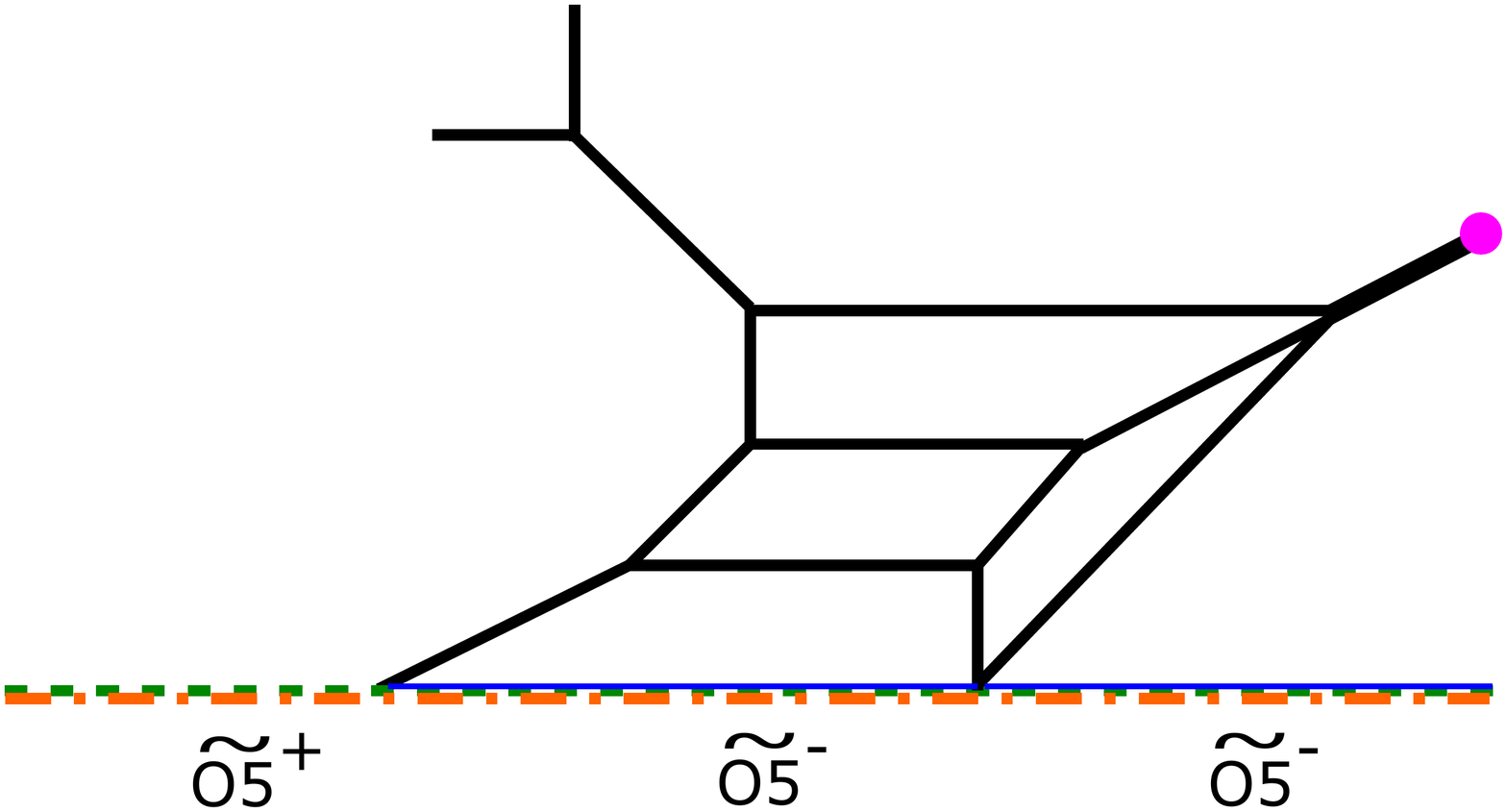} \label{fig:G2w1flvra}}
\subfigure[]{
\includegraphics[width=7cm]{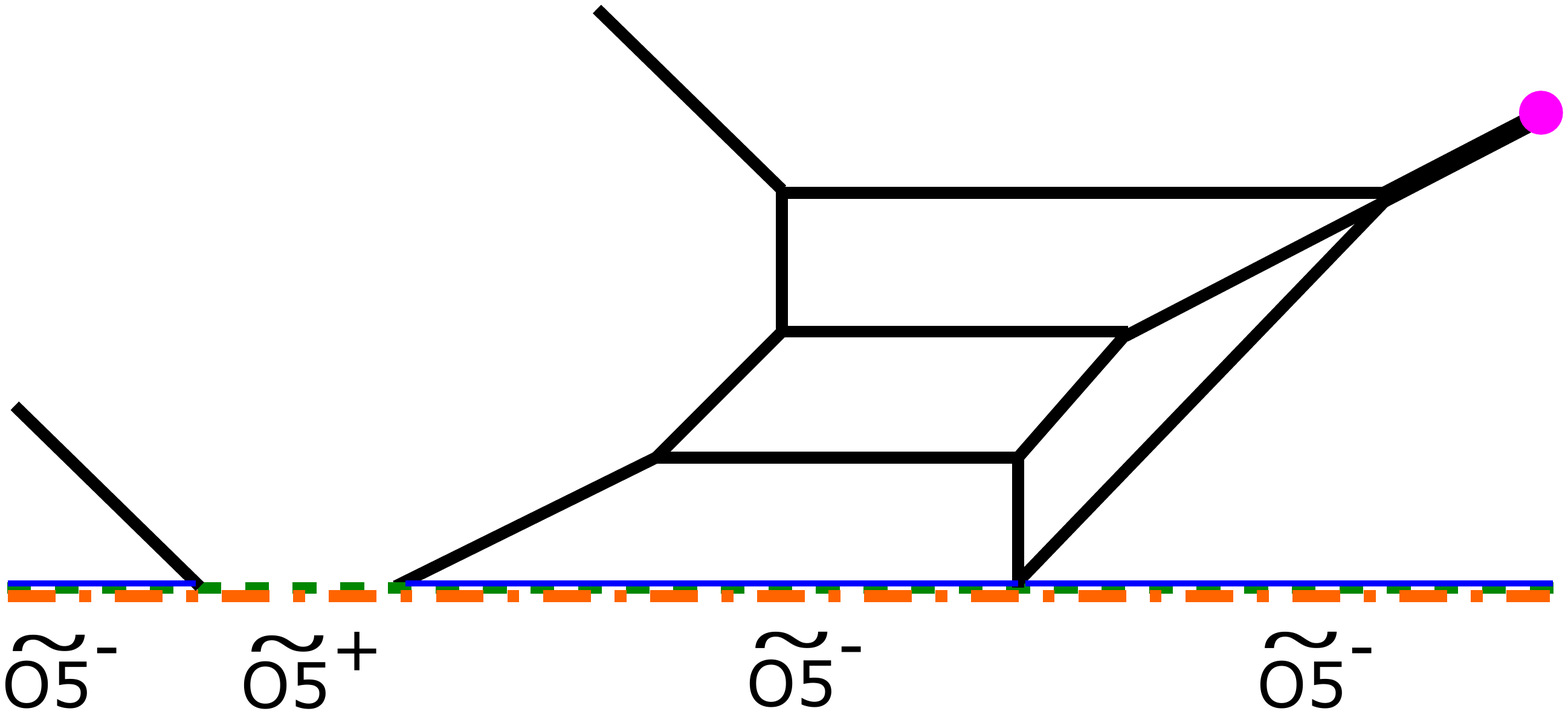} \label{fig:G2w1flvrb}}
\caption{5-brane web diagrams of the $G_2$ gauge theory with one flavor. (a): The diagram is obtained by Higgsing the $SO(7)$ gauge theory with one vector and one spinor. The resulting theory is $G_2$ gauge theory with one flavor. (b): The diagram is obtained by Higgsing the $SO(7)$ gauge theory with two spinors. The resulting theory is the $G_2$ gauge theory with one flavor and a singlet.}
\label{fig:G2w1flvr}
\end{figure}

\begin{figure}
\centering
\includegraphics[width=8cm]{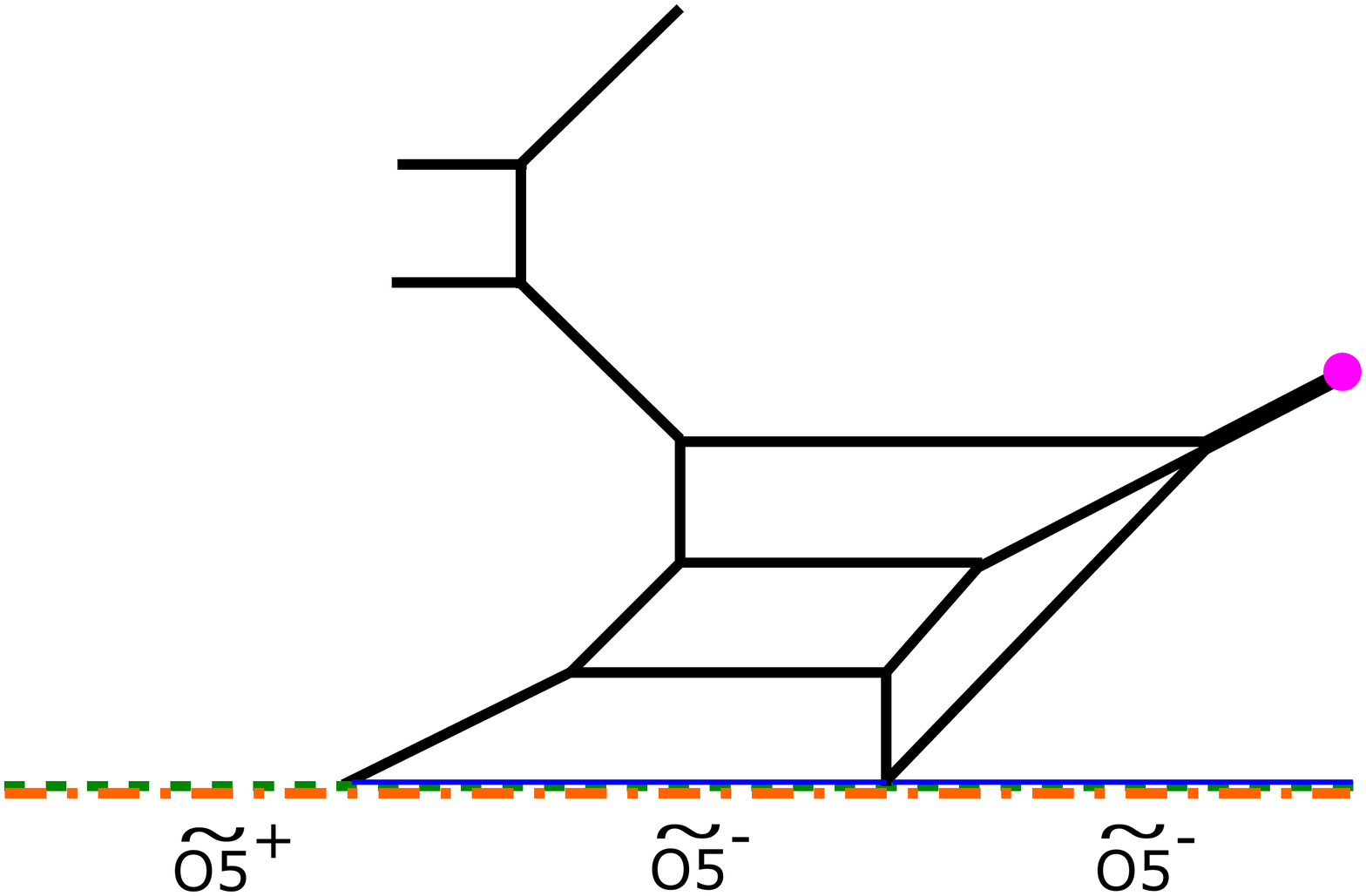}
\caption{A 5-brane diagram for the $G_2$ gauge theory with two flavors.}
\label{fig:G2w2flvr}
\end{figure}

5d $G_2$ gauge theories may have hypermultiplets in the fundamental representation and the maximal number flavors for a $G_2$ gauge theory to have a 5d UV fixed point is five \cite{Zafrir:2015uaa, Jefferson:2017ahm}. 
From the viewpoint of 5-brane webs, one can also add hypermultiplets in the fundamental representation of $G_2$ to the 5-brane web diagram in Figure \ref{fig:pureG2}. There are two ways to introduce flavors for the $G_2$ theory. One way uses the vector matter of the $SO(7)$ gauge theory and the other way utilizes the spinor matter of the $SO(7)$ gauge theory. After the Higgsing from $SO(7)$ to $G_2$, the former becomes the fundamental matter of the $G_2$ gauge theory and the latter becomes the fundamental matter plus a singlet hypermultiplet of the $G_2$ gauge theory. The singlet appears since the spinor representation of $SO(7)$ is the eight-dimensional representation and the fundamental representation of $G_2$ is the seven-dimensional representation. Hence, the Higgsing of the $SO(7)$ gauge theory with a hypermultiplet in the vector representation and a hypermultiplet in the spinor representation gives the $G_2$ gauge theory with one flavor, and similarly the Higgsing of the $SO(7)$ gauge theory with two hypermultiplets in the spinor representation gives the $G_2$ gauge theory with one flavor and a singlet. The two 5-brane diagrams giving the $G_2$ gauge theories with one flavor are depicted in Figure \ref{fig:G2w1flvra} and \ref{fig:G2w1flvrb}.
%%%%%%%%%%%%%%%%%%%%%%%%%%%%%%%%%%
%%%%%%%%%%%%%%%%%%%%%%%%%%%%%%%%%%
It is straightforward to add more flavors to $G_2$ gauge theories by Higgsing the 5d $SO(7)$ gauge theory with one spinor and more than one hypermultiplets either in the vector representation or the spinor representation. An example of a 5-brane web diagram for the $G_2$ gauge theory with two flavors is depicted in Figure \ref{fig:G2w2flvr}.
%%%%%%%%%%%%%%%%%%%%%%%%%%%%%%%%%%
%%%%%%%%%%%%%%%%%%%%%%%%%%%%%%%%%%

\begin{figure}
\centering
\subfigure[]{
\includegraphics[width=7cm]{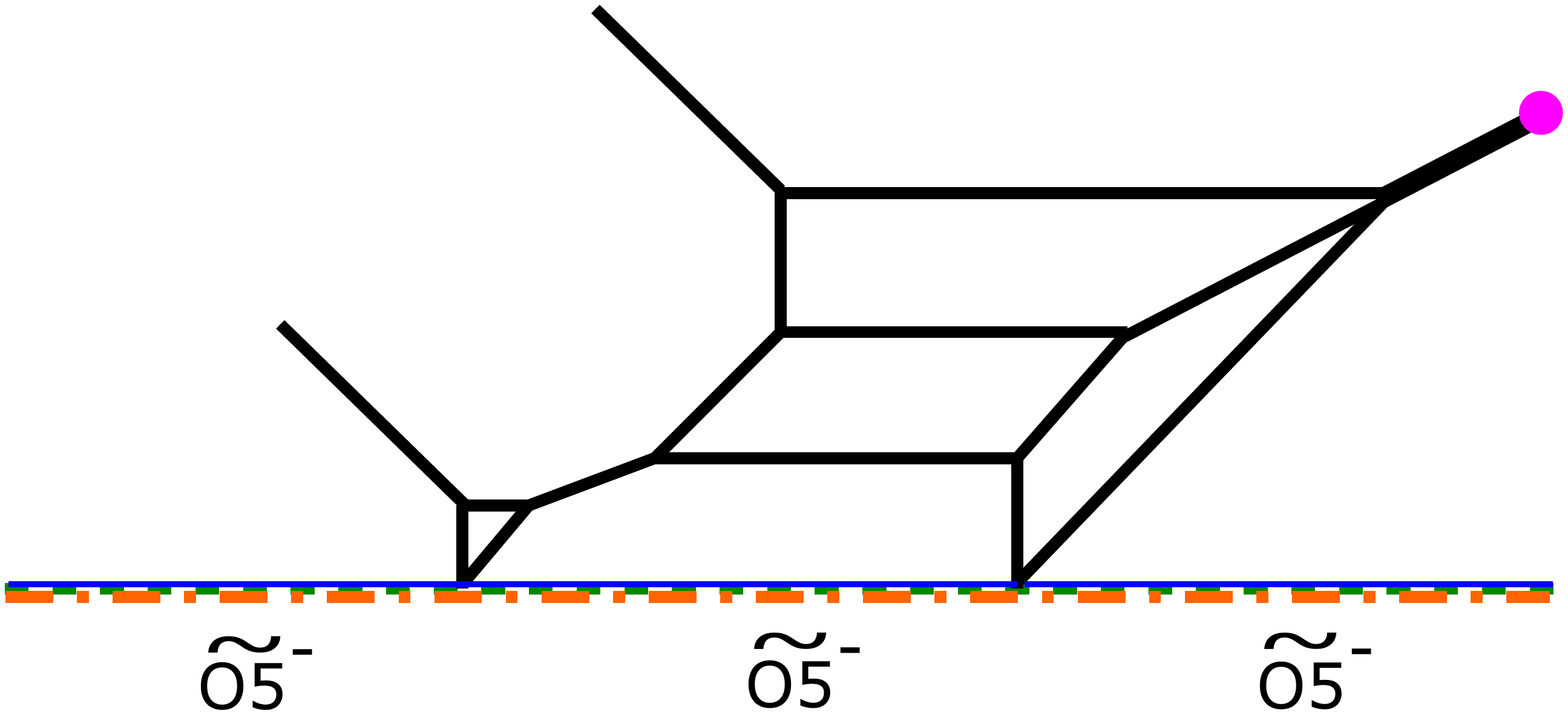} \label{fig:G2w1flvrfloppeda}}
\subfigure[]{
\includegraphics[width=7cm]{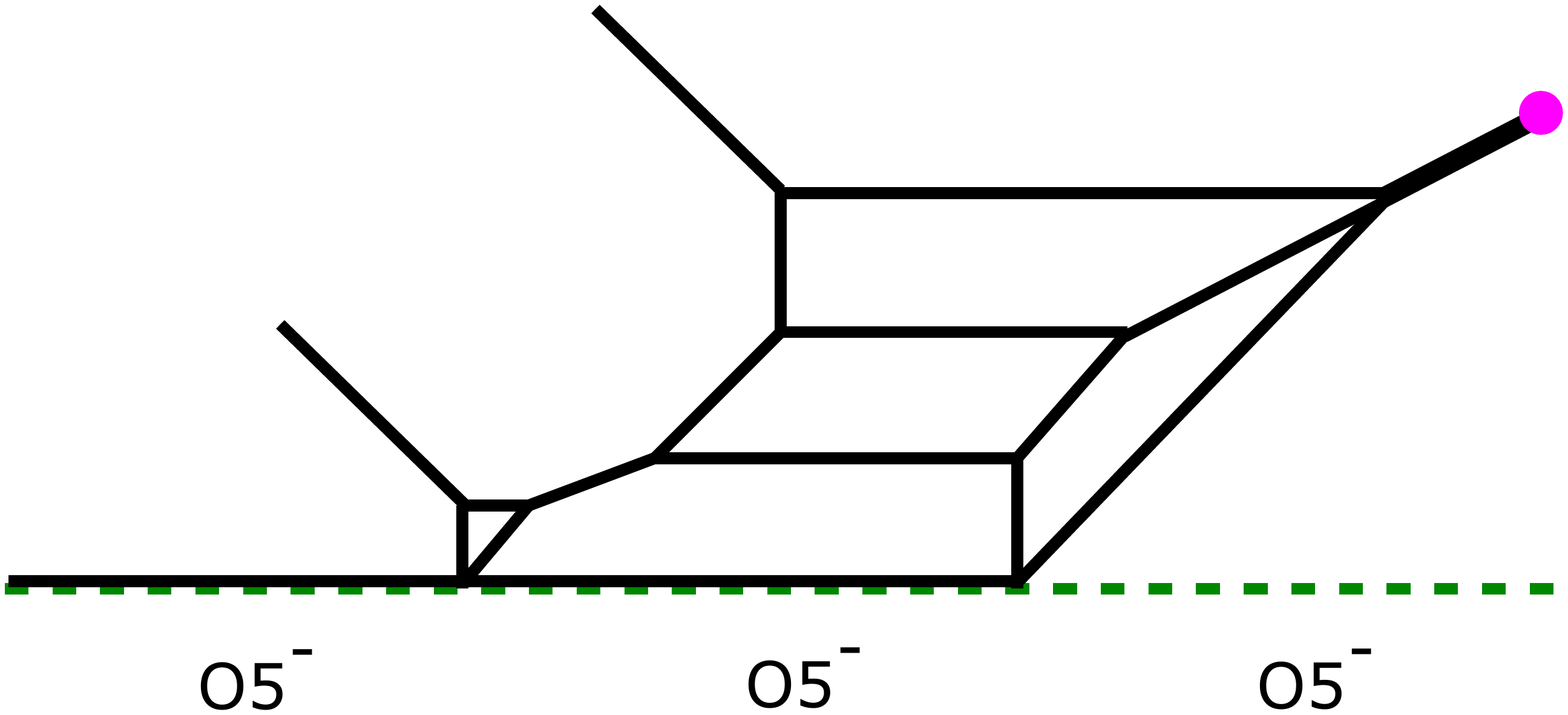} \label{fig:G2w1flvrfloppedb}}
\caption{5-brane diagrams for the $G_2$ gauge theory with one flavor and a singlet after performing the generalized flop transitions. (a): A 5-brane web for a $G_2$ gauge theory with one flavor which is obtained after performing the generalized flop transition to the diagram in Figure \ref{fig:G2w1flvrb}. The diagram contains only an $\widetilde{\text{O5}}^-$-plane. (b): An equivalent 5-brane web diagram to the one in Figure \ref{fig:G2w1flvrfloppeda}. We move the half D7-brane in the infinitely right to the infinitely left. Then the $\widetilde{\text{O5}}$-plane disappears and we only have an O5-plane. The diagram still yields the $G_2$ gauge theory with one flavor and a singlet. }
\label{fig:G2w1flvrflopped}
\end{figure}

As for the 5-brane web obtained by Higgsing the $SO(7)$ gauge theories with two spinors, one can also perform the generalized flop transition in Figure \ref{fig:flopO5tilde2} and then the 5-brane diagram becomes the one in Figure \ref{fig:G2w1flvrfloppeda}. It is also possible to obtain an equivalent diagram by moving the half D7-brane associated the monodromy cut from the infinitely right to the infinitely left in Figure \ref{fig:G2w1flvrfloppeda}, and the resulting diagram after the transition is given in Figure \ref{fig:G2w1flvrfloppedb} without any branch cut. 
%%%%%%%%%%%%%%%%%%%%%%%%%%%%%%%%%%
%%%%%%%%%%%%%%%%%%%%%%%%%%%%%%%%%%

%====================================================================
\bigskip
\section{Another 5-brane web for pure $G_2$ gauge theory}
\label{sec:G2fromO5}

In this section, we present another 5-brane web diagram for the 5d pure $G_2$ gauge theory without using an $\widetilde{\text{O5}}$-plane different to the one in section \ref{sec:G2fromO5tilde}. This diagram turns out to be useful for the topological vertex computation in section \ref{sec:Nekrasov}. 

\subsection{Higgsing 5d $SO(8)$ gauge theory with one spinor and one conjugate spinor}

In section \ref{sec:G2fromO5tilde}, we used the 5-brane web diagram in Figure \ref{fig:SO7wspinor2} for the 5d $SO(7)$ gauge theory with one spinor and the Higgsing of the diagram yielded the web diagram for the pure $G_2$ gauge theory in Figure \ref{fig:SO7wspinorfloppede}. The 5-brane diagram of the 5d $SO(7)$ gauge theory has been originally obtained by the Higgsing associated to vector matter of the 5d $SO(8)$ gauge theory. In other words, the 5-brane web diagram of the pure $G_2$ gauge theory in Figure \ref{fig:SO7wspinorfloppede} was obtained by the two successive Higgsings from the 5d $SO(8)$ gauge theory with one vector and one spinor. Due to the triality of the 5d $SO(8)$ gauge theory, the 5d $SO(8)$ gauge theory with one vector and one spinor is equivalent to the 5d $SO(8)$ gauge theory with one spinor and one conjugate spinor. Therefore, we should again obtain a 5-brane web diagram for the pure $G_2$ gauge theory by Higgsing a 5-brane web for the 5d $SO(8)$ gauge theory with one spinor and one conjugate spinor.

To introduce a hypermultiplet in the spinor representation to the 5-brane web diagram for the 5d $SO(8)$ gauge theory, we consider a ``quiver theory'' of $SO(8)- USp(0)$ and the $USp(0)$ instanton plays a role of the spinor matter \cite{Zafrir:2015ftn}. For introducing two spinors, we consider a quiver $USp(0) - SO(8) - USp(0)$. However, we need two spinors of opposite chirality. The difference between a spinor and a conjugate spinor can be realized by considering different discrete theta angles for the two $USp(0)$ gauge groups \cite{Zafrir:2015ftn}. Namely, we consider a 5-brane diagram of the quiver $USp(0) - SO(8)- USp(0)$ but the two $USp(0)$ gauge groups have different discrete theta angles\footnote{There is another 5-brane web diagram for the 5d $SO(8)$ gauge theory with one spinor and one conjugate spinor and it is given by a diagram for a quiver $SO(8) - USp(0) - [1]$. However, in order to perform a Higgsing, it is useful to consider a 5-brane web for the $USp(0) - SO(8) - USp(0)$ quiver theory.}. 

\begin{figure}
\centering
\includegraphics[width=8cm]{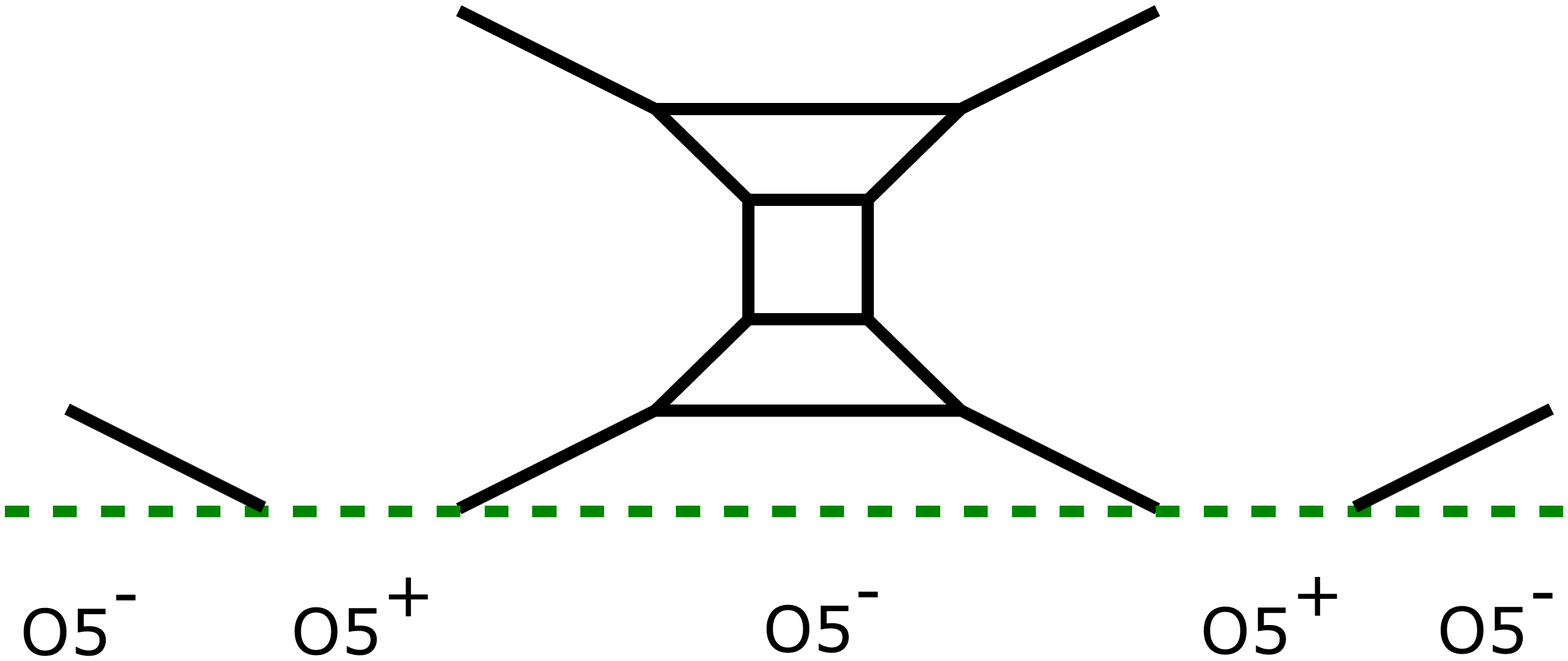}
\caption{A 5-brane web diagram for the 5d $SO(8)$ gauge theory with one spinor and one conjugate spinor. The discrete theta angle of the two $USp(0)$ theories is different from each other although the difference is not explicitly expressed in the diagram. }
\label{fig:SO8wspinorAcspinor}
\end{figure}

\begin{figure}
\centering
\subfigure[]{
\includegraphics[width=4cm]{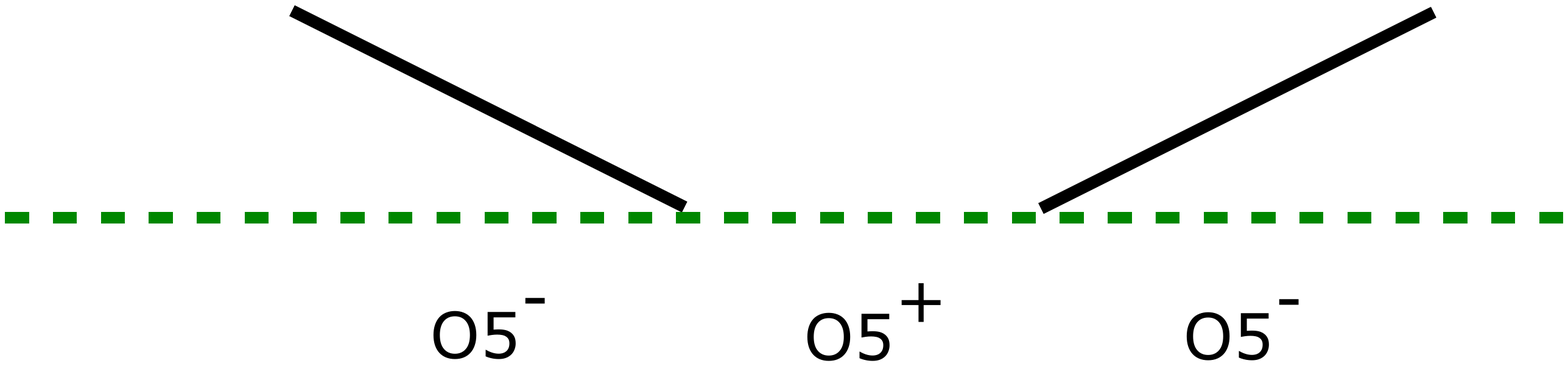} \label{fig:gflopO5a}}
\subfigure[]{
\includegraphics[width=4cm]{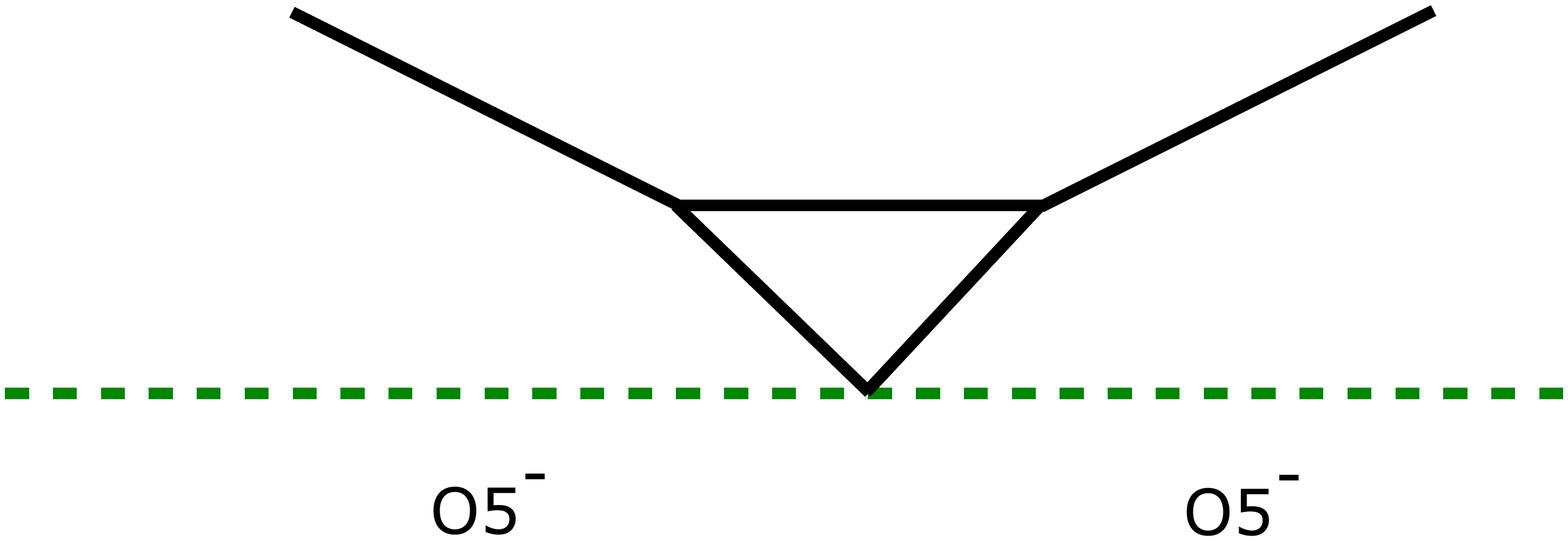} \label{fig:gflopO5b}}
\subfigure[]{
\includegraphics[width=4cm]{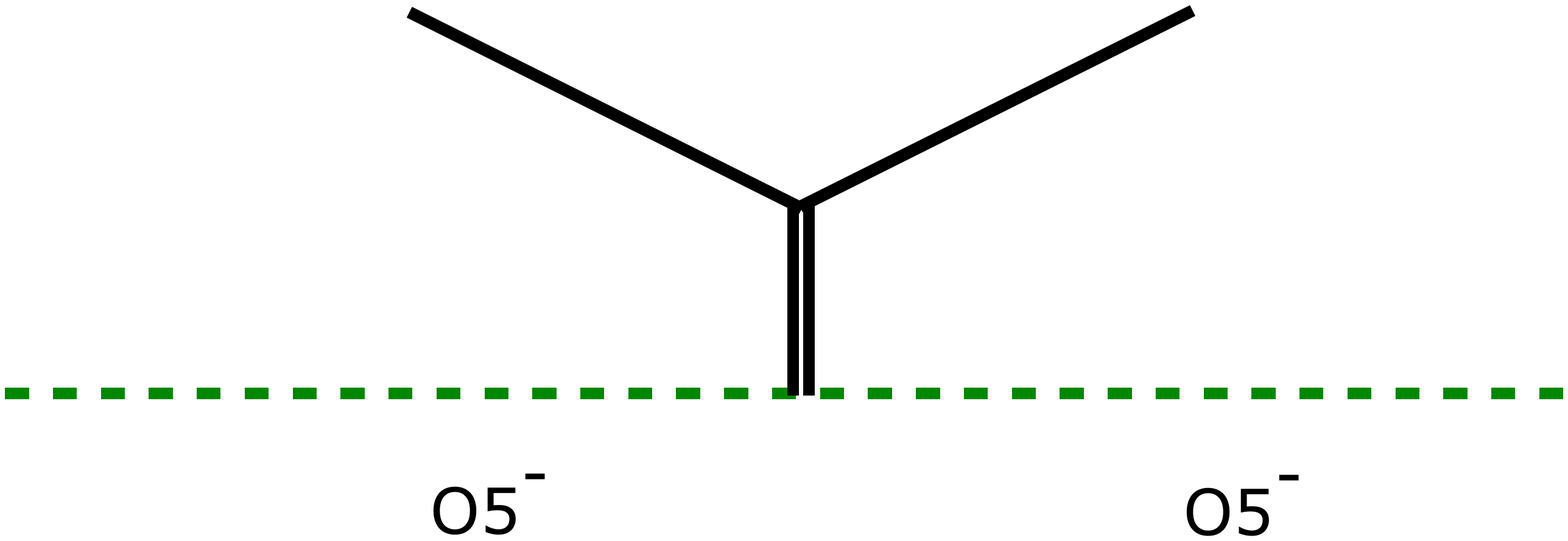} \label{fig:gflopO5c}}
\caption{The generalized flop transition for the 5-brane web for $USp(0)$. Depending on the discrete theta angle, the transition changes the figure (a) into either figure (b) or figure (c).}
\label{fig:gflopO5}
\end{figure}

\begin{figure}
\centering
\includegraphics[width=8cm]{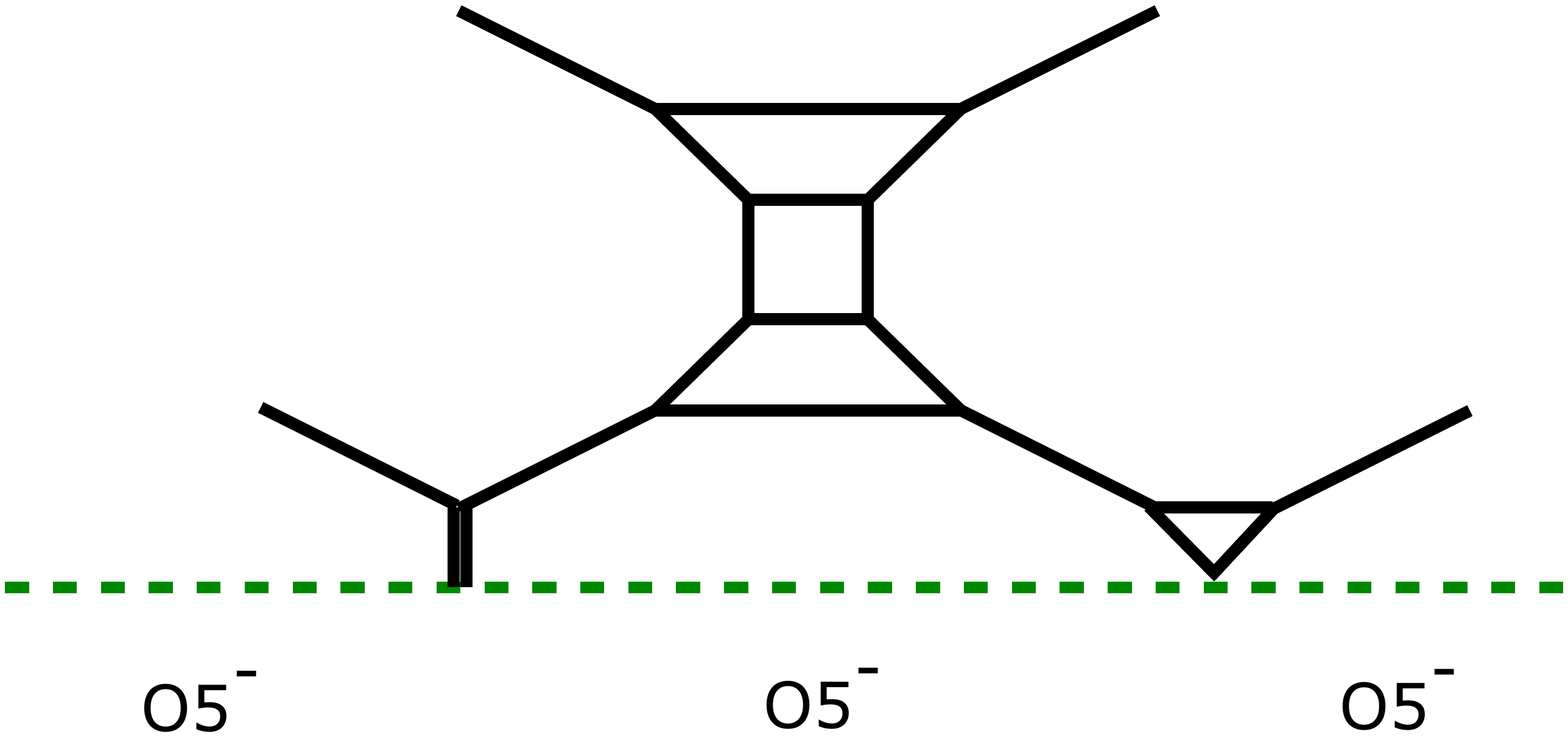}
\caption{A 5-brane web diagram for the 5d $SO(8)$ gauge theory with one spinor and one conjugate spinor after performing the generalized flop transitions. }
\label{fig:SO8wspinorAcspinor2}
\end{figure}

\begin{figure}
\centering
\includegraphics[width=8cm]{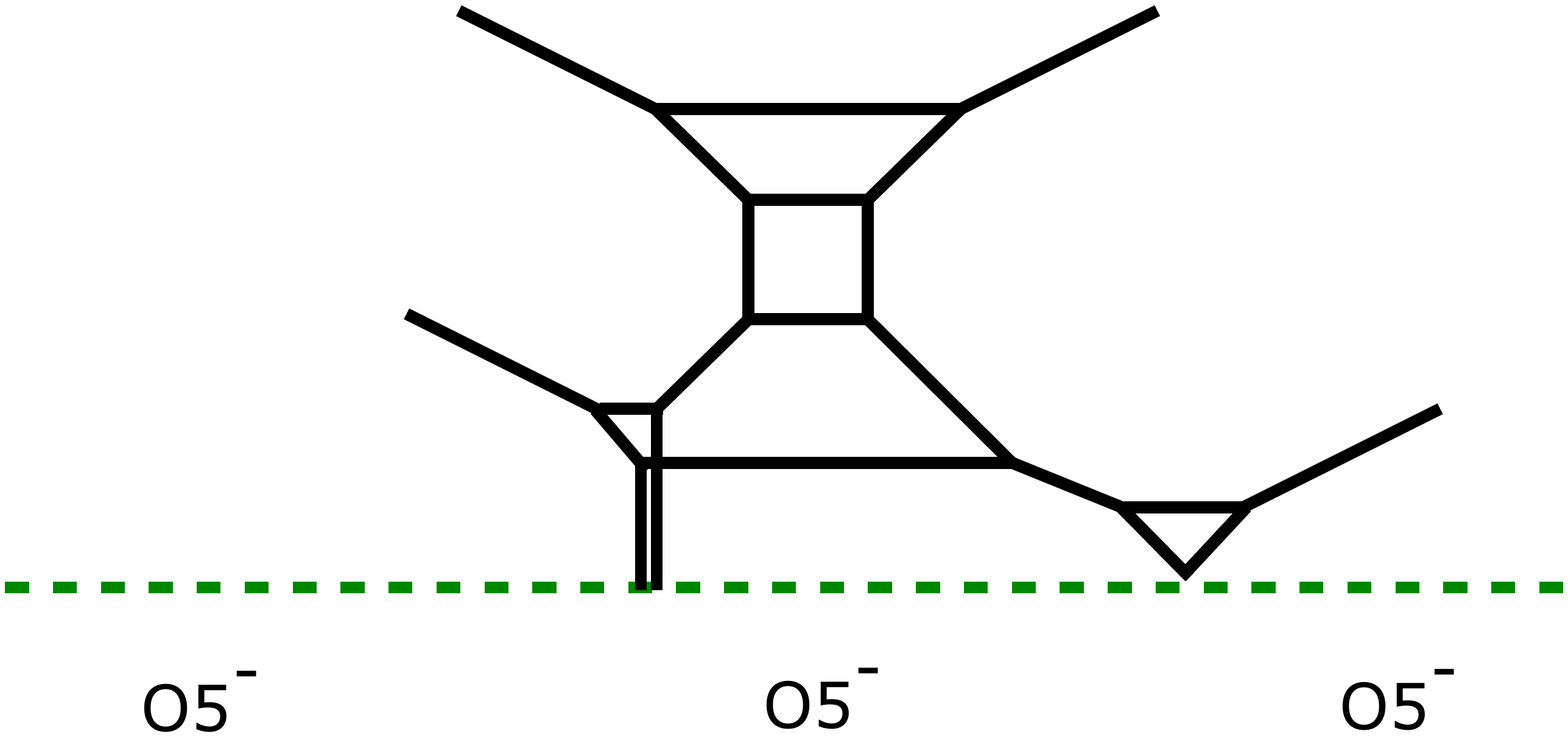}
\caption{Another 5-brane web diagram for the 5d $SO(8)$ gauge theory with one spinor and one conjugate spinor. }
\label{fig:SO8wspinorAcspinor3}
\end{figure}

A 5-brane web diagram of the 5d $SO(8)$ gauge theory with one spinor and one conjugate spinor is given in Figure \ref{fig:SO8wspinorAcspinor}. 
%%%%%%%%%%%%%%%%%%%%%%%%%%%%%%%%%
%%%%%%%%%%%%%%%%%%%%%%%%%%%%%%%%%
Note that there are two parallel external $(2, 1)$ 5-branes and two parallel external $(2, -1)$ 5-branes. Each two parallel external 5-branes implies an $SU(2)$ flavor symmetry and hence the theory shows an $SU(2) \times SU(2)$ perturbative flavor symmetry from one spinor and one conjugate spinor. At the level of the diagram in Figure \ref{fig:SO8wspinorAcspinor}, the difference of the discrete theta angle is invisible. However, the difference appears after the generalized flop transition \cite{Hayashi:2017btw}. The two different types of the flop transitions depending on the discrete theta angle are depicted in Figure \ref{fig:gflopO5}.
%%%%%%%%%%%%%%%%%%%%%%%%%%%%%%%%%
%%%%%%%%%%%%%%%%%%%%%%%%%%%%%%%%%
We can then apply the generalized flop transition in Figure \ref{fig:gflopO5} to the diagram in Figure \ref{fig:SO8wspinorAcspinor} and it yields another 5-brane web diagram for the 5d $SO(8)$ gauge theory with one spinor and one conjugate spinor. The resulting diagram is depicted in Figure \ref{fig:SO8wspinorAcspinor2}. 
%%%%%%%%%%%%%%%%%%%%%%%%%%%%%%%%%
%%%%%%%%%%%%%%%%%%%%%%%%%%%%%%%%%
In order to perform the Higgsing to the 5d pure $G_2$ gauge theory, we consider a further transition given in Figure \ref{fig:flopO5}, yielding a 5-brane web in Figure \ref{fig:SO8wspinorAcspinor3}. 
%%%%%%%%%%%%%%%%%%%%%%%%%%%%%%%%%
%%%%%%%%%%%%%%%%%%%%%%%%%%%%%%%%%

\begin{figure}
\centering
\includegraphics[width=8cm]{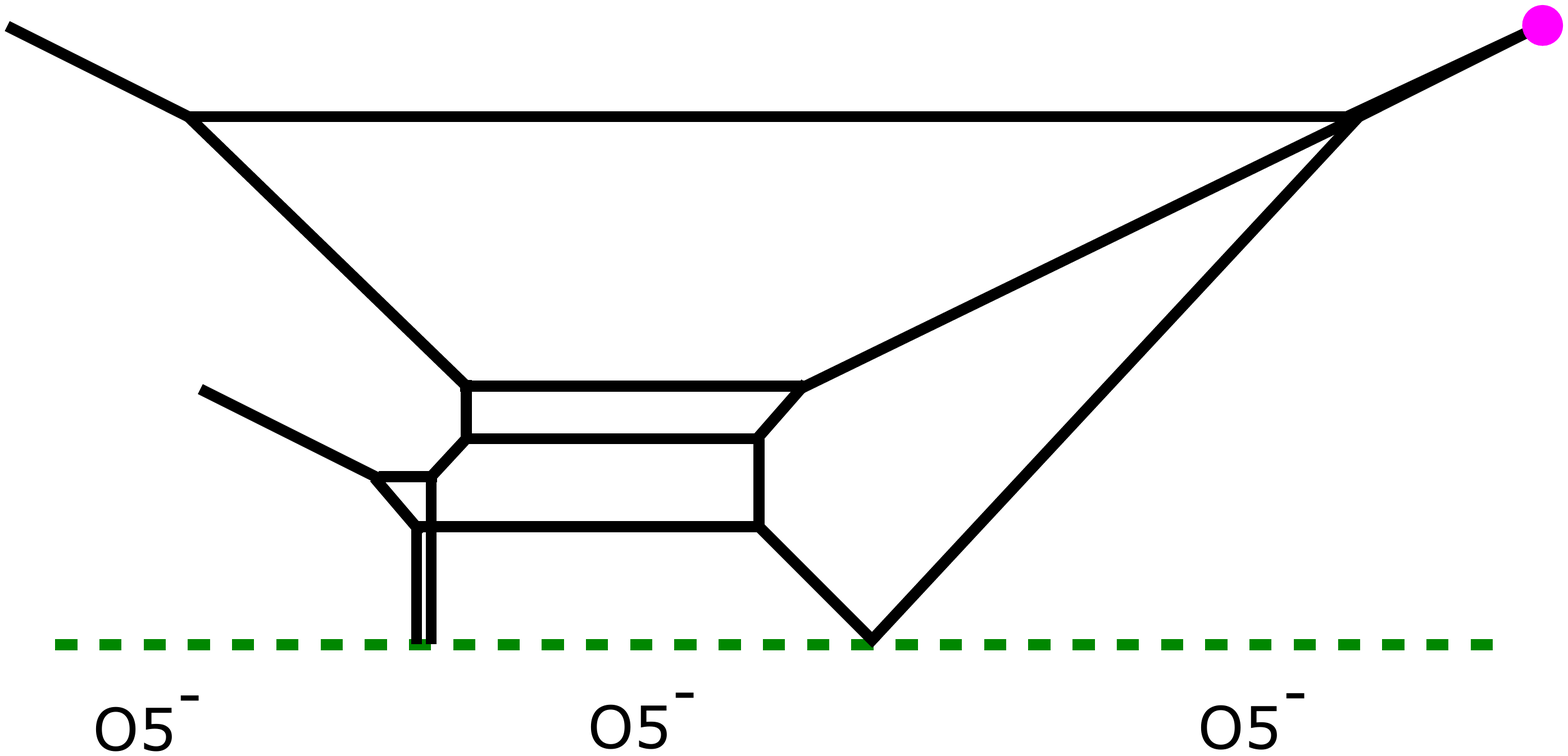}
\caption{Another 5-brane web diagram for the 5d $SO(7)$ gauge theory with one spinor.}
\label{fig:SO7wspinor3}
\end{figure}

We can use the 5-brane web in Figure \ref{fig:SO8wspinorAcspinor3} to obtain another 5-brane web diagram for the 5d pure $G_2$ gauge theory by two Higgsings. Let us first perform a Higgsing associated to the parallel external $(2, 1)$ 5-branes on the right part in Figure \ref{fig:SO8wspinorAcspinor3}. The procedure is essentially the same as what has been done in Figure \ref{fig:SO7wspinorflopped} and the resulting 5-brane web diagram is given in Figure \ref{fig:SO7wspinor3}. 
%%%%%%%%%%%%%%%%%%%%%%%%%%%%%%%%%
%%%%%%%%%%%%%%%%%%%%%%%%%%%%%%%%%
Due to the triality, the 5-brane web diagram in Figure \ref{fig:SO7wspinor3} should give rise to the 5d $SO(7)$ gauge theory with one spinor. Hence the diagram in Figure \ref{fig:SO7wspinor3} realizes the $SO(7)$ gauge group without introducing an $\widetilde{\text{O5}}$-plane different from the diagram in Figure \ref{fig:SO7wspinor}. 

\begin{figure}
\centering
\includegraphics[width=8cm]{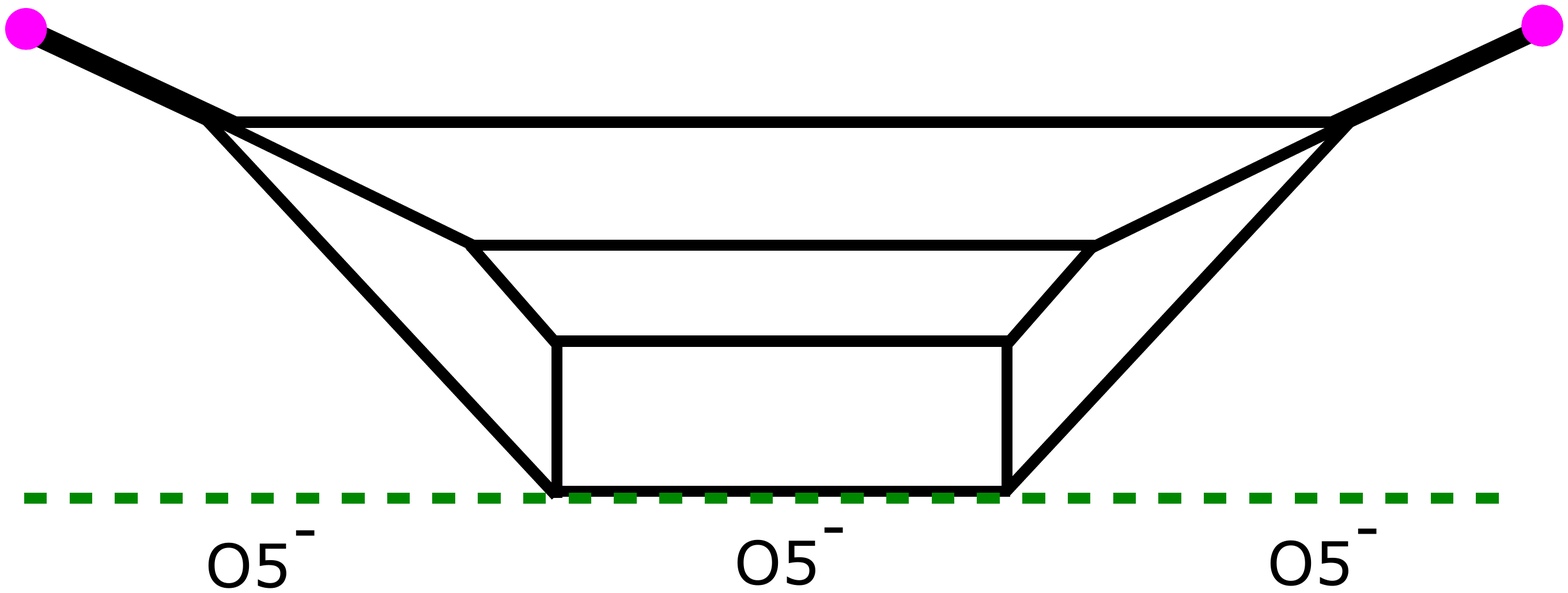}
\caption{Another 5-brane web diagram for the 5d pure $G_2$ gauge theory.}
\label{fig:G2pure2}
\end{figure}

Since the diagram in Figure \ref{fig:SO7wspinor3} still contains parallel external $(2, -1)$ 5-branes, we can perform a Higgsing associated to them. Note that after the Higgsing, the consistency of the diagram restricts the position of the lowest color D5-brane to the location of the O5-plane. The resulting diagram is depicted in Figure \ref{fig:G2pure2}.
%%%%%%%%%%%%%%%%%%%%%%%%%%%%%%%%%
%%%%%%%%%%%%%%%%%%%%%%%%%%%%%%%%%
The 5-brane web diagram of the pure $G_2$ gauge theory is given without an $\widetilde{\text{O5}}$-plane in this case. In fact, it turns out that this diagram is more useful to apply the topological vertex technique to compute the partition function than the diagram in Figure \ref{fig:pureG2}.

\subsection{Check from effective prepotential}
\begin{figure}
\centering
\includegraphics[width=8cm]{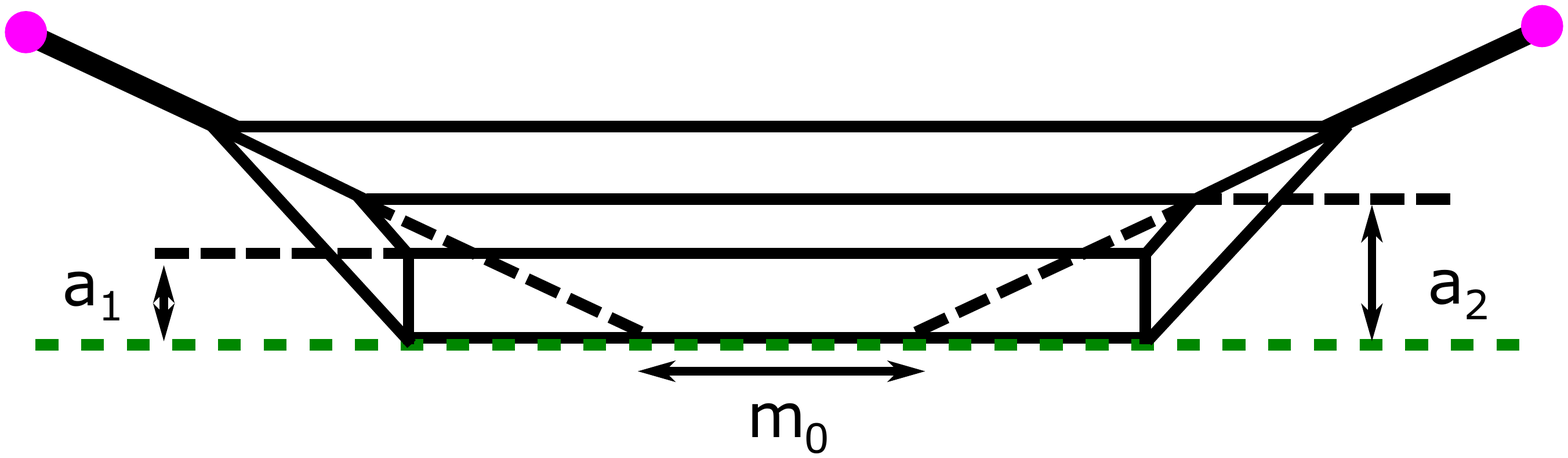}
\caption{A gauge theory parameterization for the pure $G_2$ gauge theory for the diagram in Figure \ref{fig:G2pure2}. $a_1, a_2$ are the Coulomb branch moduli and $m_0$ is related to the inverse of the squared gauge coupling.}
\label{fig:pureG2parameter2}
\end{figure}

As we have done in section \ref{sec:G2prep1}, we can give %an 
evidence that the diagram in Figure \ref{fig:G2pure2} yields the pure $G_2$ gauge theory by computing the tension of a monopole string from the diagram. For that, we first associate the gauge theory parameters, the Coulomb branch moduli $a_1, a_2$ and the inverse of the squared gauge coupling $m_0$ to some lengths of 5-branes. $a_1$ is the height of the lowest color D5-brane and $a_2$ is the height of the second lowest D5-brane. $m_0$ is determined by extrapolating the external $(2, 1)$ 5-brane and the external $(2, -1)$ 5-brane to the location of the O5-plane. The relations between the gauge theory parameters and the lengths of 5-branes are depicted in Figure \ref{fig:pureG2parameter2}. 
%%%%%%%%%%%%%%%%%%%%%%%%%%%%%%%%%%
%%%%%%%%%%%%%%%%%%%%%%%%%%%%%%%%%%

\begin{figure}
\centering
\includegraphics[width=8cm]{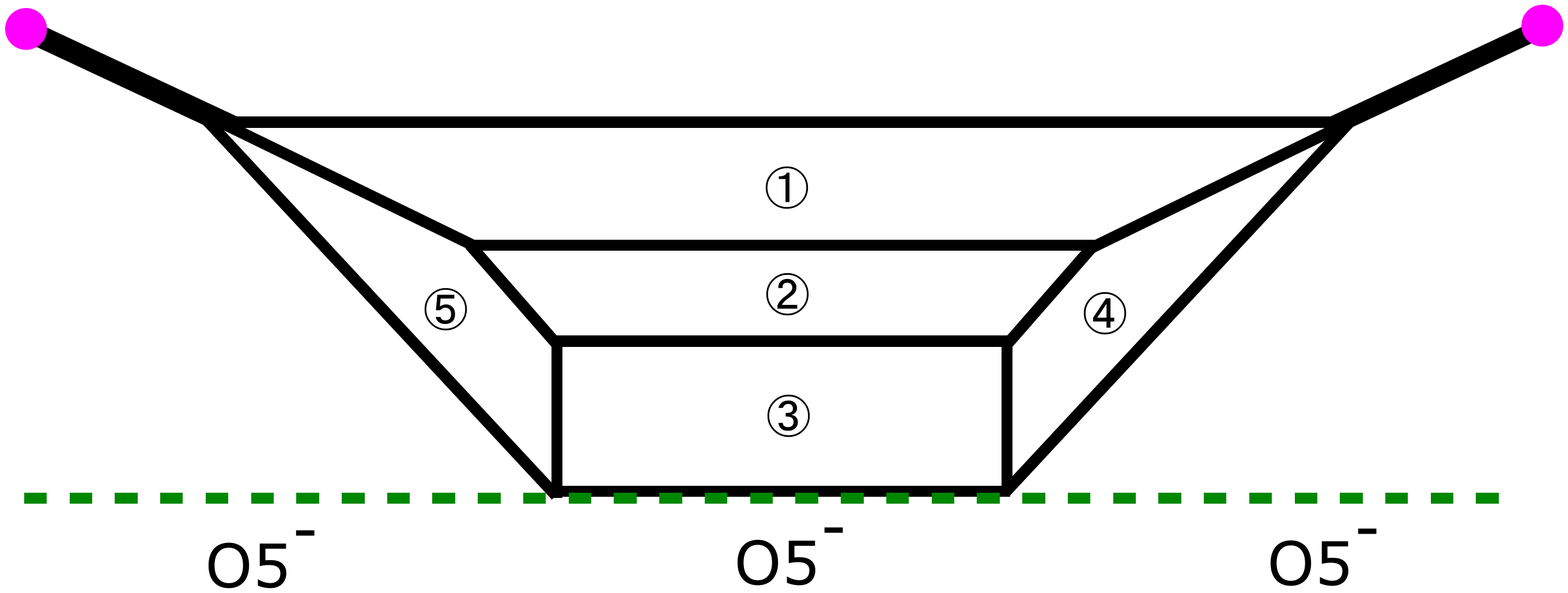}
\caption{Labeling for the five faces in the pure $G_2$ diagram. Note that the regions $\textcircled{\scriptsize 1}$, $\textcircled{\scriptsize 4}$ and $\textcircled{\scriptsize 5}$ are not separated by a 5-brane and are connected to each other.}
\label{fig:pureG2monopole2}
\end{figure}

The tension of a monopole string is given by the area of a face where a D3-brane stretch. There are five faces in the 5-brane web in Figure \ref{fig:G2pure2} and we denote the five faces by $\textcircled{\scriptsize 1}$ - $\textcircled{\scriptsize 5}$ as in Figure \ref{fig:pureG2monopole2}. 
%%%%%%%%%%%%%%%%%%%%%%%%%%%%%%%%%%
%%%%%%%%%%%%%%%%%%%%%%%%%%%%%%%%%%
%
The areas of the five faces are respectively 
\begin{eqnarray}
\textcircled{\scriptsize 1} &=& a_1(m_0 + 2a_1 + 4a_2),\\
\textcircled{\scriptsize 2} &=& (a_2 - a_1)(m_0 + a_1 + 3a_2),\\
\textcircled{\scriptsize 3} &=& a_1(m_0 + 2a_1 + 2a_2),\\
\textcircled{\scriptsize 4} &=& a_1a_2,\\
\textcircled{\scriptsize 5} &=& a_1a_2.
\end{eqnarray}
One face where a D3-brane stretch is $\textcircled{\scriptsize 2}$. As for the other case, note that we needed to double the area of the region $ \textcircled{\scriptsize 3}$ and further add the area $ \textcircled{\scriptsize 4}$ as in Figure \ref{fig:pureG2monopole1}. 
Two sequences of Higgsing connect the region $ \textcircled{\scriptsize 1}$, $ \textcircled{\scriptsize 4}$ and $ \textcircled{\scriptsize 5}$. 
Hence, in this case the other face should be $\textcircled{\scriptsize 1} + 2\times \textcircled{\scriptsize 3} + \textcircled{\scriptsize 4} + \textcircled{\scriptsize 5}$. Then, the area of the faces corresponding to the monopole string tension is 
\begin{eqnarray}
\textcircled{\scriptsize 2} &=& (a_2 - a_1)(m_0 + a_1 + 3a_2), \label{G2monopole5}\\
\textcircled{\scriptsize 1} + 2\times\textcircled{\scriptsize 3} + \textcircled{\scriptsize 4} + \textcircled{\scriptsize 5} &=& a_1(3m_0 + 6a_1 + 10a_2), \label{G2monopole6}
\end{eqnarray}
which exactly reproduce the area \eqref{G2monopole3} and \eqref{G2monopole4} calculated from the diagram in Figure \ref{fig:pureG2}. Therefore the monopole tension computation gives another evidence that the diagram in Figure \ref{fig:G2pure2} yields the 5d pure $G_2$ gauge theory.

%====================================================================
\bigskip
\section{5d Nekrasov partition functions of $G_2$ gauge theories}
\label{sec:Nekrasov}

In section \ref{sec:G2fromO5tilde} and section \ref{sec:G2fromO5}, we have constructed 5-brane web diagrams which realize 5d $G_2$ gauge theories. In section \ref{sec:G2fromO5tilde} we presented 5-brane webs for $G_2$ gauge theories using an $\widetilde{\text{O5}}$-plane and in section \ref{sec:G2fromO5} we presented the 5-brane web for the pure $G_2$ gauge theory with an O5-plane. 
One of the applications of these 5-brane webs is to compute the BPS partition functions of 5d theories realized by the webs. Since 5-brane webs can be reinterpreted as toric diagrams \cite{Leung:1997tw}, we can apply the topological vertex formalism \cite{Aganagic:2003db, Iqbal:2007ii,Awata:2008ed}. 
In \cite{Kim:2017jqn}, the topological vertex formalism for webs with an O5-plane has been developed and hence we can utilize the technique to compute the partition function for the pure $G_2$ gauge theory realized by the 5-brane web with an O5-plane. Although the 5-brane web diagrams in section \ref{sec:G2fromO5tilde} are realized with an $\widetilde{\text{O5}}$-plane, we extend the formalism so that it can apply to webs with an $\widetilde{\text{O5}}$-plane in section \ref{sec:vertexO5tilde}. Then we apply the method to compute the partition functions of the 5d pure $G_2$ gauge theory and the 5d $G_2$ gauge theory with one flavor. 

\subsection{Vertex formalism with an O5- and $\widetilde{\text{O5}}$-plane}
\label{sec:vertexO5tilde}

\begin{figure}
\centering
\includegraphics[width=11cm]{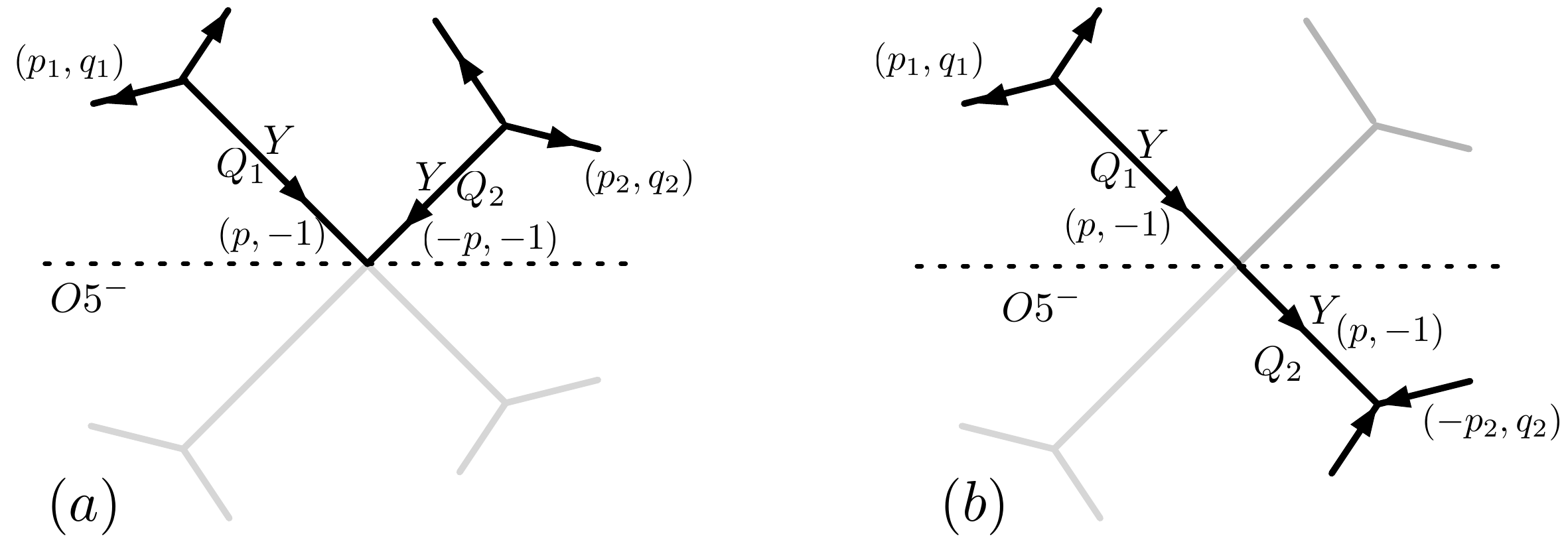}
\caption{
(a): The $(p,-1)$ and $(-p,-1)$ 5-branes which intersect with each other on the O5-plane.
(b): An equivalent diagram to (a) but we use the mirror image for a part of the diagram.
}
\label{fig:NewRule}
\end{figure}

Here, we first briefly review the topological vertex formalism 
for 5-brane webs with an O5-plane proposed in \cite{Kim:2017jqn}. 
As for the 5-branes which do not touch the O5-plane,
the rule is exactly the same as the conventional topological vertex formalism \cite{Aganagic:2003db}:
We first assign different Young diagrams $Y_i$ to different $(p,q)$ 5-branes. 
Then, we introduce the edge factor $(-Q)^{|Y|} f_Y{}^n$ to each edge,
where $Q$ is given by the exponential of the length of the corresponding $(p,q)$ 5-brane.
Here, $f_Y$ is the framing factor defined as
\begin{align}\label{eq:framing}
f_{Y} = (-1)^{|Y|} g^{\frac{1}{2} (||Y^t||^2 - ||Y||^2)},
\end{align} 
with $|Y|=\sum_i Y_i$ and $||Y||^2=\sum_i Y_i{}^2$. $n$ is a certain integer which is associated to the relative difference of the framing when we glue two topological vertices. 
The parameter $g$ is related to the Omega deformation parameters by 
$g=e^{-\epsilon_1} = e^{+\epsilon_2}$. We also introduce the topological vertex $C_{Y_1 Y_2 Y_3}$ to each vertex of a diagram, where
the Young diagrams $Y_1, Y_2, Y_3$ are ordered clockwise. 
On top of that, the additional rule is given for $(p,-1)$ and $(-p,-1)$ 5-branes which intersect with each other on the O5-plane
as depicted in Figure \ref{fig:NewRule} (a).
The point is to assign the {\it identical} Young diagram to these two 5-branes and to assign an edge factor 
\begin{align}
( + Q_1 Q_2)^{|Y|} f_Y{}^\mathfrak{n},
\qquad
( \mathfrak{n} = p_1 q_2 + p_2 q_1 + 1 ),
\end{align}
corresponding to this part,
where $Q_1$ and $Q_2$ are the exponential of the (rescaled) length of the two 5-branes respectively.
By multiplying all these factors and by summing over all the possible Young diagrams,
we obtain the topological string partition function.

Although we can compute the topological string partition function directly using the rules above with a configuration in Figure \ref{fig:NewRule} (a), 
it turns out to be more convenient to cut the D5-branes with a finite length 
and to use the mirror image for a part of the diagram %along the O5-plane 
so that the two 5-branes which originally intersected with each other on the O5-plane become a single
edge as in Figure \ref{fig:NewRule} (b) \cite{Kim:2017jqn}. This operation corresponds to using the identity
\begin{align}\label{eq:identityC}
C_{Y_1 Y_2 Y_3} = (-1)^{|Y_1|+|Y_2|+|Y_3|} f_{Y_1}^{-1}  f_{Y_2}^{-1}  f_{Y_3}^{-1}
C_{Y_3^t Y_2^t Y_1^t}.
\end{align}
to all the vertices reflected along the O5-plane
because a clockwise direction is mapped to a counter-clockwise direction under this reflection.
Note that the Young diagrams assigned to the edges reflected along the O5-plane
should be transposed.
After the proper reflection, each sub-diagram can be seen as a strip diagram. 
Such sub-amplitudes are already computed in \cite{Iqbal:2004ne}.
The strip amplitudes are written in terms of the product of 
the factor defined as 
\begin{align}
Z_{\nu}(g) = \prod_{(i,j) \in \nu} (1-g^{\nu_i+\nu_j^t-i-j+1}),
\end{align}
and the factor defined as 
\begin{align}
\R_{XY}(Q) =\M(Q)^{-1} \,\N_{X^t Y}(Q),
\end{align}
with 
\begin{align}
\M(Q) = {\rm PE} \left[ \frac{g}{(1-g)^2} Q \right],
\end{align}
where $\rm PE$ is the Plethystic exponential defined as 
\begin{align}
{\rm PE}[f(\cdot)] = \exp \bigg[ \sum_{i=1}^{\infty} \frac1n f(\cdot^n)\bigg],
\end{align}
and 
\begin{align}
\N_{\lambda \mu}(Q) = 
\prod_{(i,j) \in \lambda} (1-Q g^{\lambda_i+\mu_j^t-i-j+1} )
\prod_{(i,j) \in \mu} (1-Q g^{-\lambda_j^t-\mu_i+i+j-1} ).
\end{align}

What remains is to glue each strip at the edges where we cut before the reflection
by multiplying corresponding edge factors 
and by summing over Young diagram assigned to these edges.
In this process, we also need to take into account the additional factors of the form $(-1)^{|Y|} f_{Y}^{-1}$ in \eqref{eq:identityC}.
If two vertices connected to the same edge are both reflected along the O5-plane,
such contribution cancels with each other as 
$(-1)^{|Y|} f_{Y}^{-1} \times (-1)^{|Y^t|} f_{Y^t}^{-1} = 1$.
Only when one vertex is reflected while the other vertex is not, we need to multiply such a factor .

 \begin{figure}
\centering
\includegraphics[width=12cm]{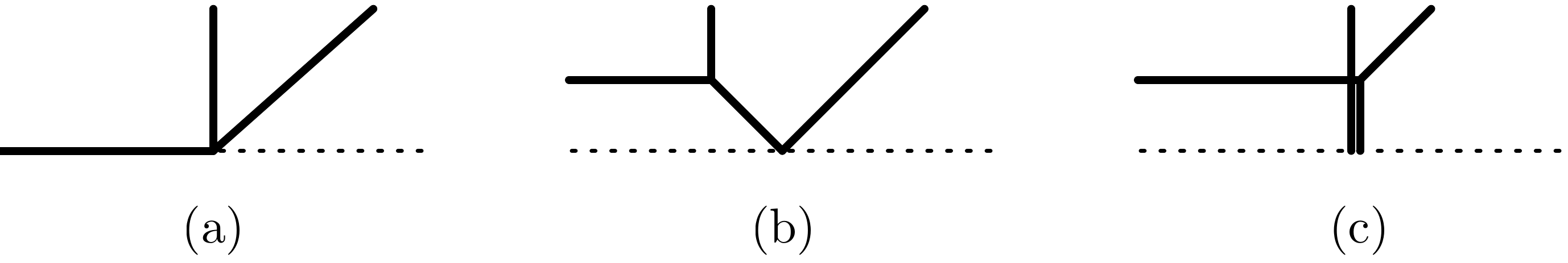}
\caption{Configurations with an O5-plane}
\label{fig:O5config}
\end{figure}
\begin{figure}
\centering
\includegraphics[width=12cm]{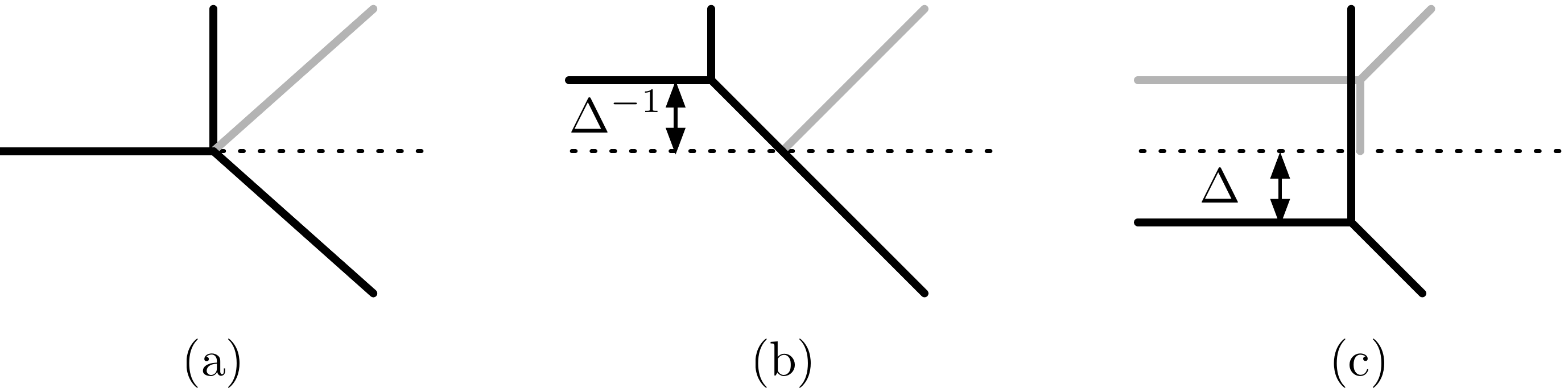}
\caption{Configurations with an O5-plane with mirror image.}
\label{fig:O5mirror}
\end{figure}

This method can be generalized to webs with an $\widetilde{\text{O5}}$-plane.
As we did in section \ref{sec:SO7Spinor},
we can interpret that an $\widetilde{\text{O5}}$-plane is realized between two fractional D7-branes
on top of the O5-plane placed at infinitely left and infinitely right, respectively \cite{Zafrir:2015ftn, Feng:2000eq}.
The point is to convert an $\widetilde{\text{O5}}$-plane to an O5-plane by moving 
one of these fractional D7-branes to the other side by using Hanany-Witten transition 
as discussed in Figure \ref{fig:SO7wspinor2}.
After that, it can be seen as a special case of a 5-brane web with an O5-plane.
For example,
as is also mentioned in section \ref{sec:G2fromO5tilde}, 
the diagram in Figure \ref{fig:O5config} (a) can be understood 
as a special case of the ones in Figure \ref{fig:O5config} (b) or (c).
%In \cite{Kim:2017jqn}, 
%it is discussed that the topological vertex formalism 
%can be applied to the web diagram with O5-plane 
As discussed above, 
by considering the mirror image as in Figure \ref{fig:O5mirror} (b) or (c),
we can interpret the configurations as a part of trip diagrams.
Therefore, a configuration with Figure \ref{fig:O5config} (a) should be 
also considered as a special case of Figure \ref{fig:O5mirror} (b) or (c),
where we tune the K$\ddot{\text{a}}$hler parameter $\Delta\to1$ so that the position of the D5-brane 
comes to exactly the place where the O5-plane exists.

\subsection{Pure $G_2$ gauge theory}
\label{sec:NekpureG2}
We apply the technique of the topological vertex for webs involving an O5-plane or an $\widetilde{\text{O5}}$-plane described in section \ref{sec:vertexO5tilde} to the 5-brane web of 5d $G_2$ gauge theories. We first consider a 5-brane web for the 5d pure $G_2$ gauge theory. So far we have presented two types of the 5-brane webs for the pure $G_2$ gauge theory. One is depicted in Figure \ref{fig:pureG2} with an $\widetilde{\text{O5}}$-plane and the other is depicted in Figure \ref{fig:G2pure2} with an O5-plane. For applying the topological vertex formalism it is appropriate to use the one in Figure \ref{fig:G2pure2} since the diagram in Figure \ref{fig:pureG2} has a configuration where a single 5-brane directly intersects with an O5-plane and we have not yet known the vertex rule for such a configuration. 

\begin{figure}[t]
\centering
\subfigure[]{\includegraphics[width=6cm]{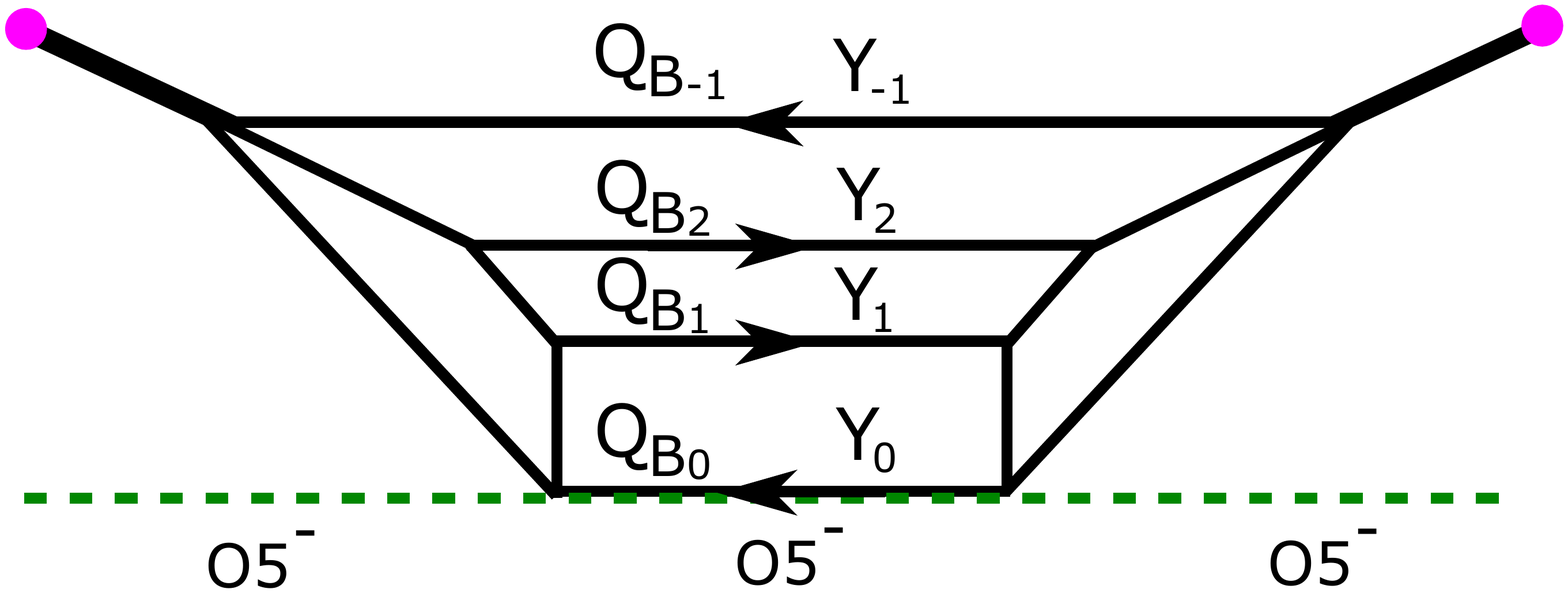}
 \label{fig:fig01left} }
\hspace{20mm}
\subfigure[]{\includegraphics[width=6cm]{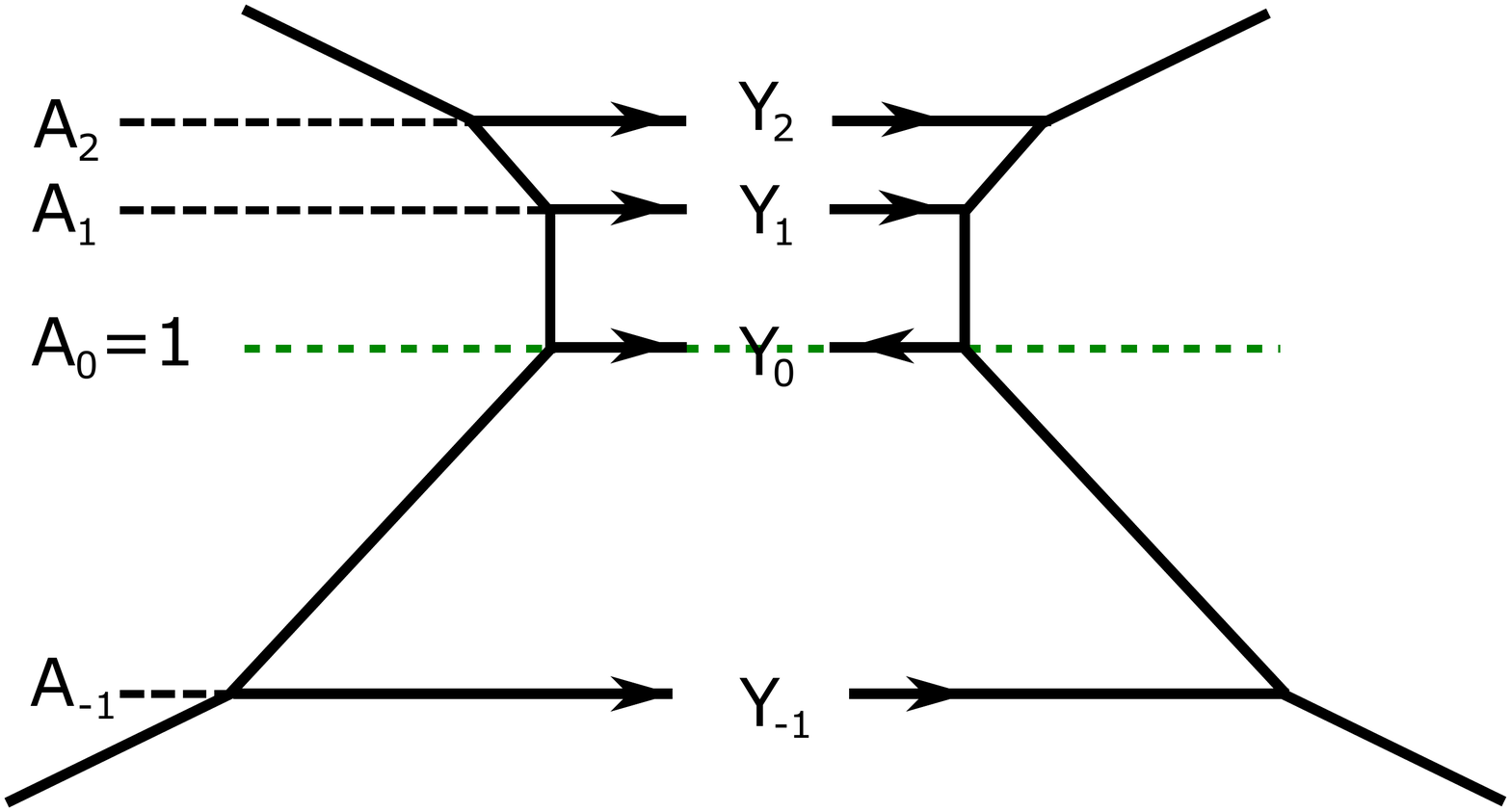}
 \label{fig:fig01right} }
\caption{(a): The assignment of the Young diagrams $Y_i, i=-1, 0, 1, 2$ and the K$\ddot{\text{a}}$hler parameter $Q_{B_i}, i=-1, 0, 1, 2$ for the diagram in Figure \ref{fig:G2pure2} of the pure $G_2$ gauge theory. (b): An equivalent diagram to the one in Figure \ref{fig:fig01left}. We use the mirror image for the outmost 5-brane in Figure \ref{fig:fig01left}. As for the bottom 5-brane in Figure \ref{fig:fig01left}, we use the mirror image for the left part and use the original one for the right part.}
\label{fig:pureG2top}
\end{figure}

We use the 5-brane diagram in Figure \ref{fig:G2pure2} to compute the partition function of the 5d pure $G_2$ gauge theory. First we introduce the gauge theory parameters for the diagram as in Figure \ref{fig:pureG2parameter2}. The Coulomb branch moduli $a_1, a_2$ of the pure $G_2$ gauge theory are given by the height of the bottom and the second bottom color D5-brane respectively. We then define $A_1, A_2$ by
\begin{align}
A_1 = e^{-  a_1}, \qquad 
A_2 = e^{- a_2}, \qquad 
\end{align}
which are the parameters directly appearing in the partition function. For the convenience of the later computation, we also introduce 
\begin{align}\label{eq:A3=1}
A_0 = 1, \qquad 
A_{-1} = A_1^{-1} A_2^{-1},
\end{align}
and define
\begin{align}
Q_{ij} = A_i A_j{}^{-1},
\qquad
(-1 \le j < i \le 2).
\end{align}
The inverse of the squared gauge coupling is given by $m_0$ in Figure \ref{fig:pureG2parameter2} and we define the instanton fugacity by
\begin{align}
q = e^{-m_0}.
\end{align}
Then the K$\ddot{\text{a}}$hler parameters associated to the length of the horizontal lines in Figure \ref{fig:fig01left} are\footnote{The length of the 5-branes in the diagram is given by a linear combination of $a_1, a_2$ and $m_0$. On the other hand, the instanton partition function is written by the exponentiated parameters $A_1, A_2, q$. Hence when we put for example $A_1$ to some length in a diagram, it means that the length is $a_1$. We will make use of this notation for the topological string partition function computation.}
\begin{align}
Q_{B_2} = q A_2{}^{4}, \qquad
Q_{B_1} = Q_{B_0} = q A_1{}^{2} A_2{}^{2}, \qquad
Q_{B_{-1}} = q A_1{}^{4} A_2{}^{4}.
\end{align}
%where
%\begin{align}
%A_1 = e^{- \beta a_1}, \qquad 
%A_2 = e^{- \beta a_2}, \qquad 
%A_3 = 1, \qquad 
%A_4 = A_1^{-1} A_2^{-1}.
%\end{align}
%where $q$ is the instanton factor.

With the parameterization, we compute the topological string partition function by applying the topological vertex to the 5-brane web in Figure \ref{fig:G2pure2}.
%In order to compute it, we 
Since the diagram involves an O5-plane, we utilize the method in section \ref{sec:vertexO5tilde}. %discussed in \cite{Kim:2017jqn}.
First, we cut the color D5-branes in Figure \ref{fig:fig01left} in the middle and %invert a part of the diagram along the O5$^-$ plane 
use the mirror image for the outmost 5-brane  for describing the diagram as in Figure \ref{fig:fig01right}.
%After this reflection, this diagram can be split into two strip diagrams.
The diagram now consists of two strip diagrams. Note that the diagram in Figure \ref{fig:fig01right} is almost the same as the diagram of the pure $SU(4)$ gauge theory of CS level zero with $A_0$ set to $1$. A difference comes from the assignment of the Young diagram for $Y_0$. In order to see it, recall the diagram of the 5d $SO(7)$ gauge theory with one spinor in Figure \ref{fig:SO7wspinor3}, which is the diagram before Higgsing to the one in Figure \ref{fig:G2pure2}. Let us cut then the diagram in Figure \ref{fig:SO7wspinor3} in the middle into the left part and the right part and use the mirror image for a part of the diagram to apply the topological vertex computation in section \ref{sec:vertexO5tilde}. Then we use the mirror image for the bottom color D5-brane in the left part while we use the original bottom color D5-brane for the right part in the upper half plane. After the Higgsing, the bottom color D5-brane in Figure \ref{fig:SO7wspinor3} becomes the color D5-brane on top of an O5$^-$-plane in Figure \ref{fig:G2pure2}. Therefore, when we consider the diagram in Figure \ref{fig:fig01right}, we should transpose the Young diagram $Y_0$ in the left strip while we keep the Young diagram $Y_0$ in the right strip. This is essentially the only difference from the 5-brane web of the pure $SU(4)$ gauge theory. The Young diagram assignment is summarized in Figure \ref{fig:fig01right}. 
%Taking into account that Figure \ref{fig:fig01left} is obtained by Higgsing the diagram in Figure \ref{fig:SO7wspinor3}, the D5-brane on top of the O5$^-$-plane 
%we should interpret that the D5-brane on top of the O5$^-$ plane is reflected in the strip diagram on the left hand side while not reflected on the right hand side.
%When we reflect the sug-diagram, the Young diagram associated to the edge is transposed.

It is now straightforward to apply the topological vertex formalism to the diagram in Figure \ref{fig:fig01right}. It is useful to compute the partition function for the left strip and the right strip separately first and glue them together later. The partition function for the left strip is given by 
\begin{align}\label{strip}
Z^{\rm strip1} (\{ Q_{ij} \};  Y_{-1}, Y_{0},Y_{1},Y_{2})  = 
\prod_{i=-1}^2 g^{\frac{1}{2} ||Y_i||^2} 
\prod_{i=-1}^2  Z_{Y_i}(g) 
\prod_{-1 \le  j < i  \le 2} \R_{Y_i Y_j^t} (Q_{ij})^{-1}.
\end{align}
The partition function of the right strip can be expressed as 
$Z^{\rm strip2} (\{ Q_{ij} \};  Y_{-1}, Y_{0}^t ,Y_{1},Y_{2})$ with
\begin{align}
Z^{\rm strip2} (\{ Q_{ij} \};  Y_{-1}, Y_{0},Y_{1},Y_{2})  = 
\prod_{i=-1}^2 g^{\frac{1}{2} ||Y_i^t||^2} 
\prod_{i=-1}^2  Z_{Y_i}(g) 
\prod_{-1 \le  j < i  \le 2} \R_{Y_i Y_j^t} (Q_{ij})^{-1}.
\end{align}
The rest is gluing the contribution of the left strip and the right strip to each other. When we glue these two sub-diagram, we should also take into account the effect of flipping the Young diagram, which is given by the extra factor of the form $(-1)^{|Y_0|} f_{Y_0}$ coming from the identity \eqref{eq:identityC}.
%\begin{align}
%C_{\lambda \mu \nu} = (-1)^{|\mu|+|\nu|+|\rho|} f_{\mu}  f_{\nu}  f_{\rho} 
%C_{\nu^t \mu^t \lambda^t}.
%\end{align}
Note that such factor for the upper most D5-brane cancels out as discussed in section \ref{sec:vertexO5tilde} 
since the whole edge is reflected.
%becomes trivial as $(-1)^{|Y_4|} f_{Y_4} \times (-1)^{|Y_4^t|} f_{Y^t_4} = 1$. 
Then, the topological string partition function for the diagram in Figure \ref{fig:fig01right} is 
\begin{align}\label{part.pureG2}
Z_{G_2} = & \sum_{\{Y_i\}}
(-1)^{|Y_0|} f_{Y_2}^{3} f_{Y_1} f_{Y_0} f_{Y_{-1}}^{-3} \prod_{i=-1}^2 (-Q_{B_i}){}^{| Y_i |} 
\cr
& Z^{\rm strip1} (\{ Q_{ij} \};  Y_{-1}, Y_{0},Y_{1},Y_{2})
Z^{\rm strip2} (\{ Q_{ij} \};  Y_{-1}, Y_{0}^t ,Y_{1},Y_{2}).
\end{align}
We claim that this is the partition function of the 5d pure $G_2$ gauge theory up to the perturbative contribution of the Cartan subalgebra of the pure $G_2$ which should be added by hand in the topological vertex computation. Notice that the only difference between this $G_2$ partition function and  the pure $SU(4)$ partition function with $A_0=1$ in \eqref{eq:A3=1} is as follows: The Young diagram $Y_0$ in $Z^{\rm strip2}$ is transposed, the power of framing factor $f_{Y_0}$ is 1 instead of $-1$,
and also the associated sign factor $(-)^{|Y_0|}$ is present.

\paragraph{Perturbative part.}
Let us look at the partition function \eqref{part.pureG2} in more detail. The perturbative part of the partition function is obtained by considering a limit $q \to 0$. The limit corresponds to the restriction of the Young diagrams $Y_i = \emptyset$ for $i=-1, 0, 1, 2$. Then the partition function \eqref{part.pureG2} is simplifies and is given by
%\begin{align}
%{\rm PE} \left( 
%- \frac{2g}{(1-g)^2} 
%(A_1{}^{-1} + A_2{}^{-1} 
%+ A_1{}^{-1} A_2{}^{-2} + A_1{}^{-2} A_2{}^{-1}
%+ A_1{}^{-1} A_2{}^{-1})
%\right)
%\end{align}
\begin{align}\label{pert.pureG2}
Z_{\text{pert}} =&\M(A_1)^2\M(A_1^{-1}A_2)^2\M(A_2)^2\M(A_1A_2)^2\M(A_1^2A_2)^2\M(A_1A_2^2)^2\cr
=&{\rm PE} \left( 
\frac{2g}{(1-g)^2} 
(A_1 + A_2 + A_1 A_2 + A_1{}^{-1} A_2
+ A_1 A_2{}^{2} + A_1{}^{2} A_2)
\right).\
\end{align}

We can compare \eqref{pert.pureG2} with the field theory result of the perturbative part of the partition function of the pure $G_2$ gauge theory, which is given by
%This is consistent with 
\begin{align}\label{locpert.pureG2}
Z'_{\text{pert}} = {\rm PE} \left( 
 \frac{g}{(1-g)^2} \chi_{14}
\right),
\end{align}
where $\chi_{14}$ is the character of the adjoint representation,
\begin{align}
\chi_{14}{} =& 
A_1 + A_1{}^{-1} + A_2 + A_2{}^{-1} 
+ A_1 A_2 + A_1{}^{-1} A_2{}^{-1} 
\cr
& 
+ A_1 A_2{}^{-1} + A_1{}^{-1} A_2
+ A_1 A_2{}^2 + A_1{}^{-1} A_2{}^{-2}
+ A_1{}^2 A_2 + A_1{}^{-2} A_2{}^{-1} 
+ 2.
\end{align}
The partition function \eqref{pert.pureG2} obtained from the topological vertex is indeed consistent with \eqref{locpert.pureG2} up to the terms independent of the Coulomb moduli, namely the Cartan part,
and up to the procedure corresponding to the ``flop transition''
\begin{align}
{\rm PE} \left( \frac{g}{(1-g)^2} Q \right)
\to 
{\rm PE} \left( \frac{g}{(1-g)^2} Q^{-1} \right). 
\end{align}

\paragraph{Instanton part.}
Let us then look at the instanton part of the partition function \eqref{part.pureG2}. The instanton part is given by removing the perturbative part 
\begin{align}\label{inst.pureG2}
Z_{G_2,\text{inst}} = \frac{Z_{G_2}}{Z_{\text{pert}}} = \sum_{k}Z_kq^k.
\end{align}
The $k$-instanton contribution is given by the function $Z_k$ for the $q^k$ order.  The explicit form of the one-instanton contribution is
%
%\begin{align}
%Z_1= 
%&
%\frac{1}{A_1{}^2 \left(1-\frac{1}{A_2{}}\right)^2 A_2{}^2 (1-g)^2 \left(1-\frac{1}{A_1{} A_2{}^2}\right)^2 \left(1-\frac{A_2{}}{A_1{}}\right)^2}
%\cr
%&
%- \frac{g}{\left(1-\frac{1}{A_1{}}\right)^2 A_1{}^2 \left(1-\frac{1}{A_2{}}\right)^2 A_2{}^2 (1-g)^2 \left(1-\frac{1}{A_1{} A_2{}}\right)^2}
%\cr
%&
%+\frac{g}{A_1{}^4 A_2{}^4 (1-g)^2 \left(1-\frac{1}{A_1{} A_2{}^2}\right)^2 \left(1-\frac{1}{A_1{}^2 A_2{}}\right)^2 \left(1-\frac{1}{A_1{} A_2{}}\right)^2}
%\cr
%&
%+\frac{g}{\left(1-\frac{1}{A_1{}}\right)^2 A_1{}^4 (1-g)^2 \left(1-\frac{1}{A_1{}^2 A_2{}}\right)^2 \left(1-\frac{A_2{}}{A_1{}}\right)^2},
%\end{align}
%
%or equivalently,
\begin{align}\label{1inst.pureG2}
Z_1 =
\frac{
2 g A_1{}^3 A_2{}^3 (1 + A_1 + A_1 A_2) (1 + A_2 + A_1 A_2)
}{
(1 - g)^2
(A_1 - A_2)^2 
(1 - A_1{}^2 A_2)^2 
(1 - A_1 A_2{}^2)^2 
}.
\end{align}
The explicit form of the two-instanton contribution is 
\begin{align}\label{2inst.pureG2}
Z_2 = g^5 A_1{}^{10} A_2{}^{10}  \frac{({\rm Numerator})}{({\rm Denominator})},
\end{align}
with
\begin{align}
({\rm Denominator}) = 
&(1 - g)^4 
(1 + g)^2 
(A_1 - A_2)^2 
(1 - A_1{}^2 A_2)^2 
(1 - A_1 A_2{}^2)^2 
\cr
&\times 
(1 - g A_1{}^2 A_2)^2 
(1 - g A_1 A_2{}^2)^2
(1 - g^{-1} A_1{}^2 A_2 )^2 
(1 - g^{-1} A_1 A_2{}^2 )^2 
\cr
& \times 
(1 - g A_1 A_2{}^{-1})^2 
(1 - g A_1{}^{-1} A_2)^2,
\end{align}
and
\begin{align}
& \!\!\!\!\!\!\!\!\!\! ({\rm Numerator})
\cr
= &
(-2 g_1 + 6) \chi_{14}{}^3 
+ (3 g_1 - 1) \chi_{7}{}^2 \chi_{14}{}^2  
+ (10 g_1 + 2) \chi_{7}{} \chi_{14}{}^2 
+ (g_3 + 5 g_2 - g_1 + 3) \chi_{14}{}^2 \cr
&
+ (2 g_2 + 16 g_1 + 22) \chi_{7}{}^3 \chi_{14}
+ (38 g_2 + 66 g_1 + 12) \chi_{7}{}^2 \chi_{14}\cr
&+ (26 g_3 + 66 g_2 + 38 g_1 + 14) \chi_{7} \chi_{14}
+ (2 g_4 + 24 g_3 + 18 g_2 + 12 g_1 + 8) \chi_{14}\cr
&
+ (-8 g_1 - 8) \chi_{7}{}^5 
+ (-18 g_2 - 11 g_1 + 3) \chi_{7}{}^4 
+ (-12 g_3 + 4) \chi_{7}{}^3 \cr
&+ (13 g_3 + 7 g_2 - 16 g_1 - 38) \chi_{7}{}^2 
+ (6 g_4 + 12 g_3 - 22 g_2 - 60 g_1 - 8) \chi_{7}
\cr
&
+ (2 g_4 - 9 g_3 - 19 g_2 - 21 g_1 - 21),
\end{align}
where 
\begin{align}
g_n =& \sum_{k=-n}^n g^k,
\\
\chi_{7}{} =& 
A_1 + A_1{}^{-1} 
+ A_2 + A_2{}^{-1} 
+ A_1 A_2 + A_1{}^{-1} A_2{}^{-1} 
+ 1, 
\\
\chi_{14}{} =& 
A_1 + A_1{}^{-1} + A_2 + A_2{}^{-1} 
+ A_1 A_2 + A_1{}^{-1} A_2{}^{-1} 
\cr
& 
+ A_1 A_2{}^{-1} + A_1{}^{-1} A_2
+ A_1 A_2{}^2 + A_1{}^{-1} A_2{}^{-2}
+ A_1{}^2 A_2 + A_1{}^{-2} A_2{}^{-1} 
+ 2.
\end{align}

We can compare the result \eqref{1inst.pureG2} and \eqref{2inst.pureG2} with the field theory result. The explicit expression of the one-instanton partition function of the pure $G_2$ gauge theory is obtained in \cite{Benvenuti:2010pq, Keller:2011ek}, and it is generalized to higher instantons in \cite{Hanany:2012dm, Keller:2012da, Cremonesi:2014xha}. We checked that the one-instanton partition function \eqref{1inst.pureG2} and the two-instanton partition function \eqref{2inst.pureG2} perfectly agree with the known results.

\subsection{$G_2$ gauge theory with one flavor}

%In section \ref{sec:vertexO5tilde}, we have developed a topological vertex formalism that is applicable to 5-brane web diagrams with an $\widetilde{\text{O5}}$-plane. We will apply the formalism to an interesting example of a 5-brane web diagram for a $G_2$ gauge theory obtained in section \ref{sec:G2fromO5tilde}. 

In section \ref{sec:NekpureG2}, we have computed the partition function of the pure $G_2$ gauge theory from the topological vertex. In this section, we apply the method to a diagram for the 5d $G_2$ gauge theory with one flavor. We have two types of the diagram for the $G_2$ gauge theory with one flavor in Figure \ref{fig:G2w1flvr}. In order to apply the topological vertex, we need to avoid a configuration where a single 5-brane intersects with an O5-plane. Hence we will use the diagram in Figure \ref{fig:G2w1flvrb} or equivalently the one in Figure \ref{fig:G2w1flvrfloppedb}, which yields the 5d $G_2$ gauge theory with one flavor and a singlet, for the application of the topological vertex 

%In order to use the topological vertex formalism, we basically change an $\widetilde{\text{O5}}$-plane into an O5-plane with a flavor D5-brane by the Hanany-Witten transition and then it is possible to apply the method developed in \cite{Kim:2017jqn}. A limitation of this method is that a 5-brane web should be written only by an O5$^-$-plane after performing generalized flop transitions. Therefore, the simplest case when we can apply the method to $G_2$ theories is the case of the $G_2$ gauge theory with one flavor and a singlet given in Figure \ref{fig:G2w1flvrflopped} (b). 

\begin{figure}
\centering
%\begin{minipage}{7cm}
\subfigure[]{
\includegraphics[width=6cm]{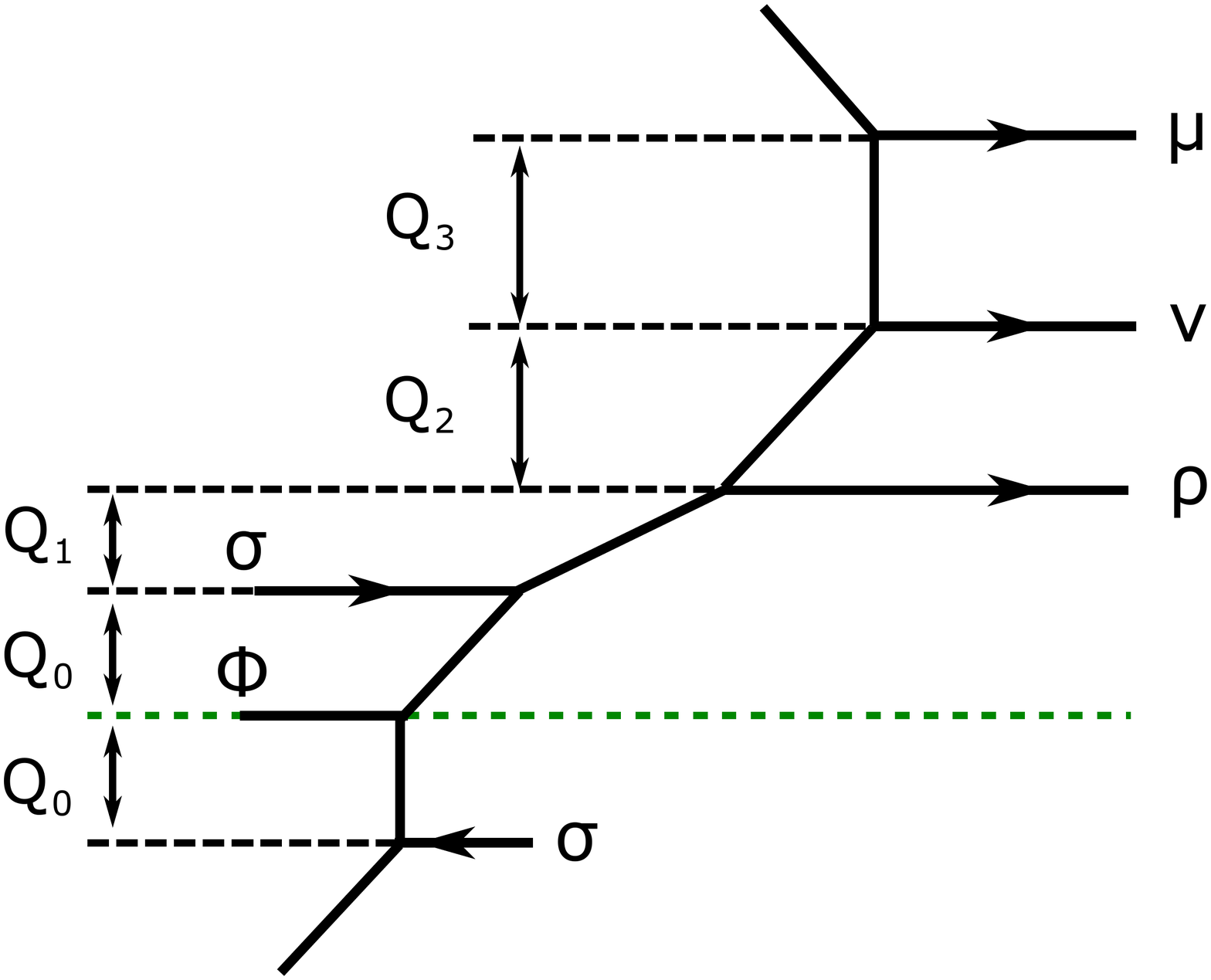} \label{fig:G2w1flvrTopa}}
%\caption{Left strip}
%\end{minipage}
%\begin{minipage}{7cm}
\subfigure[]{
\includegraphics[width=6cm]{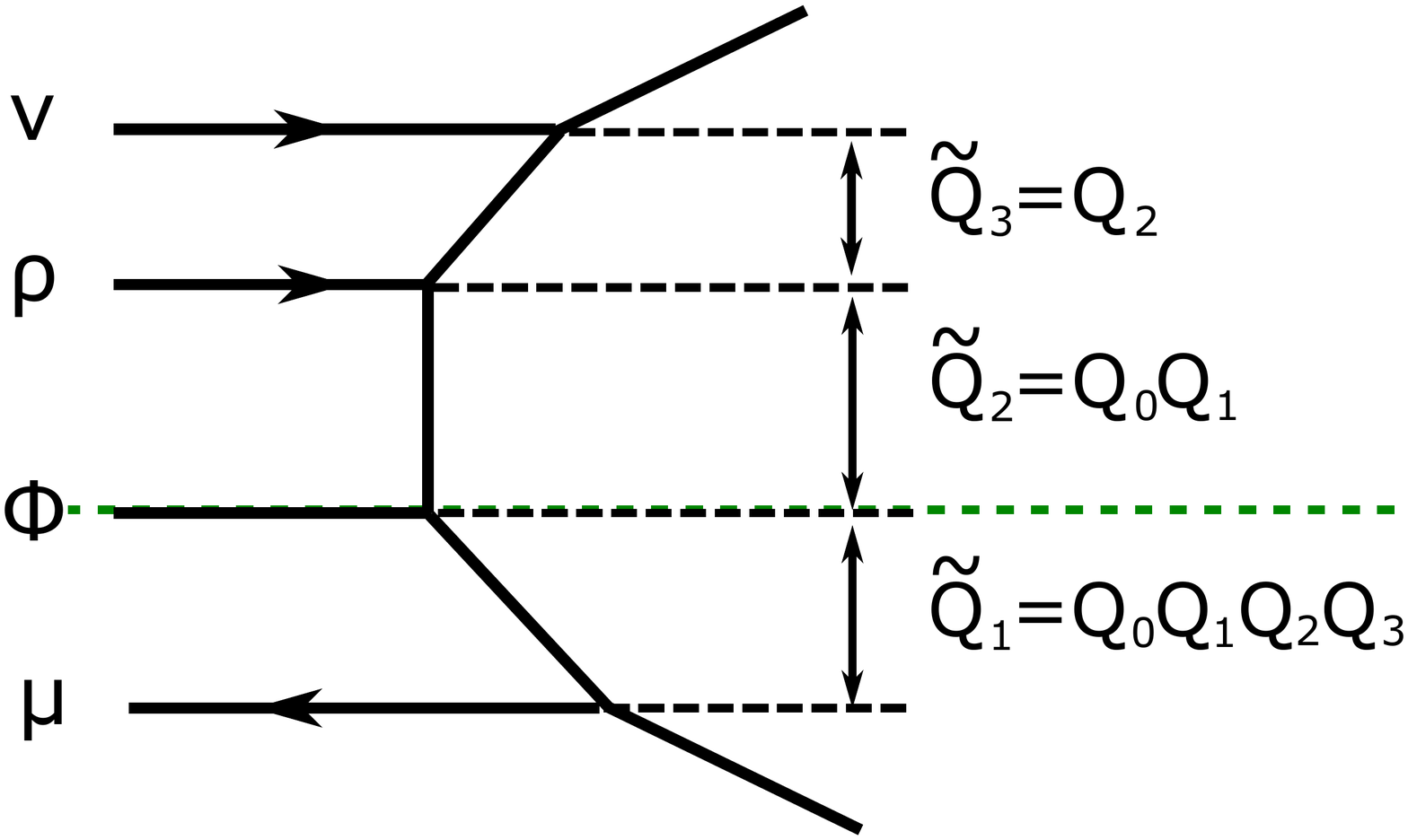} \label{fig:G2w1flvrTopb}}
%\caption{Right strip}
%\end{minipage}
\caption{(a): The left strip. (b): The right strip. }
\label{fig:G2w1flvrTop}
\end{figure}

In order to apply the topological vertex to the 5-brane web diagram in Figure \ref{fig:G2w1flvrfloppedb}, we first divide the diagram into the left strip and the right strip and assign the K$\ddot{\text{a}}$hler parameters as in Figure \ref{fig:G2w1flvrTop}.
%%%%%%%%%%%%%%%%%%%%%%%%%%%% 
%%%%%%%%%%%%%%%%%%%%%%%%%%%%
The diagram in Figure \ref{fig:G2w1flvrTopa} gives the left strip and the one in Figure \ref{fig:G2w1flvrTopb} gives the right strip. Instead of using the upper half plane with an O5-plane, we write the diagram as a strip diagram by using the mirror image as discussed in section \ref{sec:vertexO5tilde}. Then, the application of the topological vertex to the diagram in Figure \ref{fig:G2w1flvrTop} yields the result 
\begin{align}\label{leftstrip}
&Z^{\rm left strip} (\{ Q_i \}; \mu,\nu,\rho,\sigma)  = 
g^{\frac{1}{2} ( || \mu ||^2 + || \nu ||^2 + || \rho ||^2) + || \sigma^t ||^2}
Z_{\mu} (g) Z_{\nu} (g) Z_{\rho} (g) Z_{\sigma} (g){}^2
\cr
&\frac{ 
\R_{\rho \sigma^t} (Q_1) 
%R_{\emptyset \sigma}(Q_0) 
\R_{\nu \sigma^t} (Q_{1,2})
\R_{\rho \emptyset} (Q_{0,1})
\R_{\sigma \sigma} (Q_{0,0})
\R_{\mu \sigma^t} (Q_{1,2,3})
\R_{\nu \emptyset} (Q_{0,1,2} )
\R_{\mu \emptyset} (Q_{0,1,2,3})
}{ 
\R_{\mu \nu^t}(Q_3) 
\R_{\nu \rho^t} (Q_2) 
%R_{\sigma \emptyset} (Q_0)
\R_{\mu \rho^t} (Q_{2,3})
\R_{\rho \sigma} (Q_{0,0,1} )
\R_{\nu \sigma} (Q_{0,0,1,2})
\R_{\mu \sigma} (Q_{0,0,1,2,3})
},
\end{align}
where we denote
\begin{align}
Q_{i,j}=Q_i Q_j, \qquad Q_{i,j,k}=Q_i Q_j Q_k, \qquad Q_{i,j,k,\ell}=Q_i Q_j Q_k Q_{\ell}, \quad \cdots
\end{align}
for the left strip. 
As for the right strip, we obtain
\begin{align}\label{rightstrip}
&Z^{\rm right strip}  (\{ \tilde{Q}_i \}; \mu,\nu,\rho) =  
g^{\frac{1}{2} ( || \mu ||^2 + || \nu^t ||^2 + || \rho^t ||^2) }
Z_{\mu} (g) Z_{\nu} (g) Z_{\rho} (g) 
\cr
&\frac{1}{
\R_{\nu \rho^t}(\tilde{Q}_3)
\R_{\rho \emptyset}(\tilde{Q}_{2})
\R_{\emptyset \mu}(\tilde{Q}_{1})
\R_{\nu \emptyset} (\tilde{Q}_{2,3})
\R_{\rho \mu} (\tilde{Q}_{1,2} )
\R_{\nu \mu} (\tilde{Q}_{1,2,3})
},
\end{align}
with
\begin{align}
\tilde{Q}_1 = Q_0 Q_1 Q_2 Q_3, \qquad
\tilde{Q}_2 = Q_0 Q_1, \qquad
\tilde{Q}_3 = Q_2.
\end{align}
%\begin{align}
%Q_4=Q_0 Q_1, \qquad Q_5 = Q_0 Q_1 Q_2 Q_3
%\end{align}
Note also that the web diagram gives a constraint 
\be
Q_3 = Q_0Q_1. \label{Q3}
\ee
The full partition function is given by gluing the left strip \eqref{leftstrip} to the right strip \eqref{rightstrip} with framing factors and the final result is given by 
%The partition function is
\begin{align}\label{G2part1}
\tilde{Z}_{G_2, N_f=1} = &\sum_{\mu,\nu,\rho,\sigma} 
(+Q_B Q_3{}^2)^{|\mu|} (-Q_B)^{|\nu|+|\rho|} (+Q_0)^{|\sigma|} 
f_{\mu}{}^{3} f_{\nu}{}^{} f_{\rho}{}^{-1} f_{\sigma}
\cr
&Z^{\rm left strip} (\{ Q_i \}; \mu,\nu,\rho,\sigma) 
Z^{\rm right strip}  (\{ \tilde{Q}_i \}; \mu,\nu,\rho),
\end{align}
where $Q_B$ is the K$\ddot{\text{a}}$hler parameter for the bottom color D5-brane in Figure \ref{fig:G2w1flvrflopped}. Since the 5-brane web in Figure \ref{fig:G2w1flvrfloppedb} has parallel external legs. We need to remove an extra factor \cite{Bergman:2013ala, Bao:2013pwa, Hayashi:2013qwa, Bergman:2013aca, Hwang:2014uwa}. Hence the partition function after removing the extra factor is 
\be
Z_{G_2, N_f=1} = \frac{\tilde{Z}_{G_2, N_f=1}}{\M(Q_0Q_1^3Q_2^2Q_3)}. \label{G2part}
\ee
We claim that the partition function \eqref{G2part} is the partition function of the $G_2$ gauge theory with one flavor and a singlet.

Since the partition function \eqref{G2part} is still written by the K$\ddot{\text{a}}$hler parameter of the 5-brane web in Figure \ref{fig:G2w1flvrTop}, we will rewrite it by the gauge theory parameters. First the $G_2$ gauge theory should have two Coulomb branch moduli $A_1, A_2$ %\footnote{More precisely $A_1, A_2$ are $A_1 = \exp(-a_1), A_2 = \exp(-a_2)$ and $a_1, a_2$ are the Coulomb branch moduli. All the gauge theory parameters are exponentiated. } 
and one instanton fugacity $q$. There is also a mass parameter $M$ for the flavor. Since the one flavor and the singlet both come from the same spinor matter of the $SO(7)$ gauge theory, the mass parameter for the singlet is the same as the mass parameter for the one flavor.

The Coulomb branch parameterization is essentially the same as the parameterization in the case of the pure $G_2$ gauge theory. Namely, the Coulomb branch moduli are the height of the color D5-branes. Hence we impose
\be
Q_0Q_1 = A_1, \quad Q_0Q_1Q_2 = A_2. \label{CB}
\ee
Note that the web diagram in Figure \ref{fig:G2w1flvrfloppedb} has parallel $(1, -1)$ 5-branes. The parallel $(1, -1)$ 5-branes imply an $SU(2)$ flavor symmetry for the one flavor of the $G_2$ gauge theory. Therefore, the distance between the two parallel $(1, -1)$ 5-branes is the mass parameter. This leads to a condition 
\be
Q_0Q_1^3Q_2^2Q_3 = M^2. \label{Mass}
\ee

The instanton fugacity can by computed similarly to the way when we calculated the instanton fugacity for the 5d pure $G_2$ gauge theory. We extrapolate the leftmost external $(1,-1)$ 5-brane and the external $(2, 1)$ 5-brane on the right to the position of the O5-plane in Figure \ref{fig:G2w1flvrfloppedb} and compute the length between the two extrapolated 5-branes on the O5-plane. In fact it turns out that we need to extrapolate the leftmost external $(1, -1)$ 5-brane instead of the other external $(1, -1)$ 5-brane. This can be justified by the comparison with the instanton fugacity obtained by another 5-brane web digram for the $G_2$ gauge theory with one flavor in Figure \ref{fig:G2w1flvra}. Then the instanton fugacity is given by  
\be
Q_BQ_0^{-1}Q_1Q_2^{-1}Q_3^{-1} = q. \label{instfugacity}
\ee
Hence the relations \eqref{CB}, \eqref{Mass}, \eqref{instfugacity} with the condition \eqref{Q3} yields
%Kahler parameters are give as
\begin{align}\label{gaugeparameters}
Q_0 = M^{-1} A_1 A_2, \quad
Q_1 = M A_2{}^{-1}, \quad
Q_2 = A_1{}^{-1} A_2, \quad
Q_B = q M{}^{-2} A_1 A_2{}^3.
\end{align}
Namely, we can compare the partition function \eqref{G2part} with the parameterization \eqref{gaugeparameters} with the Nekrasov partition function of the $G_2$ gauge theory with one flavor and a singlet.

\paragraph{Perturbative part.}
Since $Q_B$ is proportional to the instanton fugacity $q$, the perturbative part of the partition function can be obtained by setting $Q_B \to 0$. The limit corresponds to the restriction $\mu = \emptyset, \nu = \emptyset, \rho=\emptyset$. In this case, the sum of Young diagrams for the right strip partition function disappears and it reduces to 
\bea
Z^{\text{right}}_{\text{pert}} &=&\M(\tilde{Q}_3)
\M(\tilde{Q}_{2})
\M(\tilde{Q}_{1})
\M(\tilde{Q}_{2,3})
\M(\tilde{Q}_{1,2} )
\M(\tilde{Q}_{1,2,3})\nn\\
&=&\M(A_1^{-1}A_2)\M(A_1)\M(A_1A_2)\M(A_2)\M(A_1^2A_2)\M(A_1A_2^2).
\eea
On the other hand, one Young diagram summation remains in the left strip partition function and it becomes
\bea
Z^{\text{left}}_{\text{pert}} &=& Z^{\text{left}}_{\text{pert 1}}Z^{\text{left}}_{\text{pert 2}},\nn\\
Z^{\text{left}}_{\text{pert 1}}&=&\frac{ 
\M(Q_3) 
\M(Q_2) 
%R_{\sigma \emptyset} (Q_0)
\M(Q_{2,3})
\M(Q_{0,0,1} )
\M(Q_{0,0,1,2})
\M(Q_{0,0,1,2,3})}{ 
\M(Q_1) 
%R_{\emptyset \sigma}(Q_0) 
\M(Q_{1,2})
\M(Q_{0,1})
\M(Q_{0,0})
\M(Q_{1,2,3})
\M(Q_{0,1,2} )
\M(Q_{0,1,2,3})},\label{leftpertproduct}\\
Z^{\text{left}}_{\text{pert 2}}&=&\sum_{\sigma}Q_0^{|\sigma|}f_{\sigma}g^{||\sigma^t||^2}Z_{\sigma}^2(g)\frac{ 
\N_{\emptyset \sigma^t} (Q_1) 
%R_{\emptyset \sigma}(Q_0) 
\N_{\emptyset \sigma^t} (Q_{1,2})
%R_{\rho, \emptyset} (Q_{0,1})
\N_{\sigma^t \sigma} (Q_{0,0})
\N_{\emptyset \sigma^t} (Q_{1,2,3})
%R_{\nu \emptyset} (Q_{0,1,2} )
%R_{\mu, \emptyset} (Q_{0,1,2,3})
}{ 
%R_{\mu \nu^t}(Q_3) 
%R_{\nu \rho^t} (Q_2) 
%R_{\sigma \emptyset} (Q_0)
%R_{\nu \rho^t} (Q_{2,3})
\N_{\emptyset \sigma} (Q_{0,0,1} )
\N_{\emptyset \sigma} (Q_{0,0,1,2})
\N_{\emptyset \sigma} (Q_{0,0,1,2,3})
}.\label{leftpertsum}
\eea
By the explicit computation of the Young diagram summation of \eqref{leftpertsum}, we argue that the partition function \eqref{leftpertsum} becomes
\bea
Z^{\text{left}}_{\text{pert 2}}&=&\frac{
\M(Q_{0,0})\M(Q_{0,1})\M(Q_{0,1,2})\M(Q_{0,0,1,1,2})\M(Q_{0,1,2,3})\M(Q_{0,0,1,1,2,3})}{\M(Q_0)\M(Q_{0,0,1})\M(Q_{0,0,1,2})\M(Q_{0,1,1,2})\M(Q_{0,0,1,2,3})\M(Q_{0,1,1,2,3})}\nn\\
&&\times\frac{
\M(Q_{0,0,1,1,2,2,3})\M(Q_{0,1,1,1,2,2,3})
}{
\M(Q_{0,1,1,2,2,3})\M(Q_{0,0,1,1,1,2,3})
}.\label{leftpertsumup}
\eea
We checked the equality \eqref{leftpertsumup} until the order $Q_0^6$. Then combining \eqref{leftpertsumup} with \eqref{leftpertproduct} yields
\bea
Z^{\text{left}}_{\text{pert}}&=&\frac{
\M(Q_3)\M(Q_2)\M(Q_{2,3})\M(Q_{0,0,1,1,2})\M(Q_{0,0,1,1,2,3})\M(Q_{0,0,1,1,2,2,3})
\M(Q_{0,1,1,1,2,2,3})
}{
\M(Q_{1})\M(Q_{1,2})\M(Q_{1,2,3})\M(Q_0)\M(Q_{0,1,1,2})\M(Q_{0,1,1,2,3})\M(Q_{0,1,1,2,2,3})\M(Q_{0,0,1,1,1,2,2,3})
}\nn\\
&=&\frac{
\M(A_1)\M(A_1^{-1}A_2)\M(A_2)\M(A_1A_2)\M(A_1^2A_2)\M(A_1A_2^2)\M(M^2)
}{\M(MA_2^{-1})\M(MA_1^{-1})\M(M)\M(M^{-1}A_1A_2)\M(M)\M(MA_1)\M(MA_2)\M(MA_1A_2)
}.\nn\\
\eea
Therefore, the perturbative part of the partition function of \eqref{G2part} is
\bea
Z_{\text{pert}} &=& \frac{Z^{\text{left}}_{\text{pert}}Z^{\text{right}}_{\text{pert}}}{\M(Q_0Q_1^3Q_2^2Q_3)}\nn\\
&=&\frac{
\M(A_1)^2\M(A_1^{-1}A_2)^2\M(A_2)^2\M(A_1A_2)^2\M(A_1^2A_2)^2\M(A_1A_2^2)^2
}{\M(MA_2^{-1})\M(MA_1^{-1})\M(M)\M(M^{-1}A_1A_2)\M(M)\M(MA_1)\M(MA_2)\M(MA_1A_2)
}.\nn\\\label{pertpart}
\eea
The perturbative partition function then consists of three factors as follows:
\bea
Z_{\text{pert}} = Z_{\text{pert}}^{\text{vector}} \cdot Z_{\text{pert}}^{\text{fund}} \cdot Z_{\text{pert}}^{\text{singlet}},
\eea
where
\bea
Z_{\text{pert}}^{\text{vector}} &=&\M(A_1)^2\M(A_1^{-1}A_2)^2\M(A_2)^2\M(A_1A_2)^2\M(A_1^2A_2)^2\M(A_1A_2^2)^2,\label{pertvector}\\
Z_{\text{pert}}^{\text{fund}} &=&\frac{1}{\M(MA_2^{-1})\M(MA_1^{-1})\M(M^{-1}A_1A_2)\M(M)\M(MA_1)\M(MA_2)\M(MA_1A_2)},\nn\\\label{pertflavor}\\
Z_{\text{pert}}^{\text{singlet}} &=&\frac{1}{\M(M)}.\label{pertsinglet}
\eea
\eqref{pertvector}, \eqref{pertflavor} and \eqref{pertsinglet} are exactly equal to the perturbative part of the partition functions of the $G_2$ vector multiplets, hypermultiplets in the fundamental representation of $G_2$ and a singlet hypermultiplet up to the Cartan parts and flop transitions as in the case of the pure $G_2$ gauge theory. 

\paragraph{Instanton part.}

Let us move on to the instanton part of the $G_2$ partition function \eqref{G2part}. The instanton part is obtained after removing the perturbative part obtained in \eqref{pertpart}. Namely the instanton partition function of the $G_2$ gauge theory with one flavor is 
\be
Z_{G_2, N_f=1, \text{inst}} = \frac{Z_{G_2, N_f=1}}{Z_{\text{pert}}}, \label{instpart}
\ee
where $Z_{G_2, N_f=1}$ is \eqref{G2part} and $Z_{\text{pert}}$ is \eqref{pertpart}. The order $q^k$ of \eqref{instpart} gives the $k$-instanton contribution of the $G_2$ gauge theory with one flavor. 

Since an explicit form of the instanton partition function of the $G_2$ gauge theory with one flavor is not known, we compare the expression \eqref{instpart} with the instanton partition function of the pure $G_2$ gauge theory after decoupling the flavor. After sending $M \to 0$, then the 5-brane web diagram will become the one for the pure $G_2$ gauge theory in Figure \ref{fig:SO7wspinorfloppede}. Hence, $Q_0Q_1, Q_2, Q_B$ are still finite in the limit and the instanton fugacity for the pure $G_2$ gauge theory is given by $q' = qM^{-2}$. Since the limit $M \to 0$ should be compatible with the expansion of $Q_0$ in \eqref{instpart} for the comparison, we will take the following steps. At each order of $q'$, we first rewrite \eqref{instpart} by $M, A_1$ and $Q_2$ and then expand it by $M$ and $A_1$. In this case, the term $Q_B^kQ_0^aQ_1^b$ becomes
\be
Q_B^kQ_0^aQ_1^b = q'^k M^{-a+b}A_1^{2a-b+4k}Q_2^{a-b+3k} \label{order.comparison}
\ee
The term which survives after the limit $M \to 0$ satisfies $a=b$.\footnote{%It turns out that an explicit evaluation of \eqref{instpart} does not have the negative order terms in $M$ until $a=6, k=2$.
It turns out that an explicit evaluation of \eqref{instpart} shows that there are no terms in $M$ of the negative powers until $a=6, k=2$.} In order to obtain an reliable expression until the order $q'^kA_1^{a+4k}$, we need to sum up the Young diagrams in \eqref{instpart} until the order $Q_0^aQ_B^k$. Then we can compare the result which is expanded by $A_1$ with the gauge theory computation for the pure $G_2$ gauge theory. We checked that the partition function \eqref{instpart} after applying the limit $M \to 0$ indeed agrees with the gauge theory result until $a=6$ for the one-instanton and the two-instanton parts of the Nekrasov partition function of the pure $G_2$ gauge theory obtained in  \cite{Benvenuti:2010pq, Keller:2011ek, Hanany:2012dm, Keller:2012da, Cremonesi:2014xha}. This gives an evidence that we obtain the correct the partition function \eqref{instpart} for the $G_2$ gauge theory with one flavor.

%====================================================================
\bigskip
\section{Conclusion}
\label{sec:conclusion}
In this paper, we studied 5d $\mathcal{N}=1$ $G_2$ gauge theories from 5-brane web diagrams with an O5-plane in type IIB string theory. 
The result that we obtained is summarized as follows:
\begin{itemize}
	\item Two equivalent types of 5-brane webs for 5d pure $G_2$ gauge theory are presented, in Figures \ref{fig:SO7wspinorfloppede}, \ref{fig:pureG2}, and in Figure \ref{fig:G2pure2}. Webs for 5d $G_2$ gauge theory with matter are also discussed in subsection \ref{sec:addingF}.
	\item  Based on the 5-brane webs that we obtained, we computed the partition functions of the BPS spectrum. In particular, the partition function for the pure $G_2$ theory is given in \eqref{part.pureG2}, and the $G_2$ theory with one flavor is given in \eqref{G2part}.
\end{itemize}

For the first type of pure $G_2$ diagram, corresponding to Figures \ref{fig:SO7wspinorfloppede} and  \ref{fig:pureG2}, we started from a conventional 5-brane web diagram for the 5d $SO(7)$ gauge theory with a hypermultiplet in the spinor representation. Applying a generalized flop transition such as in Figure \ref{fig:flopO5tilde2}, we perform a Higgsing associated to the spinor matter to obtain a 5-brane web diagram for the 5d pure $G_2$ gauge theory. %Depending on where the fractional 7-branes are located at the far infinity along the orientifold, one may get Figure \ref{fig:SO7wspinorfloppede} which is of an $\widetilde{\rm O5}^-$-plane or Figure \ref{fig:pureG2}, which is of an {O5}$^-$-plane. 
The detail discussion regarding two diagrams is presented in section \ref{sec:G2fromO5tilde}. 

For the second type of pure $G_2$ diagram, corresponding to Figure \ref{fig:G2pure2}, we started with a 5-brane configuration for 5d $SO(8)$ gauge theory with a hypermultiplet in the spinor representation and a hypermultiplet in the conjugate spinor representation. We made use of triality of $SO(8)$ gauge theory so that one can interpret this configuration as 5d $SO(8)$ gauge theory with a vector and a spinor rather than  a spinor and a conjugate spinor.  With this, we performed a successive Higgsing associated with the vector and the spinor %to obtain yet another 5-brane web for 5d $SO(7)$ gauge theory with a spinor and performed one more Higgsing with the spinor 
to obtain another 5-brane web for pure $G_2$ theory, %which yields the second type diagram in Figure \ref{fig:G2pure2}, 
which is of an ${\text{O5}}$-plane only.

Our results are tested in two different ways:
\begin{itemize}
\item Based on the webs, we computed the tension of monopole strings and compared it with the tension of monopole strings obtained from derivatives of the prepotential of 5d $G_2$ gauge theories. They completely agree as shown in \eqref{Eq:G2monotension1} and \eqref{Eq:G2monotension2}.
\item For pure $G_2$ case, we explicitly compared both the perturbative and instanton parts of the obtained pure $G_2$ partition function with the literature. For one flavor case, however, as an explicit expression for the partition function for $G_2$ theory with one flavor is not known, we instead checked an important consistency. Namely, the flavor decoupling limit of the instanton part for the $G_2$ gauge theory with one flavor reproduces that for the pure $G_2$ gauge theory. % while for the partition function the $G_2$ gauge theory with one flavor, we showed that the flavor decoupling limit of the instanton part reproduces that for the pure $G_2$ gauge theory.
\end{itemize}

Regarding the 5-brane web for pure $G_2$ gauge theory, it is worth noting that the 5-brane diagram of the second type %realized without an $\widetilde{\text{O5}}$-plane (of the second type) 
has an interesting feature. The diagram is almost identical to the diagram for the pure $SU(4)$ gauge theory of Chern-Simons level zero with a restriction of the Coulomb branch moduli. From the point of view of the topological string computation, the only difference between this $G_2$ gauge theory and the pure $SU(4)$ gauge theory is whether a Young diagram is transposed or not. This means that the (unrefined) partition function computations only differ by whether the power of the framing factor is $+1$ or $-1$, and also by whether an associated sign factor is included or not. 
It would be interesting to see how this feature is modified when one fully refines the $G_2$ partition functions \cite{HKLY}.

As the topological vertex formalism is now applicable to a large class of 5-brane webs with an O5-plane or an $\widetilde{\text{O5}}$-plane, it would be interesting to see whether one can use this method to compute partition functions of 5d $SO(N)$ gauge theories with spinors \cite{HKLY}. Another direction to pursue is to obtain  5d Seiberg-Witten curves dictating M5-brane configurations based on their dual diagrams \cite{HKLY}. %that, as we have obtained 5-brane webs for 5d $G_2$ gauge theories, it would be also interesting to obtain 5d Seiberg-Witten curves dictating M5-brane configurations based on their dual diagrams \cite{HKLY}.

\acknowledgments
We thank Amihay Hanany, Hee-Cheol Kim, Seok Kim, Jaewon Song, and Gabi Zafrir for useful discussions. We also thank the authors of \cite{Kim:2018gjo} for kindly agreeing to coordinate our submission. SSK is supported by the UESTC Research Grant A03017023801317.  KL is supported in part  by the National Research Foundation of Korea Grant NRF-2017R1D1A1B06034369 and also by the National Science Foundation under Grant No. NSF PHY11-25915. FY is supported in part by Israel Science Foundation under Grant No. 352/13. We would like to thank International Workshop on Superconformal Theories 2017 (SCFT2017) in Chengdu. HH would like to thank Harvard University for kind hospitality during a part of the work. KL would like to thank KITP, UCSB for the hospitality. 

\bigskip

\appendix 

%====================================================================

%\input{secappendix.tex}

%====================================================================

\bigskip

%%%%%%%%%%%%%%%%%%%%%%%%%%%%%%%%%%%%%%%%%%%%%%%%%%
\bibliographystyle{JHEP}
\bibliography{ref}
\end{document}